\documentclass{aa}  

\usepackage{natbib}
\usepackage{graphicx}
\usepackage{txfonts}
\usepackage{multirow}
\usepackage{lipsum}
\PassOptionsToPackage{hyphens}{url}
\usepackage[colorlinks, citecolor=blue]{hyperref}

\newcommand{\kms}{km\,${\rm s}^{-1}$}

\begin{document} 
  \title{The observed multiplicity properties of B-type stars in the Galactic young open cluster NGC 6231\thanks{Based on observations collected at the ESO Paranal observatory under ESO program 099.D-0895 and 0101.D-0163.}}
  \author{G. Banyard\inst{1}
     \and
           H. Sana\inst{1}\fnmsep
     \and
          L. Mahy\inst{1,2}
     \and 
          J. Bodensteiner\inst{1}
     \and 
          J.~I. Villase\~nor\inst{3} 
     \and 
          C. J. Evans\inst{4}
          }

  \institute{Institute for Astronomy, KU Leuven,
             Celestijnenlaan 200 D, 3001, Leuven, Belgium
             \and 
             Royal Observatory of Belgium, Avenue Circulaire 3, B-1180 Brussel, Belgium
             \and
             Institute for Astronomy, University of Edinburgh, Royal Observatory, Blackford Hill, Edinburgh, EH9 3HJ, UK
             \and 
             UK Astronomy Technology Centre, Royal Observatory, Blackford Hill, Edinburgh, EH9 3HJ}

  \date{Received April 09, 2021; accepted August 16, 2021}

  \abstract
   {It is well known that massive O-stars are frequently (if not always) found in binary or higher-order multiple systems, but this fact has been less robustly investigated for the lower mass range of the massive stars, represented by the B-type stars. Obtaining the binary fraction and orbital parameter distributions of B-type stars is crucial to understand the impact of multiplicity on the archetypal progenitor of core-collapse supernovae as well as to properly investigate formation channels for gravitational wave progenitors.}
   {This work aims to characterise the multiplicity of the B-star population of the young open cluster NGC 6231 through multi-epoch optical spectroscopy of 80 B-type stars.}
   {We analyse 31 FLAMES/GIRAFFE observations of 80 B-type stars, monitoring their radial velocities (RVs) and performing a least-squares spectral analysis (Lomb-Scargle) to search for periodicity in those stars with statistically significant variability in their RVs.}
   {We constrain an observed spectroscopic binary fraction of $33\pm5$\% for the B-type stars of NGC 6231, with a first order bias-correction giving a true spectroscopic binary fraction of $52\pm8\%$. Out of 27 B-type binary candidates, we obtained orbital solutions for 20 systems: 15 single-lined (SB1) and 5 double-lined spectroscopic binaries (SB2s). We present these orbital solutions and the orbital parameter distributions associated with them.}
   {Our results indicate that Galactic B-type stars are less frequently found in binary systems than their more massive O-type counterparts, but their orbital properties generally resemble those of B- and O-type stars in both the Galaxy and Large Magellanic Cloud.}


\maketitle


\section{Introduction}
\label{s:intro}
The importance of massive stars in driving the evolution of galaxies throughout the history of the Universe has been well documented. As an example of this, their core collapse supernovae (CCSNe), through their imparted momentum and their release of processed material, can dramatically alter the dynamical and chemical evolution of their host galaxies and the star formation rates within them \citep{hopkins_galaxies_2014}. \par
 Increasingly well established observational evidence shows that the majority of massive stars are part of multiple systems that are close enough to interact during their lifetime \citep{kiminki_radial_2007,mason_high_2009,sana_binary_2012,sana_southern_2014,kobulnicky_toward_2014}. These observational studies challenge the accepted predominance of the single star evolutionary channel and complicate the exact determination of the evolutionary status and final fates of these stars. \par 
 
 Establishing the multiplicity properties of large, statistically significant samples of newly born massive stars is consequently crucial to understand the initial conditions and properly compute the evolution of these objects  \citep{langer_rotation_2008,de_mink_rotational_2009}. This will also facilitate accurate predictions on the frequency, timing and nature of the varying end-of-life stages of these massive stars, including CCSNe and progenitor systems for gravitational wave (GW) sources like compact object mergers. For example, the ratio of hydrogen deficient Ib/c supernovae to type II supernovae is altered by an increased binary fraction as a higher binary fraction leads to fewer red supergiants (RSGs), more stripped stars and consequently more H-deficient supernovae \citep{eldridge_massive_2007}. \par
 
Assuming a Kroupa initial mass function \citep[IMF,][]{kroupa_galactic-field_2003}, the majority (70\%) of CCSNe and neutron stars (NS) are produced by stars born as early B-type stars (spectral type B0 to B3, i.e., stars with initial masses $8$\,<~M/M$_{\odot}\,<\,16$). Lower mass B-type stars  can also contribute to the population of CCSNe through merging into products greater than about 8 M$_{\odot}$. Similarly, B-type stars in multiple systems may further dominate the formation of double neutron star systems \citep{de_mink_merger_2015}. Deriving the binary properties of B-type stars therefore provides valuable constraints into the calculation of neutron star merger rates and the evolution of the main progenitors of CCSNe. \par
However, most of the recent progress in investigating the multiplicity fraction and the distribution of orbital parameters of massive stars has been achieved in the O-star domain (M $> 16$~M$_{\odot}$). Several spectroscopic \citep{sana_binary_2012,sana_vlt-flames_2013,kobulnicky_toward_2014,almeida_tarantula_2017} and interferometric \citep{mason_high_2009,sana_southern_2014,aldoretta_multiplicity_2015} studies of large samples of O stars in different environments have constrained the multiplicity fraction and orbital parameter distributions of these stars. Similar constraints are scarcer for B-type stars \citep{kobulnicky_toward_2014,dunstall_vlt-flames_2015,bodensteiner_young_2021,villasenor_b-type_2021}. \citet[in a follow up to \citealt{abt_binaries_1978}]{abt_frequency_1990}  performed a study of 109 (mostly) field B stars in the Galaxy between spectral types B2 and B5. The work was based on 20 obtained spectra of each star along with a literature search for information on visual companions. They found 32 spectroscopic and 49 visual binaries in their sample, leading to an observed spectroscopic binary fraction of 32\%, and a total observed binary fraction of 74\%. After estimating how many spectroscopic binaries were missing from the sample, they found a corrected spectroscopic binary fraction of 57\%. They also provisionally came to the conclusion that most or all binaries are formed in capture processes, though it must be noted that these results were not associated with a single cluster or association of stars.\par
\defcitealias{abt_frequency_1990}{A90}
NGC 6231 is a young open cluster situated in the Sco OB2 association in the Galaxy. Estimations of the cluster's age vary between 2 and 7 Myrs  and it has a well characterised star formation history \citep{sana_massive_2008-1,sung_initial_2013,kuhn_structure_2017}. It also has a rich B-type star population, and the cluster age is young enough so that these should be unevolved stars and potentially before the onset of the majority of interaction between binary companions. Previous RV measurements through spectroscopy of different subsamples of the B-type star population of NGC 6231 have also been reported \citep{levato_rotation_1983,raboud_binarity_1996,raboud_evolution_1998,garcia_high-mass_2001}. An early estimate of the observed spectroscopic B-star binary fraction (not including correction for observational biases)  was reported at 50\% \citep{raboud_binarity_1996}. This result was based on a sample of 56 B-type stars and from two observational epochs. A further study \citep{garcia_high-mass_2001} confirmed the presence of a high frequency of short period B-star binaries, but again with a relatively small sample (32 objects) and between two and nine spectra per star\footnote{Although, also see the discussion by \citeauthor{sana_massive_2003} regarding the robustness of the results from \citeauthor{garcia_high-mass_2001}}. 
X-ray studies of NGC\,6231 reported that 35-41\% of its B-type stars have likely pre-main sequence companions \citep{sana_xmm-newton_2006, kuhn_structure_2017}. Photometric studies also exist for some stars in this cluster. Identification and characterisation of pulsating stars in this population has been reported \citep{arentoft_search_2001,meingast_pulsating_2013-1} along with the discovery and characterisation of eclipsing binaries. These include three systems that have been studied in depth before, namely HD 152200 \citep{pozo_nunez_survey_2019-1}, V1208 Sco \citep{balona_mapping_1985,balona_ccd_1995} and V1293 Sco \citep{otero_new_2005-1}.

In this paper, we use 31 epochs of optical spectroscopy to investigate the multiplicity properties of the B-star population in NGC~6231. 
This paper is organised as follows. Section~\ref{s:data} describes our observations and data reduction, while Section~\ref{s:orbital} details our RV measurements, methods for binary detection and orbital fitting. Our results are discussed in Section~\ref{s:discuss} while Section~\ref{s:summary} presents a summary and our conclusions.


\section{Observations and data reduction}
\label{s:data}
\subsection{Target selection}
\label{ss:targetselection}

Our initial target list was mostly selected from a literature census compiled in the framework of a large XMM-Newton X-ray campaign towards NGC 6231 \citep{sana_xmm-newton_2006, sana_xmm-newton_2006-1} and relying mainly on photographic, photoelectric or CCD campaigns from the 1950s through to the early 2000s.

A total of 96 stars from the input list were observed, with magnitudes in the range  $7.6 < V < 13.7$ mag: 92 B-type stars, 3 O-type binaries with B-type companions (HD 152218, HD 152219 and CD-41 11042), and HD 326319, an A0 type star (which was a candidate guide star included due to the presence of spare fibres). The O+B systems and the A0 star are excluded from the binary statistics discussed in this work. Throughout all of the aforementioned works on binarity in NGC 6231, 27 out of the 92 B-star targets have been previously suggested to be in binary systems to varying degrees of certainty \citep[e.g.][]{levato_rotation_1983,raboud_binarity_1996,garcia_high-mass_2001,pozo_nunez_survey_2019-1}. The sample size is large enough to obtain a statistical (binomial) error of less than 6\% on the measured spectroscopic binary fraction.

\begin{figure}
\centering
\includegraphics[width=\hsize]{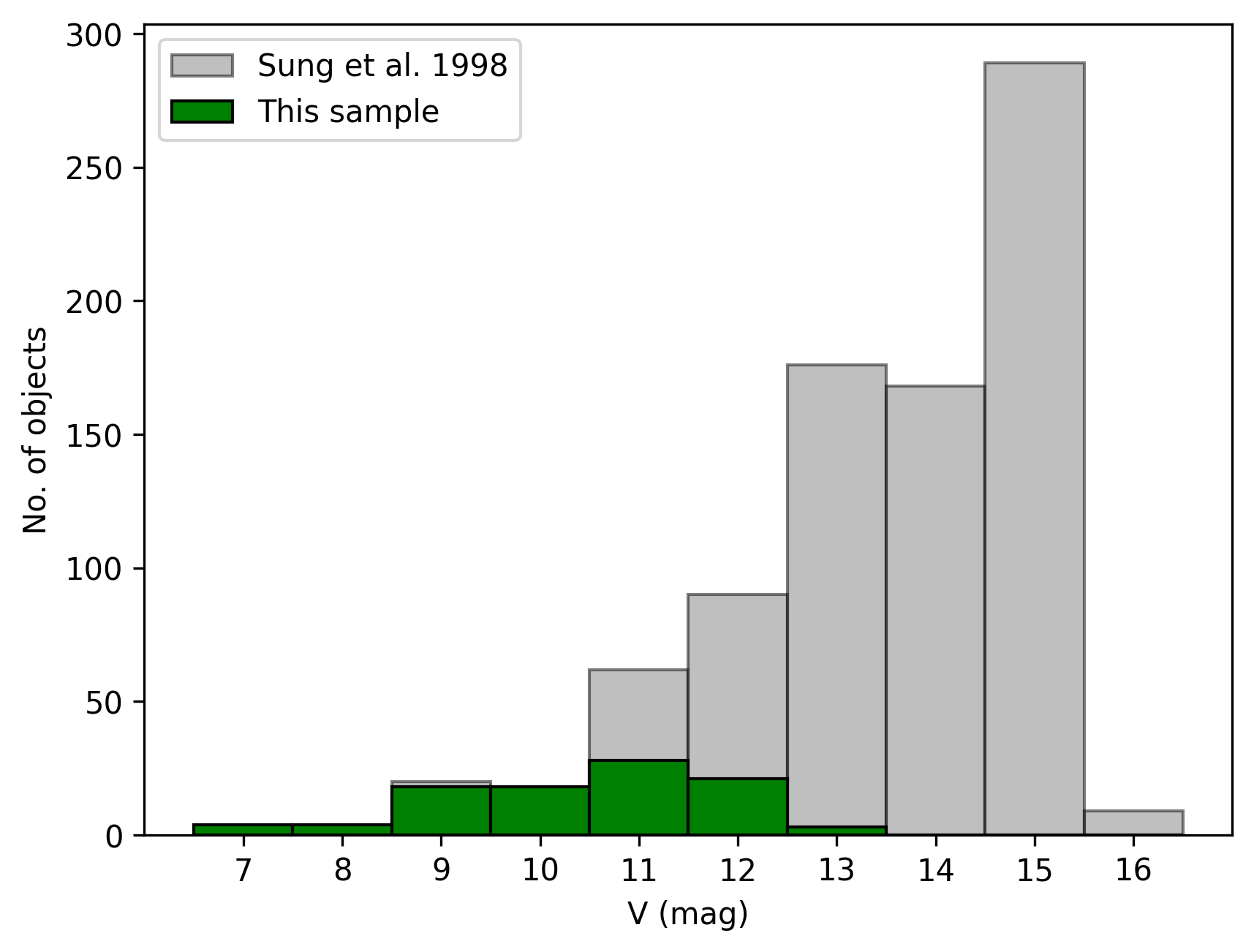}
    \caption{The brightness distribution in the $V$ band for the total sample (96 stars) compared to the brightness distribution of all targets in the photometry catalogue by \cite{sung_ubvri_1998} within a 12\farcm5 radius from the center of NGC 6231.
         }
    \label{FigCompleteness}
\end{figure}

\begin{figure}
\centering
\includegraphics[width=\hsize]{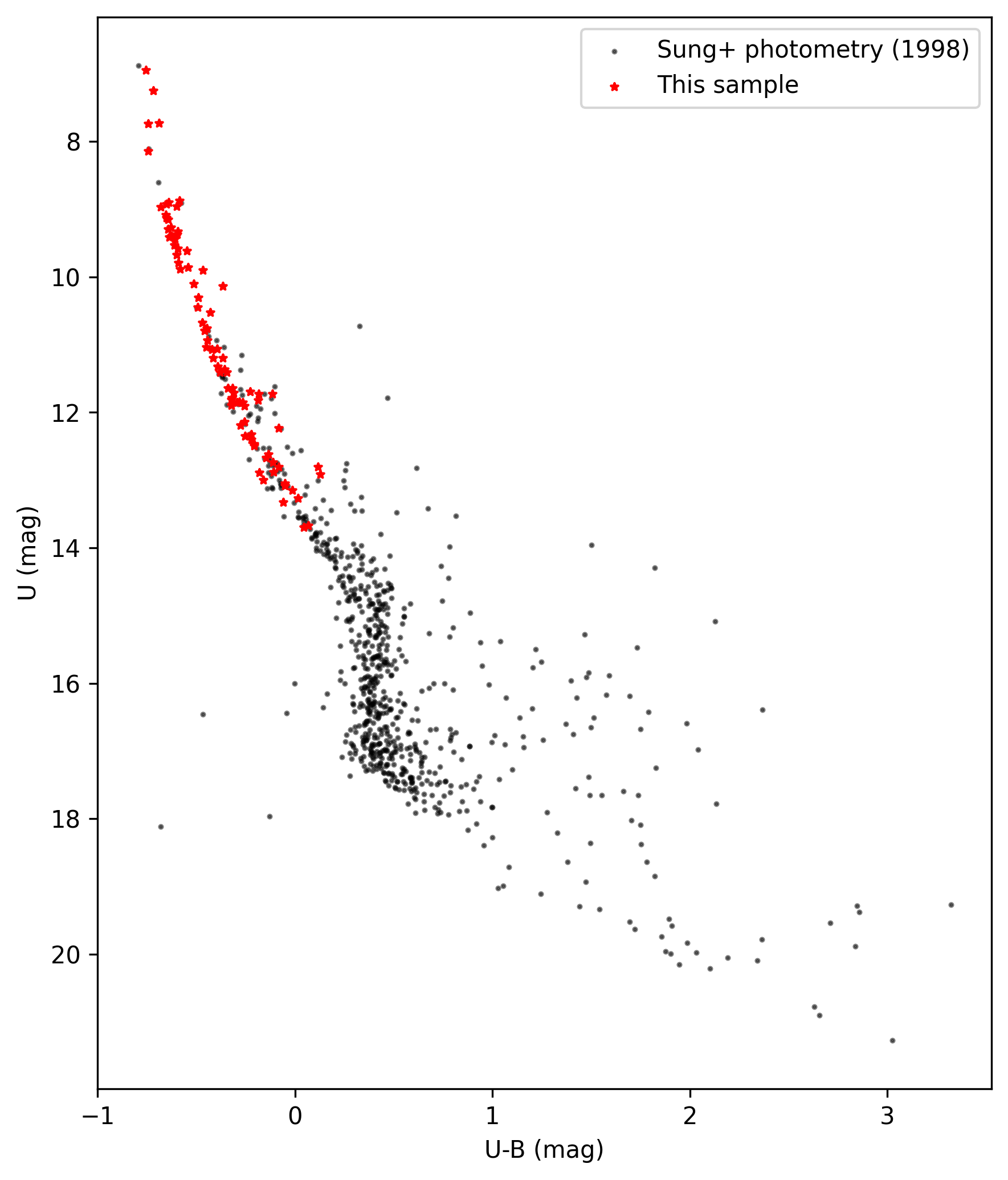}
    \caption{Colour-magnitude diagram for the stars in  the photometry catalogue by \cite{sung_ubvri_1998} within a 12\farcm5 radius from the center of NGC 6231. Targets in our sample that have an equivalent in the catalogue (82 out of 96 targets) are marked with red stars.
         }
    \label{FigCMD}
\end{figure}

To briefly review how complete our chosen sample is with regards to the B-type star population in NGC 6231, we compare the brightness distribution of the B-type stars in the sample observed by FLAMES to all the targets in the same observed window in the photometric catalogue of \citet[][see Fig.~\ref{FigCompleteness}]{sung_ubvri_1998}, in which there are matches to 82 of the 96 targets in the sample. It is clear that the sample is nearly complete down to $V = 10$ mag. Completeness drops to about 50\%\ in the range  $V = 10.5-11.5$ mag and to about 25\%\ between  $V = 11.5-12.5$ mag. Given the distance  \citep[$1710\substack{+130 \\ -110}$\,pc,][]{kuhn_kinematics_2019} and reddening to the cluster, one does not expect B-type stars to exhibit brightnesses below $V =$ 12.5 mag. The colour-magnitude diagram in Fig. \ref{FigCMD} shows that these targets lie along a main sequence as expected.
\par \par

However, the target list was based on previous spectral classifications in literature. consequently, it is possible that especially late-B stars and heavily extincted stars could be missing from the sample as they might have avoided previous spectral classification. As for the extinction itself, \cite{kuhn_kinematics_2019} state that it is relatively low compared to the other Galactic open clusters they study, and there appears to be a lack of strong variation in extinction within NGC 6231, so that no strong biases are expected. \cite{damiani_chandra_2016} further suggest that as the average reddening is similar for all stars in the cluster, and that the source of reddening is likely interstellar, and not circumstellar. However, \cite{rangwal_interstellar_2017} find a range of E(B-V) values indicating non-uniform extinction in the cluster (a 0.28 mag difference between the highest and lowest), and they find signatures of an abnormal extinction law. They do, however, state that there is no dependence of E(B-V) on either spectral type or luminosity, and they suggest that this implies little to no circumstellar material. This is discussed by \cite{sagar_study_1987}, where it is suggested that a dependence of E(B-V) on spectral type or luminosity is a result of the varying amount of circumstellar material with initial stellar mass (increasing with mass) and evolutionary phase (decreasing with age).
 \par

\subsubsection{Cluster membership of sample stars}
\label{sec:gaia}
\begin{figure}
\centering
\includegraphics[width=\hsize]{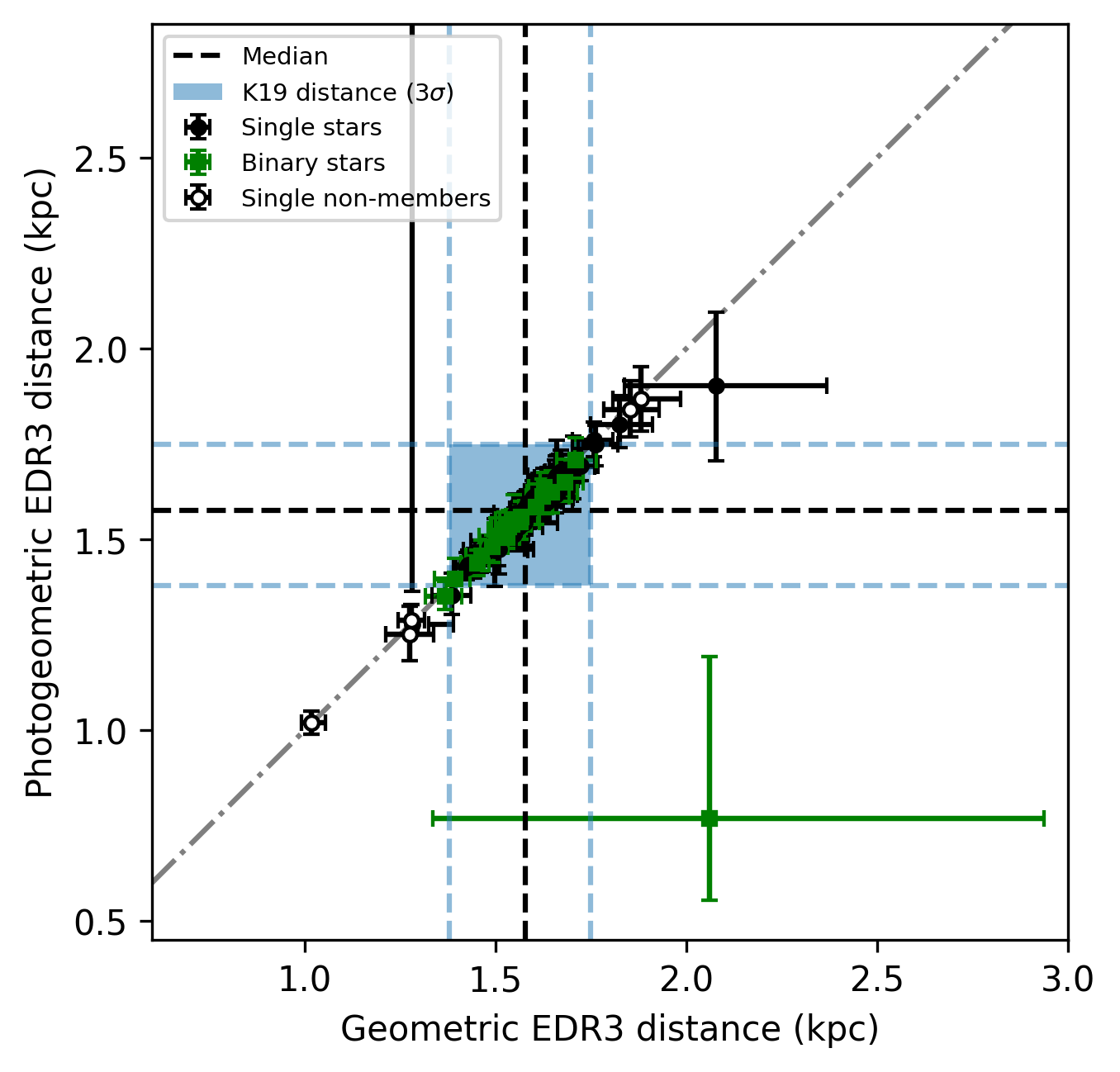}
    \caption{Geometric ($r_{geo}$) and photogeometric ($r_{pgeo}$) distances estimated by \cite{bailer-jones_estimating_2021} for the targets in NGC 6231, where the black dashed lines are the median geometric and photogeometric distances to the targets, and the blue dashed lines indicate the distance to the cluster found by \cite{kuhn_kinematics_2019} within $3\sigma$. Black points are stars found to be single, and green points are stars that are found to have binary solutions in this work. Open-faced points are stars that have an $r_{pgeo}$ that does not fall within $3\sigma$ of the quoted distance to NGC 6231.
         }
    \label{FigGaiaDist}
\end{figure}

We also verify that the selected targets are members of NGC 6231 using the estimated geometric and photogeometric distances from parallaxes in Gaia EDR3 \citep{bailer-jones_estimating_2021}. We compare this to the distance to the cluster found by \cite{kuhn_kinematics_2019}, which is $1710\substack{+13 \\ -110}\,$pc. We find Gaia distances for all but one of the stars, namely NGC 6231 172. The asymmetric uncertainties on the distances to the stars are the 16th and 84th percentile of the distance posteriors. The estimated distances can be seen in Fig. \ref{FigGaiaDist}. \par

The median geometric and photogeometric distances of the targets are 1579 and 1576 pc respectively, just out of a $1\sigma$ agreement with the \cite{kuhn_kinematics_2019} distance, but in excellent afreement with the distance of $1528\substack{+117 \\ -109}\,$pc from the eclipsing SB2 system CD-41 11042 \citep{sana_massive_2005}. The geometric and photogeometric distances are generally consistent with each other. The most extreme outlier in distance is NGC 6231 723, with by far the biggest discrepancy between the geometric and photometric distances along with the largest uncertainties ($2060\substack{+880 \\ -720}$ and $770\substack{+130 \\ -210}\,$pc respectively). It is also identified as an SB1 in this work, so erroneous values from poor fitting are to be expected, also indicated by its high RUWE value \citep{2016A&A...595A...1G,2021A&A...649A...1G} of over 23. The RUWE value corresponds to the quality of the astrometric fit of the star, and an astrometric fit of poor quality (typically with a RUWE greater than 1.4) can indicate the presence of one or more companions. As a result, we still consider this target a cluster member in this paper. In Table \ref{tab:nonmembers}, we list the 5 objects that we reject from the sample as non-members, as their photogeometric distances (which we assume are more accurate) are not within a $3\sigma$ agreement of the \cite{kuhn_kinematics_2019} distance to the cluster. One of these is HD 326319, the candidate guide star in the target list, so its status as a non-member of this cluster is unsurprising. \par

\begin{figure}
\centering
\includegraphics[width=\hsize]{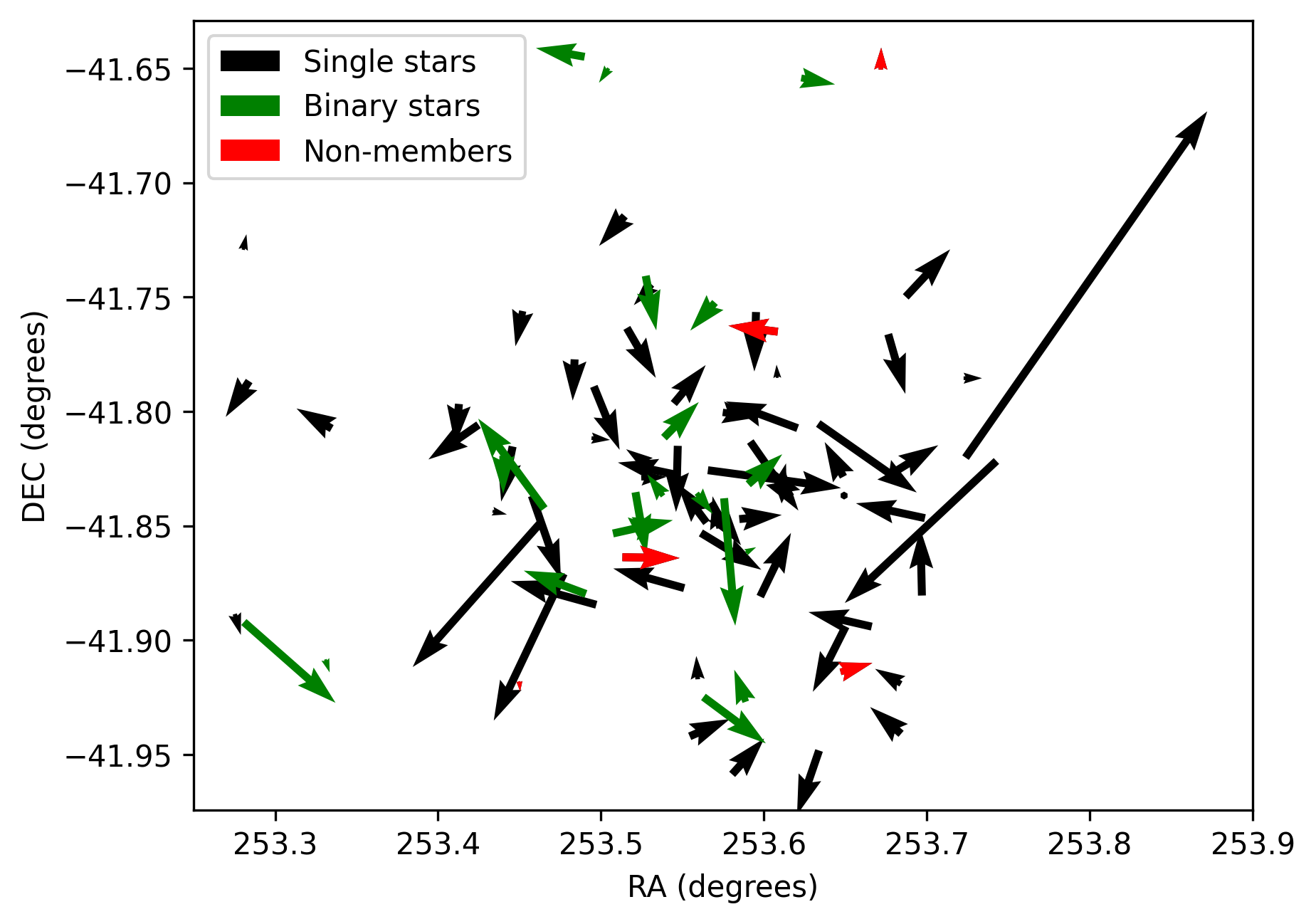}
    \caption{Gaia EDR3 median-subtracted proper motions for the B stars in NGC 6231, where the black arrows are the single stars, green arrows are the binary systems identified in this paper and the red arrows are the stars we regard as non-members due to their photogeometric distance to the cluster not being within $3\sigma$ to the distance found by \cite{kuhn_kinematics_2019}.
         }
    \label{FigGaiaPM}
\end{figure}

We also show proper motions retrieved from Gaia EDR3 (Fig. \ref{FigGaiaPM}, with the median proper motion of the sample subtracted from the stars). Proper motions seem aligned and of similar magnitude with the whole sample. The only large outlier is NGC 6231 147, a RV variable with no significant periodic signal and a peak-to-peak RV amplitude of 16.8 \kms. It is possible that this is an undetected, likely long-period binary, which could lead to erroneous proper motion measurements. Its reported (photo)geometric distance also suggests that it is a cluster member. The rest of the detected binaries do not appear to be outliers in the sample with respect to their proper motions. \par


\begin{table}
\centering
\caption{Stars in the original target list that we identify as non-members of NGC 6231 from the Gaia EDR3 results, with their geometric ($r_{geo}$) and photogeometric ($r_{pgeo}$) distances estimated by \cite{bailer-jones_estimating_2021}, along with their RUWE values  \citep{2016A&A...595A...1G,2021A&A...649A...1G} to indicate the quality of their astrometric fits in Gaia EDR3.}
\label{tab:nonmembers}
\begin{tabular}{llll}
\hline\hline
Object name  & $r_{geo}$ (pc)               & $r_{pgeo}$ (pc)             & RUWE \\
\hline
NGC 6231 54  & $1850\substack{+80 \\ -70}$  & $1840\substack{+80 \\ -70}$ & 1.28\\
NGC 6231 259 & $1270\substack{+60 \\ -60}$  & $1250\substack{+70 \\ -70}$ & 1.36\\
NGC 6231 115 & $1900\substack{+100 \\ -70}$ & $1870\substack{+80 \\ -90}$ & 1.07\\
NGC 6231 184 & $1280\substack{+30 \\ -40}$  & $1290\substack{+40 \\ -30}$ & 1.15\\
HD 326319    & $1020\substack{+40 \\ -30}$  & $1020\substack{+30 \\ -30}$ & 1.59\\
\hline
\end{tabular}
\end{table}

\subsection{Observations}
\label{sec:observations}
Observations were carried out with the ESO VLT-FLAMES instrument at Paranal Observatory, Chile. FLAMES is a multi-object positioner with a 25-arcmin field of view that can be coupled to the GIRAFFE and UVES spectrographs. In the instrument's GIRAFFE/MEDUSA + UVES mode, it offers 132 + 6 fibres, respectively, for simultaneous observations of up to 138 targets (25 of which we use to observe the sky for sky subtraction). Here we focus on the B-type stars observed with FLAMES coupled to the GIRAFFE spectrograph by the MEDUSA fibres.
FLAMES/GIRAFFE is an intermediate resolution spectrograph. We used the L427.2 (LR02) setting which provides a spectral resolving power $\lambda / \delta \lambda \sim 6500$, which is high enough to identify stars where the helium lines might suffer from nebular contamination (via inspection of the Balmer lines). The LR02 data give continuous wavelength coverage between 3964–4567\AA, providing access to several spectral lines suitable for accurate RV measurements (see Table \ref{table:lines}). 
Fibres were allocated using the FPOSS\footnote{\href{https://www.eso.org/sci/observing/phase2/SMGuidelines/FPOSS.FLAMES.html}{eso.org/sci/observing/phase2/SMGuidelines/FPOSS.FLAMES.html}} fibre allocation software, which ensures that the fibre configuration is physically possible. 
No relative priorities were given to particular targets and there were no unallocated targets.

A total of 31 observations were obtained between 2017 April 02 and 2018 June 17, giving a timebase of 441 days throughout ESO periods 99 and 101\footnote{PI: H. Sana, IDs 099.D-0895 and 0101.D-0163}, allowing us to be sensitive to orbital periods of about 1 day to over a year. 

For all but the first epoch, a total integration of 1680~s was split in four consecutive exposures of 420~s to avoid saturation and reduce the impact of cosmic rays. This strategy was implemented after the first epoch as three exposures of 600s were found to saturate the fibres. The exposure time during several epochs (the 8th, 9th, 10th, 11th, 13th, 15th, 21st, and 22nd) was also reduced to avoid saturation during excellent seeing. The strategy facilitates a typical signal-to-noise ratio (S/N) of $>$70, allowing RV measurements to be obtained with an uncertainty of 2 \kms, and allows the detection (and subsequent RV measurements) of the fainter secondary spectrum in the double-lined (SB2) systems for a wide range of mass ratios ($q>0.4$). Seven targets were only included during the first epoch of observations (NGC 6231 283, CD-41 11048, CD-41 11027B, HD 326340, CD-41 11037, V* V964 Sco and V* V1207 Sco) as the plate configuration used in the first epoch of observations was modified, and these targets were not reallocated on the follow-up observations. As such, they are not further analysed in this work. One target, CD-41 11023, was also excluded from further study due to a low signal-to-noise ratio present in its spectra (having an apparent V band magnitude of 12.854 mag, making it one of the faintest targets in the sample). Also removing the O+B systems and the A0 star, HD 326319, in the sample reduces the target list to 80 B-star objects for analysis in this work. \par

\subsection{Data reduction}
\label{ss:datareduction}
The data were reduced with the ESO CPL FLAMES/GIRAFFE pipeline v.2.16.2, applying bias and dark subtraction, flat-field correction and wavelength calibration. We coadded consecutive exposures (3 for the first epoch, 4 per night for the rest) and performed sky subtraction by taking a median of, at most, 25 sky fibre spectra. These are fibres assigned to GIRAFFE that are pointed at the field away from stellar sources. Sky fibres that were contaminated by starlight or were impacted by saturation of nearby fibres were excluded upon visual inspection of their spectra on an epoch-by-epoch basis. Cosmic rays incident on the spectra were removed through the identification of outliers with Dixon's Q test \citep{dean_simplified_1951}. The coadded spectra were normalised via semi-automatic polynomial fitting described by \citet{sana_vlt-flames_2013}.

As mentioned above,  saturation of the fibres was a potential issue in the first observations (with 600s exposures) and some of the subsequent observations that were truncated due to excellent seeing - in some extreme cases, the seeing was as low as 0.3". To mitigate the effects of the saturation caused, the wavelength range of the saturated fibres were reduced, as only the extremes of the entire range were impacted. This limited access to three lines past 4450 \AA{}  in two of these cases (\ion{He}{i}~$\lambda$4471, \ion{Mg}{ii}~$\lambda$4481 and \ion{Si}{iii}~$\lambda$4553 in stars CD-41 7736 and NGC 6231 361) but as these were limited to just three epochs, the impact to the quality of RV measurements is negligible.

\section{Orbital properties}
\label{s:orbital}

\subsection{RV measurements}
\label{sec:rv}
We follow closely the methodology of the VLT-FLAMES Tarantula Survey \citep[VFTS,][]{evans_vlt-flames_2011,sana_vlt-flames_2013} and Tarantula Massive Binary Monitoring \citep[TMBM,][A17 from here on]{almeida_tarantula_2017}, both multi-epoch spectroscopic studies of O (and several B) stars in the 30 Doradus region of the Large Magellanic Cloud (LMC). RVs were measured by fitting either Gaussian or Lorentzian profiles to suitable absorption lines in the wavelength range of the LR02 setup, fitting one Gaussian/Lorentzian per line for single-lined (SB1) systems and two Gaussian/Lorentzian lines for double-lined (SB2) systems. The spectra of each target were visually inspected for signs of double-lined or asymmetric features to identify SB2 candidates. The choice between Gaussian and Lorentzian profiles to fit the absorption lines was made on which profile resulted in the smallest statistical error in RV. We choose to implement this over other methods, like using a cross-correlation function (CCF) for example, as the absorption lines in these spectra can be well approximated with either a Gaussian or Lorentzian profile. Errors and $\chi^2$-statistics given by noise in the spectra can also be lost upon performing a CCF and these errors are necessary for correcting for observational biases (see Section \ref{sec:biascorrection}). Gaussian/Lorentzian fitting preserves these errors.\par
\defcitealias{almeida_tarantula_2017}{A17}
\begin{table}
\centering
\caption{Lines used to measure RVs (described in Section \ref{sec:rv}) and the fraction of the sample that has RV measurements with the listed lines.
         }
\label{table:lines}
\begin{tabular}{l c l c}
\hline\hline
Line                    & Rest wavelength & Reference & Frequency \\
                    &   (\AA{}) & & (\%) \\
\hline
\ion{He}{i} $\lambda$4009      & 4009.256                & PvH      & 66       \\
\ion{He}{i+ii} $\lambda$4026 & 4026.070                 & CLL77     & 86     \\
\ion{He}{i} $\lambda$4143      & 4143.759                & PvH      & 8       \\
\ion{C}{ii} $\lambda$4267      & 4267.261                & NIST     & 46      \\
\ion{He}{i} $\lambda$4387      & 4387.929                & PvH      & 59      \\
\ion{He}{i} $\lambda$4471      & 4471.480                & NIST     & 8      \\
\ion{Mg}{ii} $\lambda$4481     & 4481.228                & CLL77    & 16     \\
\ion{Si}{iii} $\lambda$4553    & 4552.622                & PvH      & 24    \\
\hline
\end{tabular}\\
\flushleft
\footnotesize {\sc notes:} CLL77: \cite{conti_spectroscopic_1977}, PvH: Peter van Hoof line list\footnote{\url{http://homepage.oma.be/pvh/}};  NIST\footnote{\url{https://physics.nist.gov/PhysRefData/ASD/lines\_form.html}} 
\end{table}

\footnotetext[4]{\url{http://homepage.oma.be/pvh/}}
\footnotetext[5]{\url{https://physics.nist.gov/PhysRefData/ASD/lines\_form.html}}

The absorption lines and their adopted rest wavelengths are listed in Table \ref{table:lines}. The same Gaussian/Lorentzian profile is fitted per line through all spectra of each object - the profile's parameters (except the RV shift) are fixed over each epoch. All lines in an epoch are required to have the same RV shift (they are fit simultaneously). The \ion{He}{i+ii}~$\lambda$4026 blend is used most frequently as the reference frame (to which the RVs of subsequent lines must match) for these RV measurements as it is generally the strongest and most well behaved for these stars. However, earlier sub-types of B-type stars can exhibit an effective rest wavelength shift in this line due to the increased strength of the \ion{He}{ii}~$\lambda$4025 component of the blend. In these cases, the individual component \ion{He}{ii}~$\lambda$4025 is used as the reference frame. \ion{He}{i+ii}~$\lambda$4026 was used as a reference frame over other \ion{He}{i} lines like \ion{He}{i}~$\lambda$4143 and \ion{He}{i}~$\lambda$4387 as these were generally weaker and, in the case of \ion{He}{i}~$\lambda$4143, due to its proximity to the \ion{Si}{ii}~$\lambda$4128-4130 doublet which sometimes caused blending. Some of the lines in Table 1 are not strong enough across our full range of spectral types for reliable fits (e.g. \ion{He}{i}~$\lambda$4143 and 4387 for the later-type stars, and metallic lines such as \ion{C}{ii}~$\lambda$4267 at the earlier types), so the lines used between objects varied. The frequency of use of these lines in the RV fits can be seen in Table \ref{table:lines}. \par

A given star is then identified as RV variable if at least one of the pairs of RV measurements passes the following criteria introduced by \citet{sana_vlt-flames_2013} and also used by \citet{dunstall_vlt-flames_2015}:
\begin{equation}
\frac{|v_{i} - v_{j}|}{\sqrt{\sigma_{i}^{2} + \sigma_{j}^{2}}} > 4.0
\label{var_equation}
\end{equation}
where $v_{i}$ and $\sigma_{i}$ are the RVs and their uncertainties respectively at epoch $i$. We choose 4.0 for the confidence threshold of significant RV variability detection in order to have a maximum of one false-positive detection in our entire sample. As discussed by \citet{sana_vlt-flames_2013}, this is a more sensitive criterion to detect orbital variability than a conventional $\chi^2$ test of $v_i$ deviating from the average RV measured of an object, especially in the case of eccentric systems that may have only a few points that lie far from the average RV. Table \ref{table:nonvariable} lists the 12 objects in the target list that did not meet this criterion, which means that we classify 68 objects (85\%) as variable in RV. Three of these stars (NGC 6231 265, NGC 6231 243 and NGC 6231 127) have a RUWE greater than 1.4, indicating a poor astrometric fit in Gaia EDR3 (as mentioned in Section \ref{sec:gaia}). This suggests that these stars could be long period binary systems, but they do not show a smooth modulation of their RV during the length of this campaign (though they could still have a period longer than 441 days).

\begin{table}
\caption{Targets classified as stable in RV regarding the significance criteria in Section \ref{sec:rv} (Equation \ref{var_equation}), where dRV is the maximum difference in RV. Also shown is their RUWE values to indicate the quality of their astrometric fit in Gaia EDR3 \citep{2016A&A...595A...1G,2021A&A...649A...1G}.}
\label{table:nonvariable}
\centering
\begin{tabular}{l l l l l}
\hline\hline
Object name         & V & SpT  & dRV & RUWE\\
                    &(mag)&      & (\kms) &\\
\hline
NGC 6231 30         & 11.625  & B5 V   & 12.6   & 0.93       \\
TIC 339565755 & 11.952  & B5 V   & 11.3   & 1.28       \\
NGC 6231 265        & 12.616  & B7 V   & 46.1   & 1.65       \\
NGC 6231 243        & 12.484  & B5 V   & 17.3   & 4.19    \\
NGC 6231 234        & 12.818  & B7 V   & 10.3   & 0.96       \\
NGC 6231 227        & 12.498  & B5 V   & 15.3   & 0.83       \\
NGC 6231 152        & 12.371  & B5 V   & 9.7    & 0.94       \\
NGC 6231 123        & 12.311  & B5 V   & 15.2   & 0.93       \\
NGC 6231 127        & 12.033  & B3 V   & 16.5   & 1.49       \\
NGC 6231 24         & 11.718  & B4 V   & 10.8   & 0.90      \\
NGC 6231 194        & 11.459  & B3 V   & 31.2   & 0.85       \\
NGC 6231 165        & 12.493  & B6 V   & 37.4   & 1.01       \\
\hline
\end{tabular}
\end{table}

\subsection{Spectral classification}
\label{sec:spectraltyping}

To assess the impact of mass on the multiplicity properties of these B-type stars, we can use their spectral type as a proxy. As spectral types have not been recorded for the entire sample, we have classified the sample using the composite spectra (and disentangled spectra in the case of SB2s) for our targets (see Figs. \ref{FigNonVarSpec} to \ref{FigSB2Spec}). \par
We visually classify these spectra with two sets of criteria. We compare the composite FLAMES spectra  to HERMES spectra of stars identified as spectral standards \citep{gray_stellar_2009}. The HERMES spectra of standards are rebinned to the resolution of FLAMES and each HERMES standard spectrum is artificially rotationally broadened to 200 \kms  to also facilitate comparison with rotationally broadened stars in the NGC 6231 sample. \par
Standards for which we have spectra include subtypes O9V (10 Lacertae), B0V (HD 41753), B1V (HD 144470), B3V ($\eta$ Aurigae), B5V (HD 36936), B7V (HD 21071), B9V (HD 16046) and A0V (alf Lyr). If, upon comparison, a star visually resembles a subtype between two of the standards available, we classify with the intermediate subtype (e.g. between B5V and B7V, we assign B6V). Consequently, it is reasonable to assume an uncertainty of one spectral subtype to our classifications generally. To assign a luminosity class, we also compare to several HERMES spectra of identified standards of supergiants (Ia and Ib) and giants (II-III). For stars we class as B0-B3, we compare to stars at B1Ia ($\kappa$ Cas), B1Ib ($\xi$ Per) and B1II-III ($\beta$ CMa). For stars that we class as B4-B6, we compare to stars at B5Ia ($\eta$ CMa), B5Ib (67 Oph) and B5III ($\tau$ Ori). For stars that we class as B7-B8, we compare to stars at A0Ia (HR 1040), A0Ib ($\eta$ Leo) and A0III ($\alpha$ Dra). As to be expected for a cluster this age, the majority of the objects appear to be dwarfs (V). \par 
For stars that we classify as earlier than B2 with the spectral standards, we refer to the criteria of \cite{evans_vlt-flames_2015} for finer spectral typing, predominantly based on the strength of the relative strength of the \ion{Si}{iv} $\lambda$4089, \ion{Si}{iii} $\lambda$4553, \ion{He}{ii} $\lambda$4542 and \ion{Mg}{ii} $\lambda$4481 absorption lines. \par 
The distribution of spectral types in the sample and the impact of this on their multiplicity properties is described in Section \ref{ss:multiplicityspt}. We are unable at present to compare our sample with a standard IMF as this would require firm mass estimates of each target from atmospheric modelling; this will be investigated in a future study. 

\subsection{Period search}
\label{sec:periodsearch}
\begin{figure}
\centering
\includegraphics[width=\hsize]{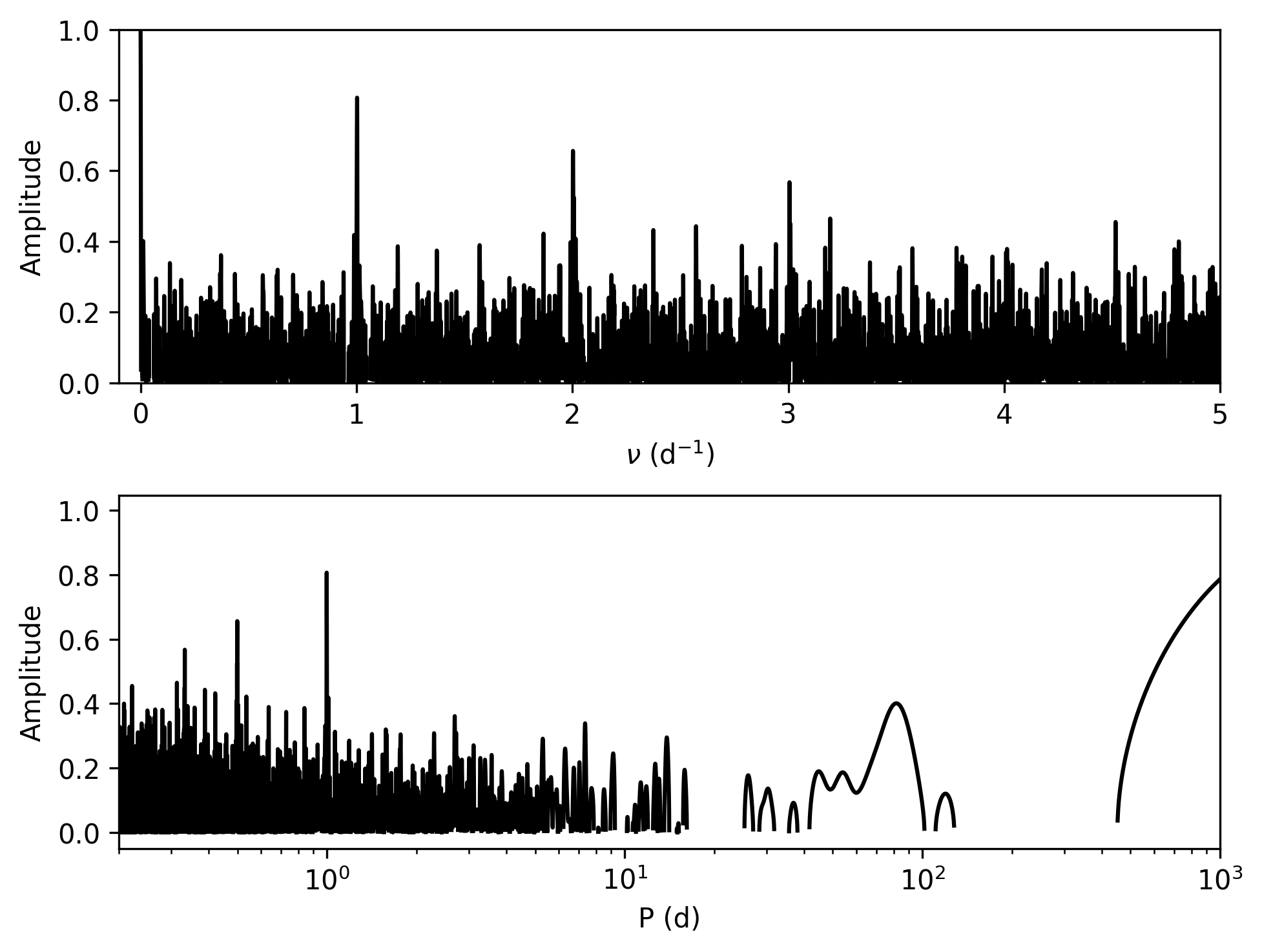}
    \caption{Power spectral window of the 31 observations of NGC 6231, with the top panel showing the power spectral window as a function of frequency and the bottom panel as a function of time.
         }
    \label{FigSpectralWindow}
\end{figure}

Lomb-Scargle periodograms \citep{lomb_least-squares_1976,scargle_studies_1982} were computed with the Astropy subpackage $timeseries$ \citep{astropy_collaboration_astropy_2013,astropy_collaboration_astropy_2018} using the RV time series of identified RV variables. The amplitude of peaks in the Lomb-Scargle periodogram that passed the 1\% false-alarm probability level were considered significant. The false-alarm probability level of the peaks were approximated as detailed by \cite{baluev_assessing_2008-1}. The power spectral window of the observations is shown in Fig.~\ref{FigSpectralWindow}. An alias in the periodogram is present at around 1 day (the minimum time between each observation). This corresponds to the Nyquist frequency of this observational campaign, which is  1.08 d$^{-1}$. \par
In Table \ref{table:variablenoperiod} we list the 41 targets that were found to be RV variables (from Equation \ref{var_equation}) but for which no significant periodicity could be found. These targets consist of objects that could either have a period outside of our detectable range (1 day to around the length of the campaign, 441 days), or objects that have RV shifts dominated by pulsations \citep{aerts_collective_2009,balona_simultaneous_1999-1,rivinius_non-radially_2003} with periods outside of the detectable range. These targets also include four stars - NGC 6231 274, NGC 6231 213, NGC 6231 96 and HD 326334 - that have periods that pass the false-alarm probability level criteria but are shorter than the period which corresponds to the Nyquist frequency of the observational campaign (0.93~d). Consequently, we do not regard these periods as significant. \par 

The number of B stars with seemingly aperiodic variability in RV is significant (51\%). Upon the inspection of their unphased RV curves, none of the stars in Table \ref{table:variablenoperiod} appear to be long period binary candidates, but many of the stars show indications of line profile variability. A likely source for this variability is pulsations in B stars. Depending on the spectral type, the detected variability of OB-type supergiants not associated with orbital motion in a binary system can reach a peak-to-peak amplitude of up to 30 \kms \citep{aerts_collective_2009,simon-diaz_iacob_2020}. A subsequent complementary study into time-series photometry of the targets would help establish how dominant pulsations are in driving this abundant variability in these (mostly) B dwarfs.\par
The 29 targets that were found to have significant periodicity were then subject to orbital solution fitting.

\begin{table}
\caption{Targets considered RV variables without a significant periodic signal, where dRV is the maximum difference in RV. Also shown is their RUWE values to indicate the quality of their astrometric fit in Gaia EDR3 \citep{2016A&A...595A...1G,2021A&A...649A...1G}.}
\label{table:variablenoperiod}
\centering
\begin{tabular}{l l l l l}
\hline\hline
Object name          & V & SpT  & dRV & RUWE \\
 &(mag)& &(\kms)& \\
\hline
CD-41 10989          & 11.08   & B1V   & 10.3 & 0.94       \\
HD 326326            & 10.931  & B3V   & 7.0 & 0.83        \\
NGC 6231 41          & 12.423  & B5V & 12.1 & 1.12       \\
NGC 6231 33          & 12.375  & B7V   & 6.6 & 0.89         \\
NGC 6231 249         & 11.876  & B4V   & 13.9 & 1.10       \\
NGC 6231 274         & 11.769  & B5V   & 8.7 & 0.88        \\
CD-41 11032          & 9.512   & B1V   & 5.4 & 0.84        \\
NGC 6231 374         & 10.157  & B2V   & 8.9 & 0.84        \\
CPD-41 7730          & 9.275   & B1V   & 8.9 & 0.89        \\
NGC 6231 75          & 12.757  & B7V   & 14.3 & 0.90       \\
CPD-41 7734          & 10.331  & B2V   & 13.1 & 0.83       \\
NGC 6231 235         & 13.292  & B8V   & 73.0 & 1.08       \\
CD-41 7736           & 10.174  & B2V   & 17.3 & 1.84       \\
NGC 6231 361         & 11.841  & B4V   & 19.2 & 0.89       \\
NGC 6231 213         & 11.059  & B2V   & 10.1 & 0.84       \\
HD 326339            & 10.108  & B2V   & 4.6 & 0.84        \\
NGC 6231 217         & 12.248  & B5V   & 13.6 & 1.01       \\
NGC 6231 108         & 11.435  & B2V   & 6.1 & 0.89        \\
TIC 339568125 & 9.959   & B0.7V & 5.2 & 1.47        \\
NGC 6231 222         & 11.829  & B4V   & 14.5 & 0.91       \\
NGC 6231 160         & 11.848  & B3V   & 22.5 & 0.90       \\
NGC 6231 96          & 11.052  & B2V   & 9.1 & 0.83        \\
HD 326332            & 9.669   & B0.5III & 5.5 & 1.10        \\
V* V947 Sco          & 9.78    & B1.5V & 11.7 & 0.94       \\
V* V920 Sco          & 9.601   & B0.5V & 8.3 & 0.99        \\
NGC 6231 121         & 11.217  & B2V   & 8.5 & 0.84       \\
NGC 6231 142  & 12.155  & B4V   & 10.2 & 1.80       \\
NGC 6231 147         & 12.319  & B9V   & 16.8 & 0.99       \\
NGC 6231 146         & 11.290   & B3V & 11.7 & 1.28       \\
HD 152076            & 8.9     & B0III   & 4.6 & 0.83        \\
CD-41 10990          & 10.852  & B0.5V & 4.3 & 0.90        \\
HD 326327            & 9.729   & B0.5V & 7.1 & 0.86        \\
NGC 6231 14          & 11.751  & B4V   & 14.6 & 0.83      \\
NGC 6231 284         & 12.338  & B8V & 15.2 & 0.82       \\
CD-41 11031          & 9.887   & B0.5V & 6.0 & 0.83        \\
NGC 6231 209         & 10.599  & B0.7V & 9.2 & 1.04        \\
HD 326317            & 10.58   & B1III   & 5.9 & 0.89        \\
NGC 6231 173         & 12.927  & B8V & 10.8 & 1.00      \\
NGC 6231 172         & 12.838  & B8V   & 33.0 & -       \\
NGC 6231 175         & 13.274  & B8V & 38.8 & 0.91       \\
HD 326334            & 10.738  & B3V   & 27.4 & 0.92 \\
\hline
\end{tabular}
\end{table}

\subsection{Orbital solutions}
\label{ss:orbitalsolutions}
Best-fitting orbital parameters for the identified SB1 and SB2 systems were obtained by fitting the orbital RV curve with the Levenberg-Marquardt algorithm, considering a single systemic velocity ($\gamma$). This is done with the code \emph{RaV FIT}, written for a previous study of a triple system \citep{mahy_triple_2018}.\par

Periods found to be significant in the Lomb-Scargle periodograms were used as an initial guess to constrain the period found by fitting the orbital RV curve. Table \ref{table:periodnoorbsol} lists the seven targets for which a significant period was found without a constrained orbital solution. Four of these (V* V945 Sco, V* V1032 Sco, NGC 6231 236, HD 152314) appear to show smooth modulation in RV over the course of the campaign (Fig. \ref{FigLongP}), so these could potentially be binary systems with periods longer than the duration of the campaign. Tables \ref{table:SB1} and \ref{table:SB2} detail the orbital parameters of the 15 SB1s and 5 SB2s respectively that have well constrained orbital solutions.\par

\begin{table}
\caption{Targets with a significant periodic signal but no or poorly constrained orbital solution. Also shown is their RUWE values to indicate the quality of their astrometric fit in Gaia EDR3 \citep{2016A&A...595A...1G,2021A&A...649A...1G}.}
\label{table:periodnoorbsol}
\centering
\resizebox{\columnwidth}{!}{
\begin{tabular}{l l l l l l}
\hline\hline
Object name  & V & SpT  & P & dRV & RUWE\\
 &(mag)& &(d)&(\kms)& \\
\hline
V* V945 Sco  & 9.575   & B0.5V & 10000.0    & 54.7  & 0.77        \\
V* V1032 Sco & 9.767   & B2V   & 305.29     & 36.4  & 1.16       \\
NGC 6231 236 & 11.385  & B2V   & 438.61     & 9.6   & 0.90        \\
HD 152314 & 7.749  & B0V   & 438.61     & 18.6  & 1.06       \\
CPD-41 7755  & 10.068  & B0.2V & 8.07       & 14.2  & 0.86       \\
NGC 6231 1   & 11.950   & B5V & 1.00       & 36.5   & 1.13       \\
NGC 6231 186 & 11.516  & B2V   & 1.00       & 11.2  & 1.24   \\
\hline
\end{tabular}
}
\end{table}

The residuals of the orbital solution fitting allow for the validation of the error bars associated with the RV measurements. In general, the root-mean-square deviation of the RV fits from the orbital solutions were larger than the errors from the RV measurements, even when quadratically added with the stability of the instrument (300 m/s)\footnote{\url{https://www.eso.org/sci/facilities/paranal/instruments/flames/doc/VLT-MAN-ESO-13700-2994_P107.pdf}}. The uncertainties on the RV measurements were also sometimes smaller than the stability of the instrument. After comparing the root-mean-square deviation to the formal normalised errors of regions of continuum flux in the normalised spectra, no trend was found to be adopted as a scaling factor for the uncertainties, with no indication of general underestimation of the uncertainties. To remedy this discrepancy between the orbital solution residuals and the RV uncertainties, we set a lower limit to the uncertainties of 1 \kms going forward, based on the impact to the residuals of the orbital fitting and on prior experience with VLT/FLAMES \citep{sana_vlt-flames_2013}. \par
It should be noted that there are no obvious signatures of triple or higher order systems in the sample, neither as SB3+ systems nor through the residuals of the orbital solution fitting of the binary systems. It is unclear whether this indicates a true lack of higher order systems in the cluster or whether this is limited by the observation strategy and the quality of the spectra. There is one SB1 system (NGC 6231 723) with a large RUWE value of over 23 (as mentioned in Section \ref{sec:gaia}, which could suggest the presence of multiple companions. It should be noted, however, that triple systems have been found with RUWE values above and below the critical value of 1.4. Examples of triple systems with a RUWE value below 1.4 include HD 150136 \citep{mahy_triple_2018}, HD 19820 and HD 148937, and examples of triple systems with a RUWE value greater than 1.4 include HD 167971 and HD 206267. An accompanying interferometric survey would assist in detecting tertiary or higher-order members in long period orbits around the observed systems.

\subsection{Bias correction for true multiplicity properties}
\label{sec:biascorrection}

\subsubsection{SB1 bias}
\label{sec:sb1bias}
The VFTS and TMBM work accounted for the impact of the observational biases on the observed multiplicity property distributions by applying a first order-of-magnitude correction, without a self-consistent modeling of the entire data set to constrain the parent distributions, which are unknown quantities. We follow this method \citep[detailed in ][]{sana_vlt-flames_2013} of finding  the sensitivity of our campaign with respect to the orbital period, the mass ratio, the eccentricity and the primary mass. This involves computing detection probabilities by randomly drawing orbital parameters from an equally sized synthetic sample (80 objects) described by set parent distributions, calculating orbits from those parameters and assessing whether the associated RVs measured would result in a binary detection given the binary detection criteria used in Section \ref{Sec:f_bin}. \par
We guide our assumptions on the underlying parent parameter distributions (period, mass ratio, eccentricity) on what was done by \cite{sana_vlt-flames_2013} and \cite{bodensteiner_young_2021}. We use power law representations of these distributions:

\begin{equation}
f($log$_{10} P/d) \sim (log_{10} P)^{\pi},    
\end{equation}
\begin{equation}
f(q) \sim q^{\kappa}, 
\end{equation}
\begin{equation}
f(e) \sim e^{\eta}     
\end{equation}

When making assumptions on the values of the indexes $\pi$, $\kappa$ and $\eta$, the impact of varying these choices on the calculated overall detection probability should be considered. It was found by \cite{sana_vlt-flames_2013} and \cite{bodensteiner_young_2021} that the dominant factor impacting the computed detection probability is the shape of the period distribution. For example, choosing $\pi$ to favour shorter-period systems results in systems that are easier to detect given our detection criteria that relies on the RV amplitude. On the other hand, varying the $q$ and $e$ distributions is a second order effect on this detection criteria and has little impact. \par
Like \cite{bodensteiner_young_2021}, we adopt measurements guided by previous studies. We take $\pi = 0.25 \pm 0.25$ to encompass (in a $\pm1\sigma$-interval) the values of $\pi$ measured for the O-stars in the Galaxy and LMC \citep{sana_binary_2012,sana_vlt-flames_2013,almeida_tarantula_2017} and the B stars in the LMC \citep{dunstall_vlt-flames_2015}. We then take $\kappa = -0.2 \pm 0.6$ and $\eta = -0.4 \pm 0.2$ from the Galactic O-star sample \citep{sana_binary_2012}, being the only study mentioned here that reliably constrained these values, though (as mentioned) these two choices have little impact on the overall detection probability. \par
To compute the bias correction factor, we draw 10\,000 artificial populations of 80 binaries, randomly assigning values of $\pi$, $\kappa$ and $\eta$ from the parent orbital distributions following Normal distributions with central values and $1\sigma$ dispersion. We then take the average of the fraction of detected binaries across the 10\,000 artificial populations as the overall detection probability, which we compute as 79\%. \par

\subsubsection{SB2 bias}
\label{sec:sb2bias}
There is an additional observational bias we must consider regarding SB2 systems that have similar line strengths in their spectra, i.e. systems with a mass ratio close to 1. A double lined system may appear single if the radial velocity separation between the two profiles is not sufficient for them to be clearly deblended. As a result, we must calibrate the bias correction discussed in Section \ref{sec:sb2bias} for this reduction in measured RV with SB2s. This observational bias is discussed and the correction for it is implemented by \citeauthor{bodensteiner_young_2021}, and we follow the same methodology. \par 
To quantify this bias, we use synthetic spectra from the TLUSTY B-star grid computed for Galactic metallicity \citep{lanz_grid_2007} to simulate spectra of both a single star and SB2 systems of several mass ratios (from $q = 0.35$ to $q = 1.00$). Atmospheric models corresponding to spectral types B0V to B5V were selected, and then the B0V atmospheric model is designated as the primary star. The spectra are then broadened to two rotational velocities, with a $v\,$sin$\,i$ of 100 and 200 \kms. The simulated spectra of each star in a system are then shifted apart in steps of 10 \kms and then co-added (accounting for the light ratio of the system), degraded to FLAMES resolution, binned to 0.2 \AA{} to correspond with the reduced FLAMES spectra, clipped to the LR02 wavelength coverage and adopted an S/N ratio typical of the studied observations. \par 
RVs are them measured for these simulated systems largely identically to what was done in Section \ref{sec:rv}, with the only difference that the same Gaussian/Lorentzian profile is not forced between different measurements of the same system to account for the variation in the line profiles as they blend at lower RV separations. It was found that measured RVs of systems with high mass ratios ($q>0.76$) were severely underestimated at low RV separations, up until the RV separation was large enough so that both lines were sufficiently deblended for an SB2 detection - this is where the SB2 bias lifts, and the particular RV separation where an SB2 classification is made varies with both mass ratio and rotational velocity. \par 
We then took this apparent reduction in RV into account when repeating the bias correction computations detailed in Section \ref{sec:sb1bias}. We adopt the measured RVs from the spectra simulated at a $v\,$sin$\,i$ of 200\,\kms as this appears to best represent most of our sample. The measured RVs are smoothed and interpolated on a fine grid with a range of 0 to 600\,\kms in 1\,\kms steps of RV separation and then across mass ratios from 0.35 to 1.00 in steps of 0.01. These interpolated RVs are then used as a correction to the simulations in Section \ref{sec:sb1bias}, where the computed primary RV shift in the simulated time series for each system is reduced as a function of the separation between the system's components and their mass ratio (with systems of a mass ratio less than 0.5 assumed to be unaffected by this bias). \par 
The SB2 bias we have detailed is expected to be stronger for campaigns that are devoted to observing more rapidly rotating stars, binary systems of lower primary mass (as these systems will have lower RV separations) and that use lower resolution spectrographs - all of these will result in greater difficulty in identifying the SB2 nature of a binary system due to the blending of their lines. The SB2 bias corresponding to our campaign has a large impact on the overall detection probability, and is reduced from 79\% when just considering the SB1 bias to 63\% when also considering the SB2 bias. This is quite surprising as this is just as significant of an impact as the SB2 bias had in the work of \citeauthor{bodensteiner_young_2021}, which was with similarly quickly rotating stars and lower resolution instrument, MUSE (between 1700 and 3700 compared to FLAMES' 6500). 

\section{Discussion}
\label{s:discuss}
\subsection{Binary fraction as a function of threshold velocity}
\label{Sec:f_bin}

With all the retrieved RV measurements, it is important to separate those objects whose RV variations are plausibly dominated by intrinsic variations such as pulsations and atmospheric activity, and those whose RV variations are dominated by orbital motion in a multiple system. In addition to the first criterion, described in Equation \ref{var_equation}, a second criterion is introduced and must be passed simultaneously with the first
\begin{equation}
\frac{|v_{i} - v_{j}|}{\sqrt{\sigma_{i}^{2} + \sigma_{j}^{2}}} > 4.0 \quad\mathrm{and}\quad |v_{i} - v_{j}| > C
\label{var_criteria}
\end{equation}
where $C$ is the imposed minimum threshold velocity that at least one pair of RV measurements attributed to an object must exceed to be classified as a binary system.

\begin{figure}
\centering
\includegraphics[width=\hsize]{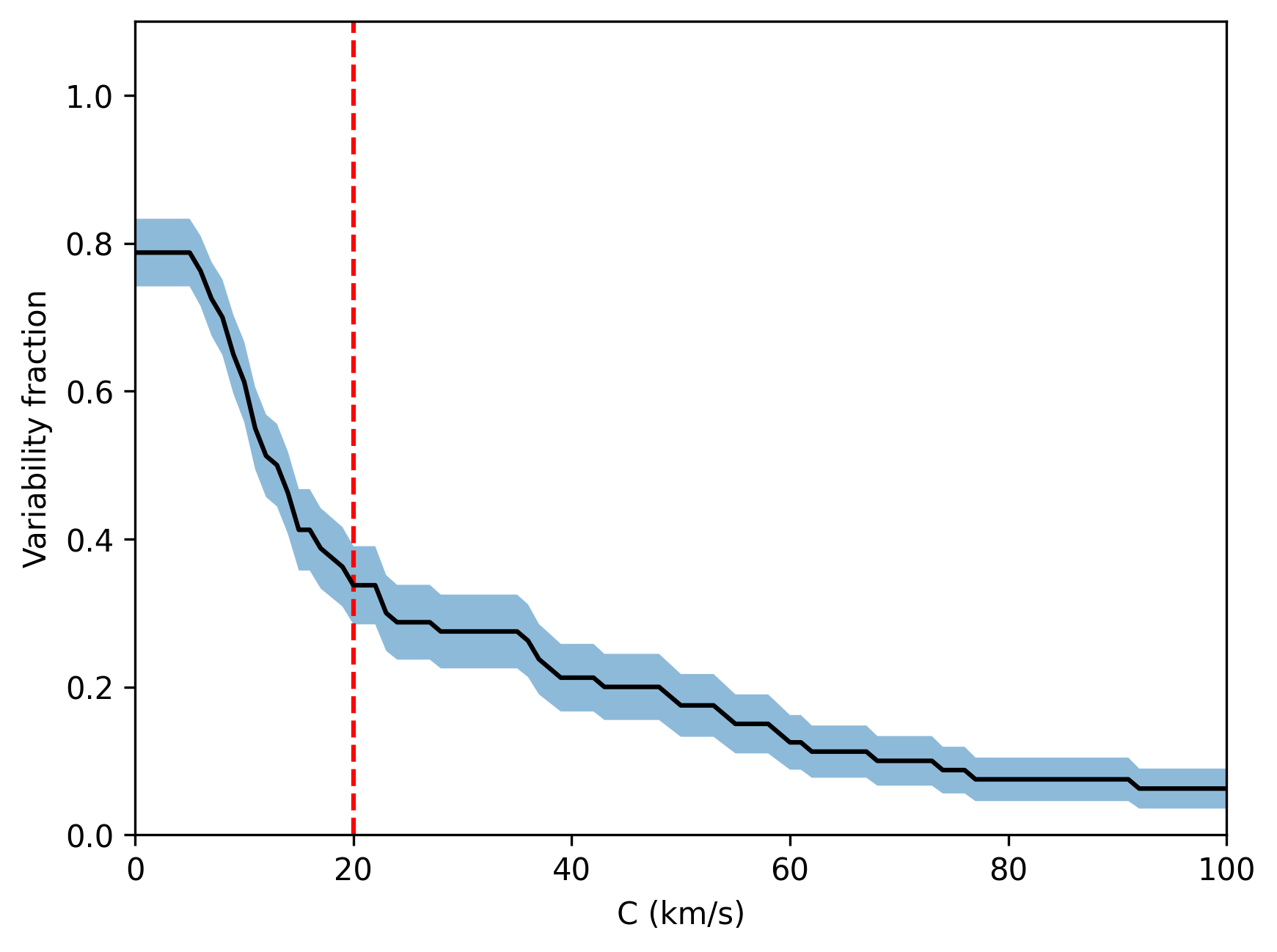}
    \caption{The inferred binary fraction as a function of the critical RV variation amplitude $C$ which separates spectroscopic binaries from variability dominated by variations from systematics and/or pulsations. The blue shaded area corresponds to the associated binomial error to the binary fraction. The vertical dashed-dotted line indicates the adopted threshold of $C = 20$ \kms.  
         }
    \label{FigBinaryFraction}
\end{figure}

The value of $C$ for various multiplicity studies has varied. \cite{sana_vlt-flames_2013} adopted  $C = 20$ km$\,$s$^{-1}$ for O stars in 30 Doradus in the LMC, \cite{abt_frequency_1990} also adopted $C = 20$ km$\,$s$^{-1}$ for (mostly) field B stars while \cite{dunstall_vlt-flames_2015} adopted $C = 16$ km$\,$s$^{-1}$ for the B-type stars in 30 Doradus. This threshold is likely to vary with spectral type and luminosity class, and this will be explored more thoroughly in a forthcoming study from \cite{simondiaz_IACOB_2020}. The impact of $C$ on the observed binary fraction in this work is shown in Fig.~\ref{FigBinaryFraction}. There appears to be a kink in the distribution of RV variables. If we adopt the previously used threshold of $C = 16$ km$\,$s$^{-1}$, the observed (and consequently, minimum) binary fraction inferred is $43 \pm 5\%$ (the uncertainty estimated as a binomial proportion confidence interval). However, to further minimise the presence of non-orbital motion related variation in RVs in the binary fraction, we adopt a slightly higher threshold of $C = 20$ km$\,$s$^{-1}$. This gives an observed binary fraction of $33 \pm 5\%$, providing a lower limit on the true binary fraction of B-type stars in NGC 6231. One system, CXOU J165421.3-415536 (a B2 star), was found to be an RV variable, have significant periodicity in its RV measurements, and a constrained orbital solution. However, its maximum threshold velocity lies just below 20 \kms (it is measured to be variable up to 19.8 \kms) so it is not included in the observed binary fraction. However, the bias correction applied to the observed binary fraction accounts for systems that lie below the cutoff. \par
To make any robust comparison to the binary fractions of other samples, we must first account for the observational biases to predict what binaries have avoided detection and where they may lie in the parameter space. This is detailed in Section \ref{sec:biascorrection}. After calculating an estimate of the sensitivity of the observational campaign, this provides us with a correction factor which can be multiplied with the observed binary fraction. We find that the overall detection fraction for periods of 1 to 3040 d is $\sim$ 63\%. This results in a true binary fraction of $52 \pm 8\%$ for the B-type stars in NGC 6231. It should also be stressed that although the method of bias-correction is similar to some of the studies cited below, the quoted fractions do not include the SB2 bias we have described in Section \ref{sec:sb2bias}. As we have demonstrated, this introduces a significant decrease in the overall detection fraction, and so quoted intrinsic binary fractions could be larger than stated. The impact of the SB2 bias, as previously noted, will also depend on the campaign - for example, the bias will be less impactful with increasing instrument resolution, and also less impactful for more massive primary stars which will tend to have higher radial velocity separations in their double-lined spectra. \par
Within the additional uncertainties of the correction for undetected SB2 systems, the true binary fraction for NGC 6231 is in agreement with that found for the 408 B-type stars in 30 Doradus \citep[$58 \pm 11\%$,][]{dunstall_vlt-flames_2015}, which was found after estimating the impact of the observational biases and the detection sensitivities on the observed binary fraction ($25 \pm 2\%$). This bias correction did not take into account the SB2 bias and so the binary fraction for the 30 Dor sample could be higher. The agreement could suggest that the formation process for B-type stars in multiple systems in different metallicity environments may be universal, although to test this further would require studies of a similarly young cluster in the SMC, for example. The NGC 6231 B-star true spectroscopic binary fraction also agrees with that found by \cite{abt_frequency_1990} for field B2-B5 stars, which, when corrected for undetected systems, was estimated as 57\%, but again this could be higher due to undetected SB2 systems. The B-type stars in the open cluster NGC 330 in the SMC have a lower bias-corrected spectroscopic binary fraction \citep[$38 \pm 6\%$,][]{bodensteiner_young_2021}, which does take the impact of undetected SB2s into account as mentioned and described in Section \ref{sec:sb2bias}. With an age of 35-40\,Myr \citep{bodensteiner_young_2020}, NGC\,330 is significantly older than NGC 6231 and so potentially contains an increased number of post-interaction products that may appear as single stars. We can also compare with the true binary fraction of massive stars in Cygnus OB2 that is predicted to be $55\%$ (without taking into account the impact of undetected SB2 systems), from observations of 45 O stars and 83 B-type stars by \citeauthor[][hereafter K14]
{kobulnicky_toward_2014} in this massive star cluster aged between 3-4 Myrs \citep{hanson_study_2003}.  \par
Regarding comparisons to O stars, the true binary fraction of the B-type stars in NGC 6231 is just in agreement with the intrinsic O-type star binary fraction in the Galaxy, $69 \pm 9\%$  \citep{sana_binary_2012}. As stated prior, this is without the SB2-related bias that we have calculated, so taking this into account is likely to increase the galactic O star binary fraction (though not to the same degree as the B-type stars). As for a direct comparison with the same cluster, only the observed (minimum) spectroscopic binary fraction for the 15 O stars in the same cluster is presented \citep[$60 \pm 10\%$]{sana_massive_2008-1}. 
\defcitealias{kobulnicky_toward_2014}{K14}
\defcitealias{sana_binary_2012}{S12}
\defcitealias{villasenor_b-type_2021}{BBC}

\subsection{Orbital parameter distributions}
\subsubsection{Orbital periods}
\label{ss:orbitalperiods}

\begin{figure}
\centering
\includegraphics[width=\hsize]{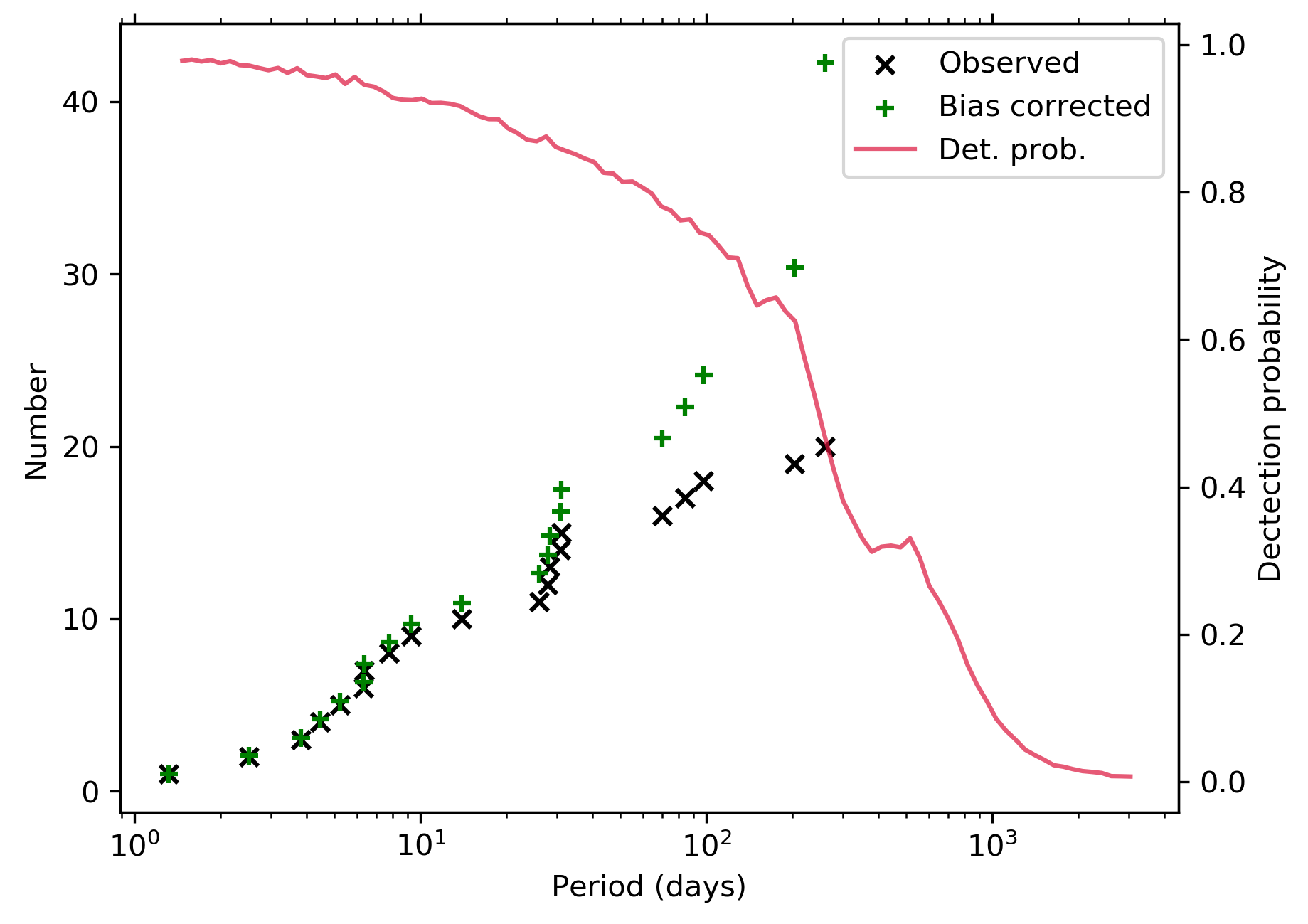}
    \caption{Observed cumulative distribution of periods of SB1 and SB2 systems in NGC 6231, with the distribution corrected for observational biases in green and the detection probability of systems as a function of period in red (as described in Section \ref{sec:biascorrection}).}
    \label{FigPDist}
\end{figure}

\begin{figure}
\centering
\includegraphics[width=\hsize]{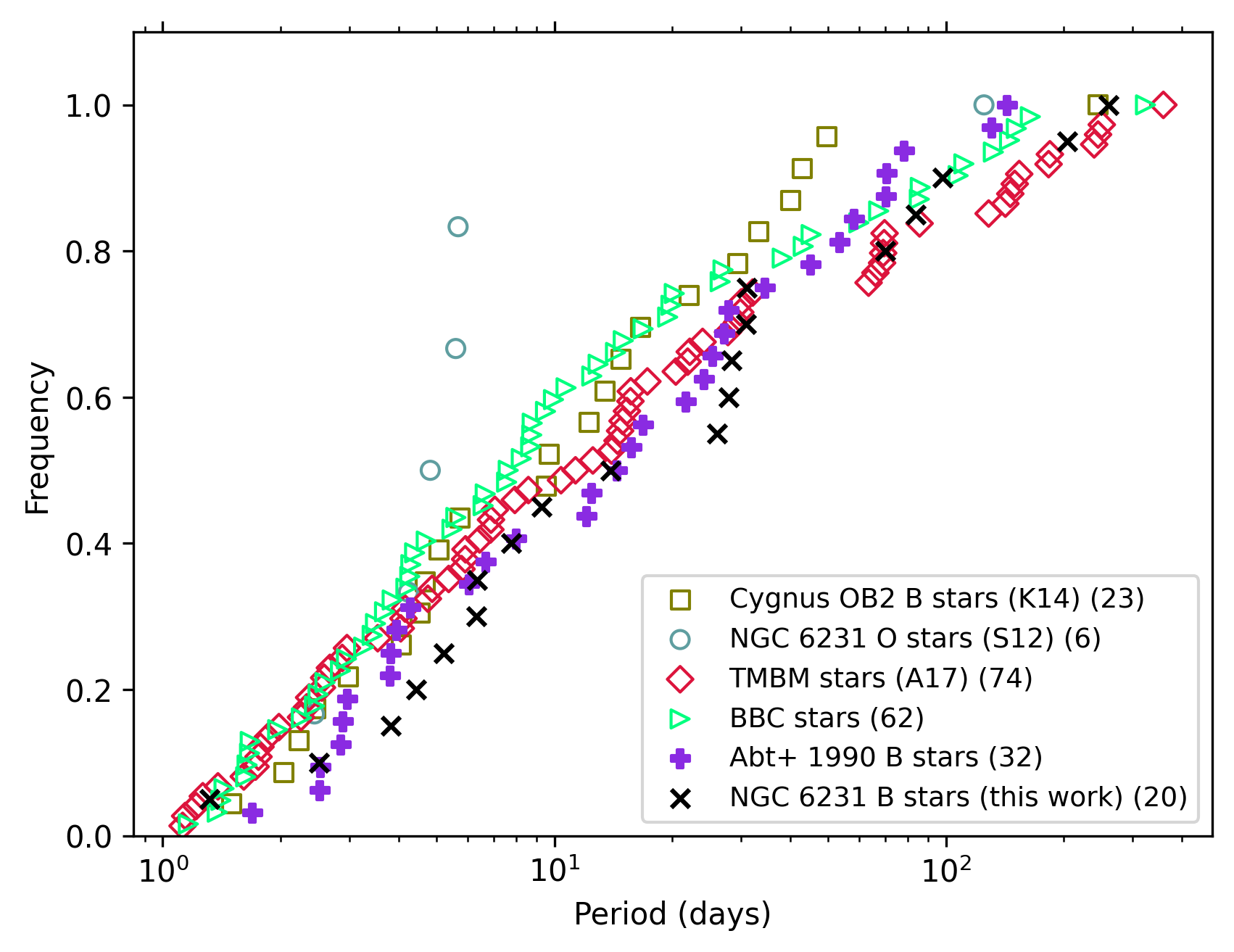}
    \caption{Observed cumulative distribution of periods of SB1 and SB2 systems in NGC 6231 and the observed period distribution found for B-type star binaries in Cygnus OB2 \citepalias{kobulnicky_toward_2014} as olive-green squares, O star binaries in NGC 6231 \citepalias{sana_binary_2012} as blue circles, O star binaries in 30 Doradus \citepalias{almeida_tarantula_2017} as red diamonds and B-type star binaries in 30 Doradus \citep{villasenor_b-type_2021} as light green triangles. For comparison with our new results, periods greater than one year were excluded from the previous studies. 
         }
    \label{FigPDistObs}
\end{figure}

The distribution of the observed orbital periods of the B-type stars in NGC 6231 ranges from around 1 to 260 days (Fig.~\ref{FigPDist}). Around $40\%$ of the systems have a period shorter than one week, about $75\%$ have periods below a month and no systems were found to have periods above a year. This lack of long period systems is likely due to limitations imposed by the total length of the observing campaign of 441 days - it becomes difficult to detect binaries with periods longer than around 220 days, especially for eccentric systems, for which we need two cycles to constrain an orbital solution. These longer period systems might appear as RV variables with no periodicity detected in this work, and there are four candidates with no orbital solution but with significant periodicity that have estimated periods of over a year. There appears to be structures in the period distribution - a lack of systems with periods around 4~d and 20~d, but an overabundance around 7~d and 30~d. This does not appear to correlate with any aliases in the power spectrum associated with the cadence of the observations (see Fig. \ref{FigSpectralWindow}). \par

The bias-corrected period distribution for the NGC 6231 B-type stars can be seen in Fig. \ref{FigPDist}, produced by calculating the probability of detecting a system of a given period, as described in Section \ref{sec:biascorrection}. It is shown that the flattening of the observed distribution at periods longer than around 100 days is reduced. At $P = 260.7$ d, the longest period detected in the campaign, the detection probability drops to around 75\%. Any system with a period of $P = 900$ d has a 50\% probability of being detected, before dropping to 0 for periods of 3000 d, highlighting the sensitivity of the campaign. However, the apparent gaps at 4- and 20-\,d periods remain. \par

\begin{table}
\centering
\caption{Results on Kuiper tests between the period and eccentricity distributions of NGC 6231 B-type stars, NGC 6231 O stars \citepalias{sana_binary_2012}, and results from the TMBM \citepalias{almeida_tarantula_2017}, BBC \citep{villasenor_b-type_2021}, Cygnus OB2 \citepalias{kobulnicky_toward_2014} and field B-type stars \citepalias{abt_frequency_1990}. D is the Kuiper D number and dpp is the probability that a value as large as D would occur if data was drawn from the given distribution. The spectral sub-type range is given in the brackets next to the ID of each sample.}
\label{tab:kuipertests}
\begin{tabular}{llll}
\hline\hline
\multirow{2}{*}{Sample 1}       & \multirow{2}{*}{Sample 2} & \multicolumn{2}{l}{Kuiper test} \\
                            &                           & D             & dpp             \\
\hline
\multicolumn{4}{l}{Period distributions:}               \\
NGC 6231 (B0-B9)              & S12 (O7-O9)               & 0.65          & 13.3\%          \\
NGC 6231 (B0-B9)              & A17 (O3-O9.7)                & 0.24          & 80.1\%          \\
NGC 6231 (B0-B9)              & K14 (B0-B2)                & 0.29          & 78.1\%          \\
NGC 6231 (B0-B9)              & BBC (B0-B5)                & 0.28          & 60.3\%          \\
NGC 6231 (B0-B9)              & A90 (B2-B5)                & 0.24          & 91.6\%          \\
\hline
\multicolumn{4}{l}{Eccentricity distributions:}               \\
NGC 6231 (B0-B9)              & S12 (O7-O9)                       & 0.30          & 99.2\%          \\
NGC 6231 (B0-B9)              & A17 (O3-O9.7)                      & 0.55          & 0.1\%          \\
NGC 6231 (B0-B9)              & K14 (B0-B2)                       & 0.28          & 80.1\%          \\
NGC 6231 (B0-B9)              & BBC (B0-B5)                       & 0.31          & 40.9\%          \\
NGC 6231 (B0-B9)              & A90 (B2-B5)                       & 0.27          & 79.2\%          \\
\hline
\end{tabular}
\end{table}

Our observed period distribution is compared with four other samples in Fig. \ref{FigPDistObs}. These samples are the B-type stars of Cygnus OB2 \citepalias{kobulnicky_toward_2014}, the O stars of NGC 6231 (the resolved binaries of which are characterised by \citealt{sana_binary_2012}, S12 from here on), the OB (mostly O) stars of 30 Dor in the LMC \citepalias{almeida_tarantula_2017}, the 64 characterised B-type star binary systems in 30 Dor from the B-type Binaries Characterisation Program \citep[][BBC from here on]{villasenor_b-type_2021} and the 32 field B2-B5 star binaries detailed in \citetalias{abt_frequency_1990}. As the orbital periods for the B-type star systems found in this work extend only to a maximum of 260.7 days (CPD-41 7727), systems with periods longer than a year in the other samples are excluded to facilitate comparison. Consequently, the three interferometric O star binaries in NGC 6231 characterised by \citet{le_bouquin_resolved_2017} are not included in our comparisons as they all have periods greater than a year.  \par

Overall, there seems to be little variation between the observed period distributions of the B-type star binaries in NGC 6231, for those in Cygnus OB2, the B2-B5 field stars and the OB-type star systems in 30 Dor. The fraction of very close binaries below $P < 1.4$ d appears similar for all samples excluding the O star binaries in NGC 6231 \citepalias{sana_binary_2012}. These distributions generally follow the same slope between 5 and 14 days, with the exception of the S12 O stars that have over 80\% of the sample lying at periods below $P < 5.6$ d before flattening out. \par

The distributions then begin to diverge after 14 d. At 19 d for example, the NGC 6231 O stars lie at 85\% (though this could also be due to the small number of resolved O star binaries in the cluster), the Cygnus OB2 stars and \citetalias{villasenor_b-type_2021} stars lie at around 75\%, the TMBM stars are at 65\%, and the NGC 6231 B-type stars are at 50\%. The TMBM, \citetalias{villasenor_b-type_2021} and NGC 6231 B-type star samples then follow a generally similar slope for periods over 30 to 300 days, with the exception of the Cygnus OB2 sample which has a dearth of systems between  around 50 and 250 d. \par

With these differences aside, the period distributions are remarkably similar, even without accounting for the individual observational biases associated with the observational campaign of each sample (which have similarities in their strategies). To test whether these differences were statistically significant, Kuiper (K) tests are computed between the period distributions (Table \ref{tab:kuipertests}), with the output being the discrepancy statistic D and the probability that two samples could be drawn from the same distribution (dpp) are calculated. This probability is an approximation, especially so for smaller samples. We take any probability below a standard threshold of 10\% to be a statistically significant indication of a difference between the samples.\par
The \citetalias{kobulnicky_toward_2014}, \citetalias{villasenor_b-type_2021}, \citetalias{almeida_tarantula_2017} and \citetalias{abt_frequency_1990} samples have a 78.1\%, 60.2\%, 80.1\% and 91.5\% probability of being drawn from the same distribution as the NGC 6231 B-type stars. Even the differences in the observed period distributions of the O- and B-type stars in NGC 6231 are found to be statistically insignificant, with the probability that the two samples are pulled from the same distribution being 13.3\%. The apparent contrast between the two samples is likely due to the small sample size of the NGC 6231 O stars. This suggests that the physics of the binary formation process produces similar populations between both B and O stars in young Galactic clusters (and perhaps in the Galactic field), and is also comparable to the O- and B-type stars in the lower metallicity environment in 30~Doradus. \par

A comparison of the bias-corrected distribution from each sample would also be necessary to validate the true nature of the differences and similitudes seen in the observed distributions. These studies each have different brightness regimes and observational cadences and strategies that need to be accounted for. \par

\subsubsection{Eccentricities}
\label{ss:eccentricities}

\begin{figure}
\centering
\includegraphics[width=\hsize]{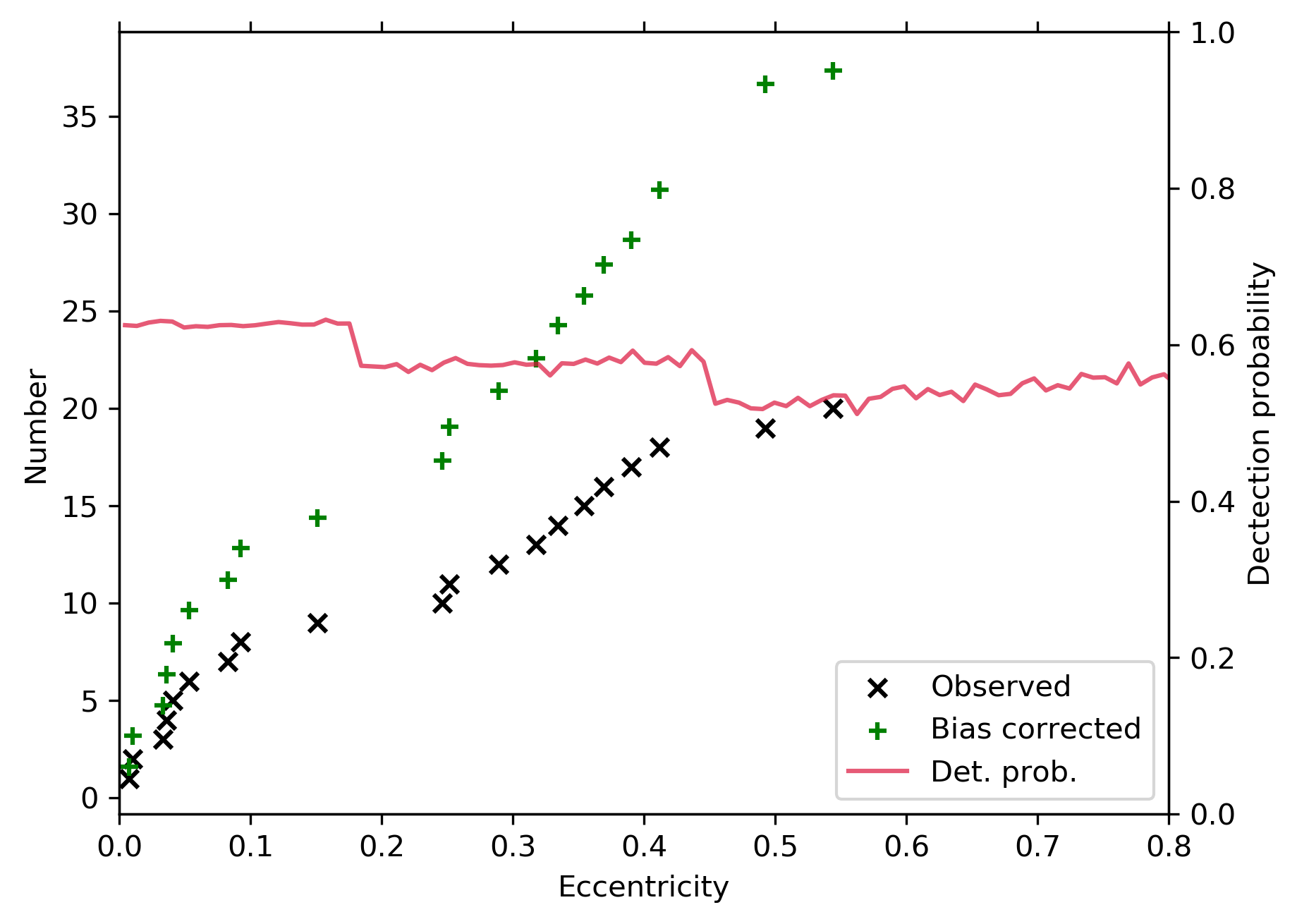}
    \caption{Observed cumulative distribution of eccentricities of SB1 and SB2 systems in NGC 6231, with the distribution corrected for observational biases in green and and the detection probability of systems as a function of eccentricity in red (as described in Section \ref{sec:biascorrection}).
         }
    \label{FigEDist}
\end{figure} 

\begin{figure}
\centering
\includegraphics[width=\hsize]{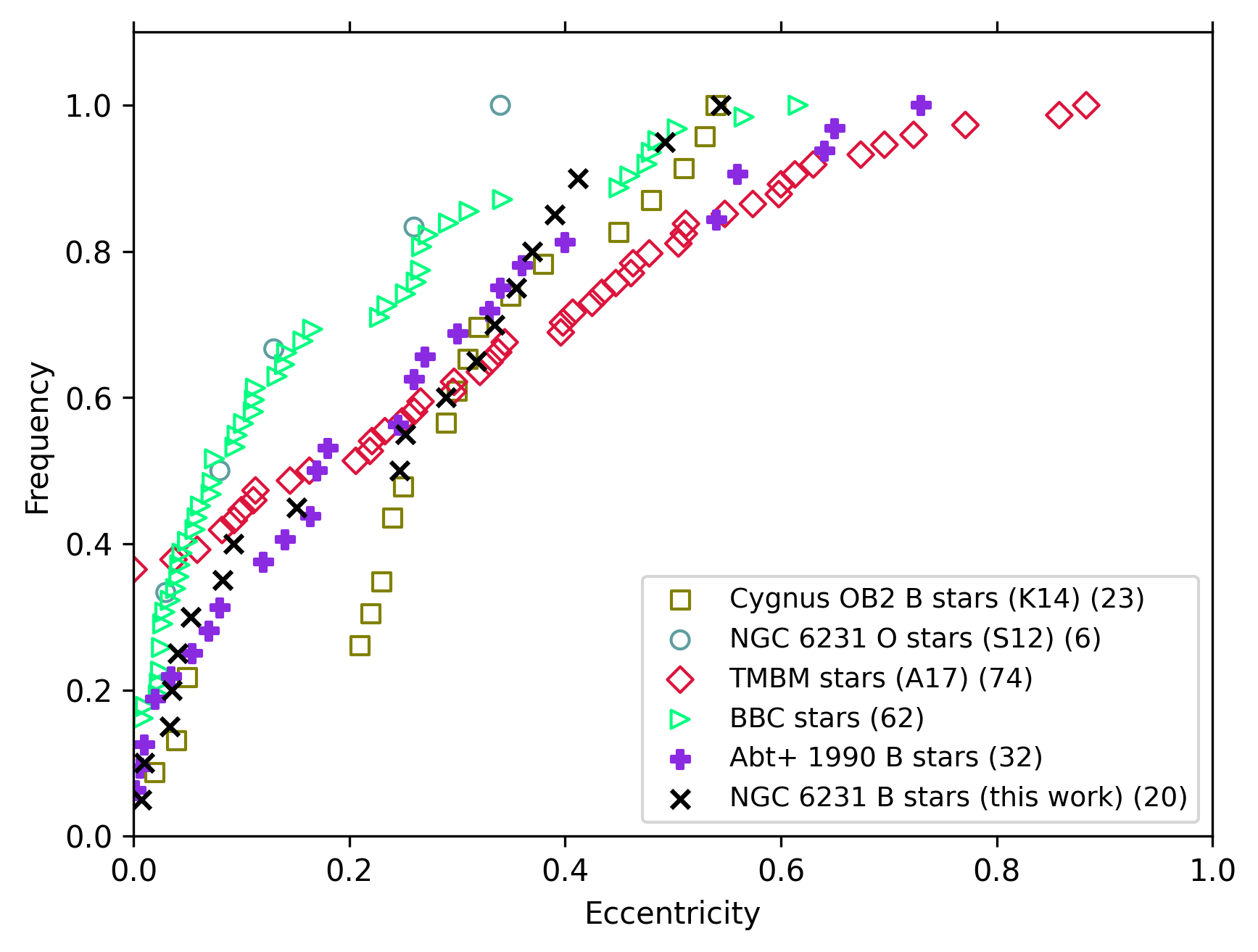}
    \caption{Observed cumulative distribution of eccentricities of SB1 and SB2 systems in NGC 6231, and the observed equivalent eccentricity distributions found for B-type star binaries in Cygnus OB2 \citepalias{kobulnicky_toward_2014} as olive-green squares, O star binaries in NGC 6231 \citepalias{sana_binary_2012} as blue circles, O star binaries in 30 Doradus \citepalias{almeida_tarantula_2017} as red diamonds and and B-type star binaries in 30 Doradus \citepalias{villasenor_b-type_2021} as light green triangles.}
    \label{FigEDistObs}
\end{figure} 

The distribution of observed eccentricities (Fig. \ref{FigEDist}) is from near zero (circular) to $e$ = 0.55, with 40\% of the systems having $e <$ 0.1. Six of the lowest eccentricity systems do not pass the Lucy-Sweeney test at the 5\% significance level ($e/\sigma_{e}\leq\,2.49$). This means that the constrained eccentricity values are not necessary in describing the orbital solution and the eccentricity lies anywhere between 0 and 2.49$\sigma_{e}$. These solutions are marked in Tables \ref{table:SB1} and \ref{table:SB2}. There does appear to be a lack of systems with eccentricities between $0.1 < e < 0.25$ and a strong flattening at $e > 0.4$. \par

The bias-corrected eccentricity distribution for the B-type stars in NGC 6231 is also shown in Fig. \ref{FigEDist}, and is also calculated in the same way as in Section \ref{ss:orbitalperiods}. The aforementioned dearth of binary systems within the eccentricity range $0.1 < e < 0.25$ is not explained by the calculated detection probability of these systems, but the flattening of the distribution at eccentricities larger than $e > 0.4$ is less severe. \par

Our observed eccentricity distribution is compared with the five observed samples in Fig. \ref{FigEDistObs}. Again, only systems with periods of less than a year are compared in this figure. \par

Comparing our results suggest that the eccentricity distribution of the \citetalias{kobulnicky_toward_2014} sample appears most similar to the one displayed by the B-type stars in NGC 6231. In both cases, 25\% of binaries have very small eccentricities ($e < 0.05$) compared to 40\% of systems for both O- and B-type stars in 30 Dor. Perhaps this might be due to the presence of older stars in 30 Dor, where the median age has been found to be 8.1 Myr \citep{schneider_vlt-flames_2018} for the stars in the wider 30 Dor field. These stars will have had more time to circularise their orbits due to ongoing Case A mass transfer. The two Galactic cluster samples then differ somewhat, where the NGC 6231 sample has more stars with an eccentricity between $0.05 < e < 0.25$. All of the O star binaries in NGC 6231 \citepalias{sana_binary_2012} with periods below a year have eccentricities below $e < 0.34$, though this is a significantly smaller sample of six stars. Both the \citetalias{sana_binary_2012} and \citetalias{kobulnicky_toward_2014} samples have a probability of over 80\% of being drawn from the same distribution as the NGC 6231 B-type star results. Again, this supports our previous conclusion that the physics of the binary formation process produces similar populations between both B and O stars in young clusters in the Galaxy. In the LMC, the \citetalias{villasenor_b-type_2021} B-type stars appear most similar to the \citetalias{sana_binary_2012} O stars, with 80\% of their eccentricities falling below $e < 0.2$. The K test results for the B-type stars between 30 Dor and NGC 6231 also indicate similarity between the two. \par 
The TMBM distribution \citepalias{almeida_tarantula_2017} generally appears to differ the most, with it being the sample with the largest eccentricities, specifically those greater than $e > 0.6$. The probability that the NGC 6231 eccentricities are sampled from the same distribution as the \citetalias{almeida_tarantula_2017} eccentricities is 0.1\%. The \citetalias{abt_frequency_1990} sample of B2-B5 Galactic field binaries also contains systems with $e > 0.6$, but the distribution of eccentricities is much more similar to the NGC 6231 B stars below $e > 0.4$, and so the two samples are statistically similar (with a probability of 79\% of the two being drawn from the same parent distribution).  \par

\begin{figure}
\centering
\includegraphics[width=\hsize]{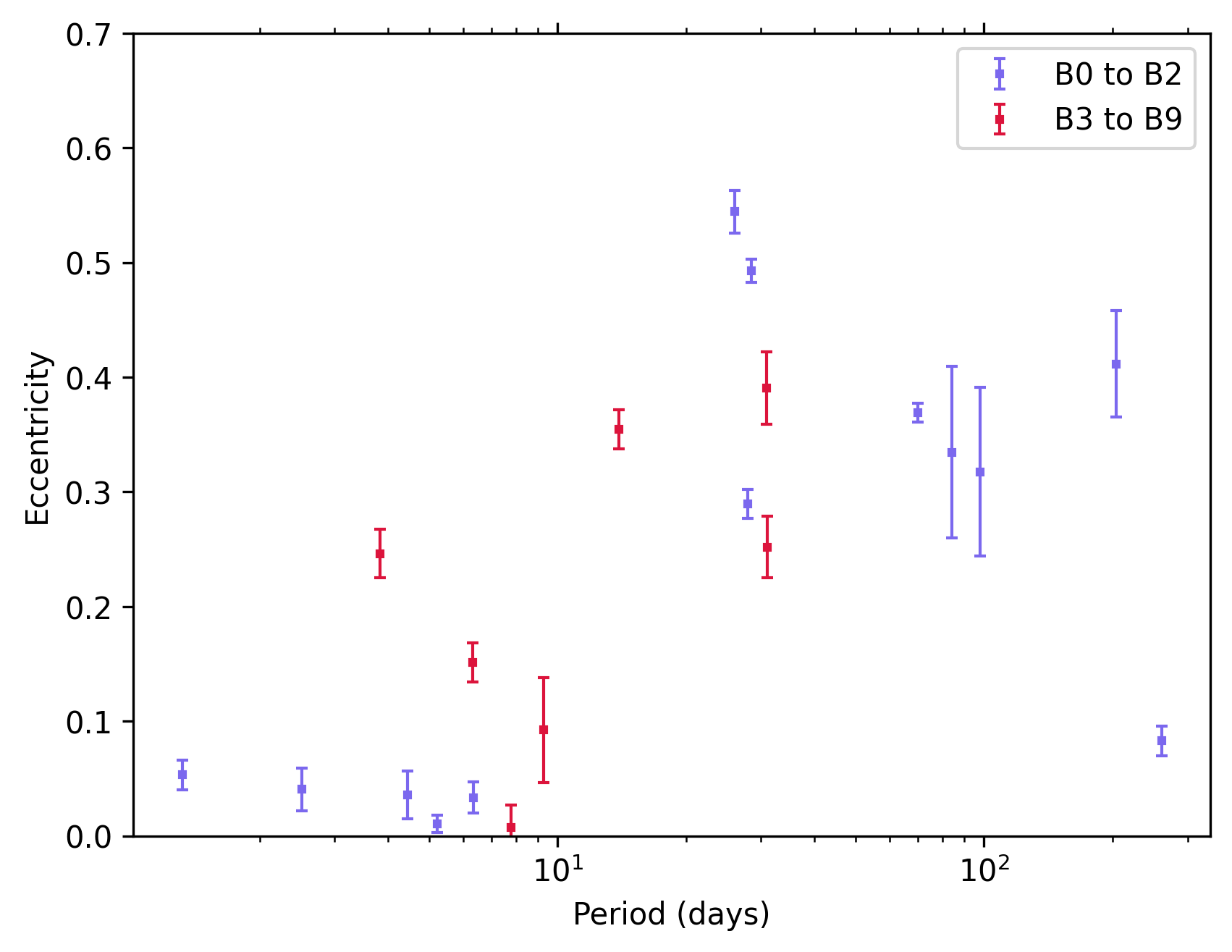}
    \caption{Period versus eccentricity for the orbital solutions found in the NGC 6231 sample. Systems classified as B0 to B2 are marked in blue, while those classified as B3 to B9 are marked in red.
         }
    \label{FigPvE}
\end{figure}

The eccentricity as a function of period is shown in Fig.~\ref{FigPvE} for the B-type stars in NGC 6231 earlier and later than spectral type B2. It shows that the majority of very low eccentricity / circular systems with periods below 10 days are for stars earlier than spectral type B2. This might be due to the increased efficiency of tidal circularisation of orbits of close binaries that have more massive stars in them. The longest period system in Fig.~\ref{FigPvE}, CPD-41 7727 (with P = $261\pm0.4$\,d, $e = 0.068\pm0.008$) is particularly interesting. The spectra of this system reveal a rapidly-rotating B2-type dwarf and a giant/supergiant B1-type secondary. The low eccentricity, the fast rotation of the primary and the evolved nature of the secondary suggest that this may be a system actively transferring mass through stable, slow Case A Roche lobe overflow, or a merging of the primary in a previously triple system. The separation and IR magnitude of this system makes it suitable for interferometric follow-up.

\subsubsection{Mass ratios}

\begin{figure}
\centering
\includegraphics[width=\hsize]{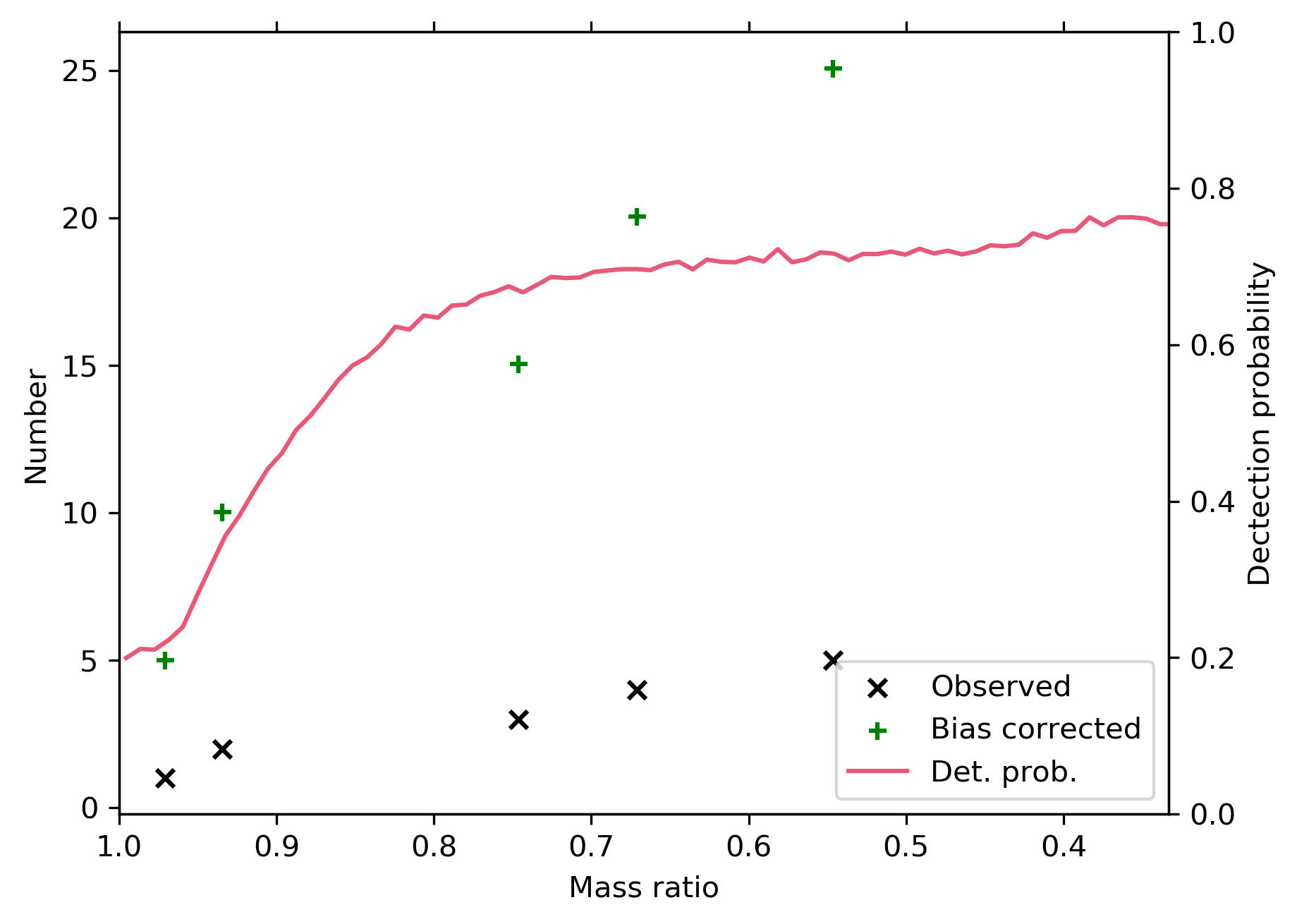}
    \caption{Observed cumulative distribution of mass ratios of SB2 systems in NGC 6231, with the distribution corrected for observational biases in green and and the detection probability of systems as a function of mass ratio in red (as described in Section \ref{sec:biascorrection}).
         }
    \label{FigQDist}
\end{figure}

\begin{figure}
\centering
\includegraphics[width=\hsize]{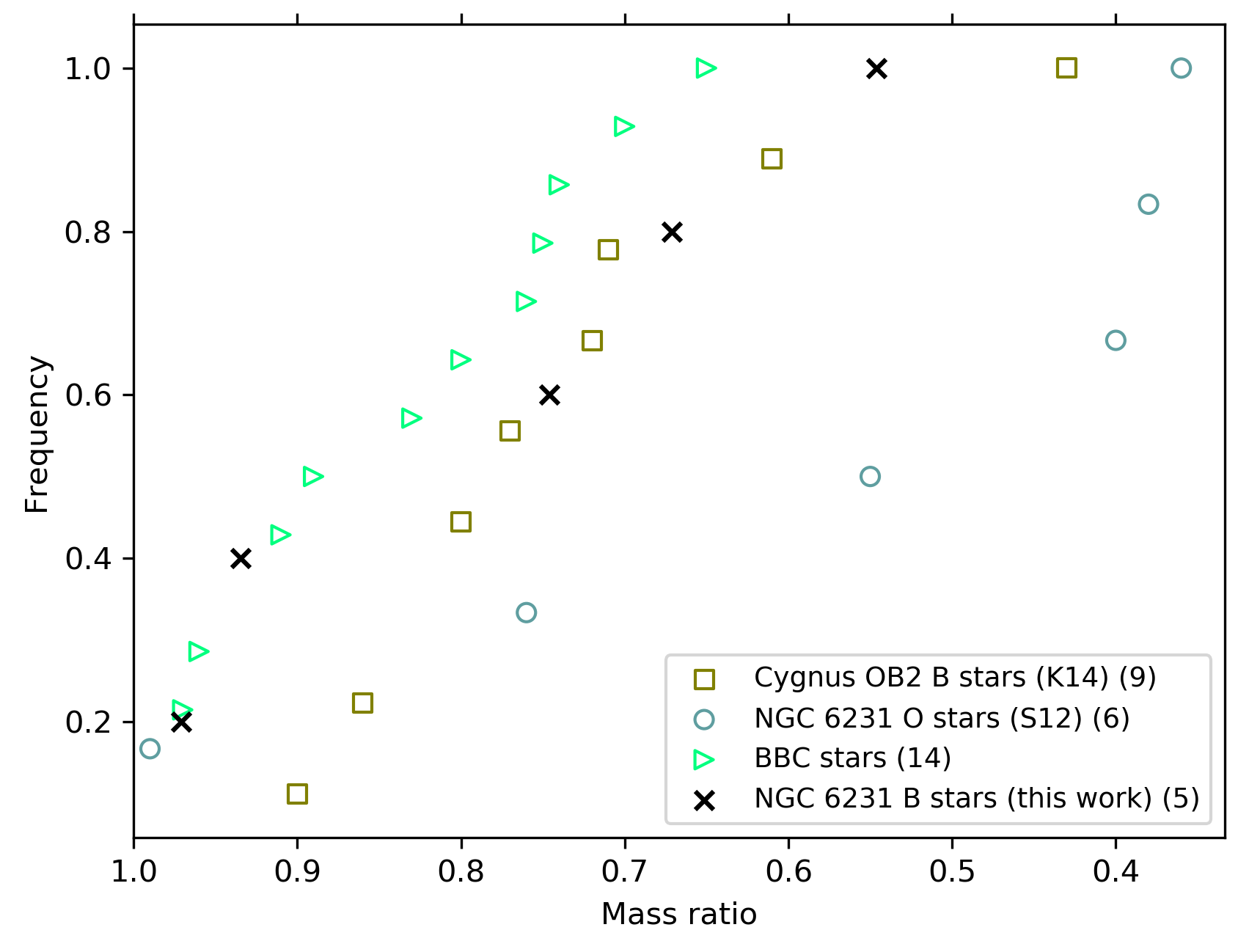}
    \caption{Observed cumulative distribution of mass ratios of SB2 systems in NGC 6231 and the equivalent observed distribution found for B-type star binaries in Cygnus OB2 \citepalias{kobulnicky_toward_2014} as olive-green squares, for O star binaries in NGC 6231 \citepalias{sana_binary_2012} as blue circles and for B-type star binaries in 30 Doradus \citepalias{villasenor_b-type_2021} as light green triangles.
         }
    \label{FigQDist2}
\end{figure}

The distribution of observed mass ratios in the 5 SB2s containing a primary B-type star found in this work is shown in Fig.~\ref{FigQDist}, and is more difficult to properly assess and compare with other results (Fig.~\ref{FigQDist2}) as the SB2s are only a small fraction of the binaries characterised (5 out of 20). For this reason, we do not present K test statistics comparing the mass ratio distributions. \par 
Nearly all of the systems have high ($q > 0.5$) mass ratios. This is not surprising, as low mass-ratio systems are difficult to find in spectroscopic SB2 systems due to the associated observational biases. There is a strong possibility that most of the low mass ratio systems are found in the sample of SB1s (the large difference in brightnesses resulting in only one signature being seen) and consequently do not contribute to this distribution. This dearth of binaries with low mass ratios is also seen in the B-type stars of Cygnus OB2 \citetalias{kobulnicky_toward_2014} and in the O-type stars of NGC 6231 \citepalias{sana_binary_2012}. The effect is even more pronounced in the \citetalias{villasenor_b-type_2021} B-type stars, with the observational bias towards equal-mass binaries likely stronger due to the increased distance to these extragalactic stars and, as such, a lower sensitivity to less massive secondary stars in the spectra. The correction for observational biases (as described in Section \ref{sec:biascorrection}) for the mass ratio distribution for the stars of NGC 6231 suggests that that there is not necessarily an overabundance of near- or equal-mass systems ('twin' systems).

\subsection{Multiplicity as a function of spectral type}
\label{ss:multiplicityspt}

\begin{figure}
\centering
\includegraphics[width=\hsize]{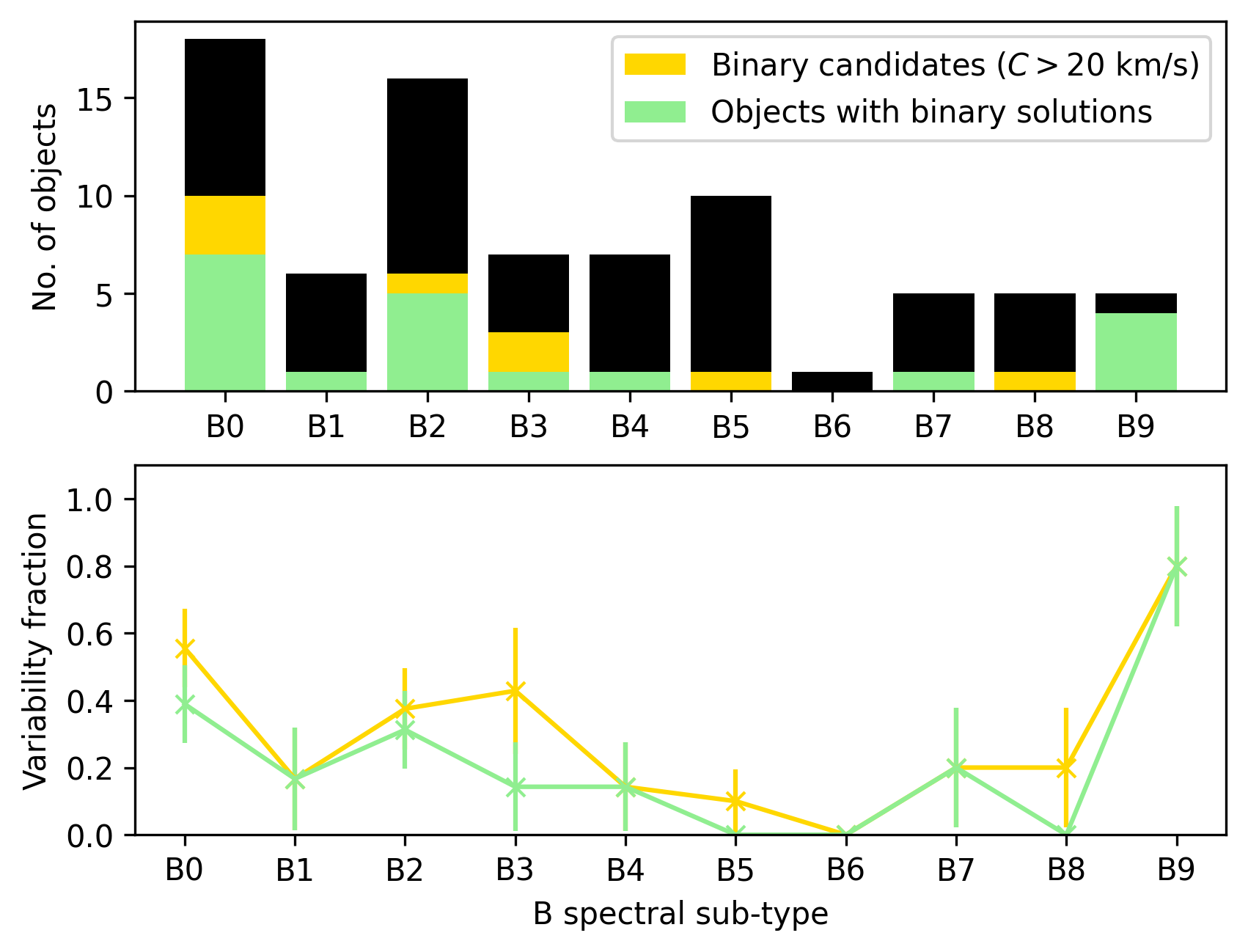}
    \caption{Frequency of spectral subtypes of B-type stars in the sample (top), where black corresponds to the total number of targets, yellow corresponds to identified binaries that meet the variability criteria in Fig.~\ref{FigBinaryFraction}, and green corresponds to systems that have an orbital solution. The variability and binary fractions as a function of the spectral sub-type are shown underneath, with their associated binomial counting uncertainties.
         }
    \label{FigSpTDist}
\end{figure}

\begin{figure}
\centering
\includegraphics[width=\hsize]{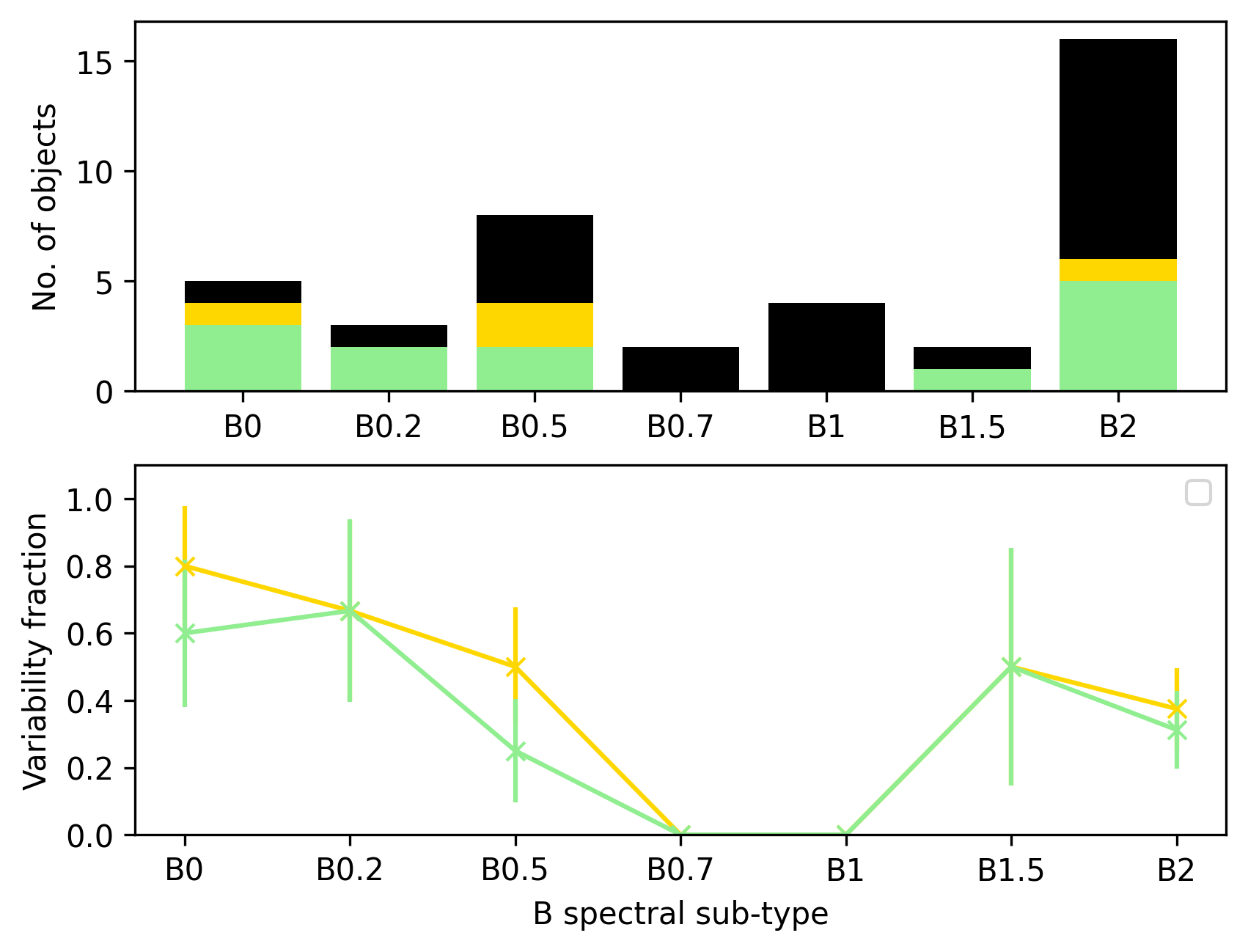}
    \caption{Frequency of spectral subtypes of B-type stars in the sample at B2 and earlier, where yellow corresponds to identified binaries that meet the variability criteria in Fig.~\ref{FigBinaryFraction}, and green corresponds to systems that have an orbital solution. The variability and binary fractions as a function of the spectral sub-type are shown underneath, with their associated binomial counting uncertainties.
         }
    \label{FigSpTDistsubB2}
\end{figure}

We have observed systems across the entire mass range of B-type stars, including massive stars and intermediate-mass objects, so we can investigate if multiplicity changes as a function of stellar mass. As mentioned in Section \ref{sec:spectraltyping}, our spectra were classified using the HERMES observations of standard stars drawn from \cite{gray_stellar_2009}, supplemented by use of the criteria from \cite{evans_vlt-flames_2015} at earlier than B2. The distribution of spectral subtypes in the sample is shown in Fig.~\ref{FigSpTDist}. We tend to observe more early B-type stars as they are brighter, with our monitoring completeness lower for later-type B-type stars in NGC 6231 (see Fig.~\ref{FigCompleteness}). The lack of B1 stars relative to their neighbouring subtypes may also be explained by the classification method, as there are several subtypes in the Evans criteria between B0 and B2 (see Fig.~\ref{FigSpTDistsubB2}), and given that the \ion{Si}{iv} $\lambda$4089 line is found in the wings of H$\delta$, it is usually difficult to use for classification, especially in fast rotators in which the line completely blends into H$\delta$. This means that the classification then heavily relies on the ratio of \ion{Mg}{ii}~4481 and \ion{Si}{III}~4553, which are of similar strength between subtypes B1 and B2. The apparent deficit in B6 stars in the sample may also be partly influenced by the lack of direct comparison of the spectra with a B6 standard - B6 stars are classified by a visual interpolation between the B5 and B7 standard spectra used. As such, we expect there to be a minimum uncertainty of one spectral subtype in our classification. \par

\begin{figure}
\centering
\includegraphics[width=\hsize]{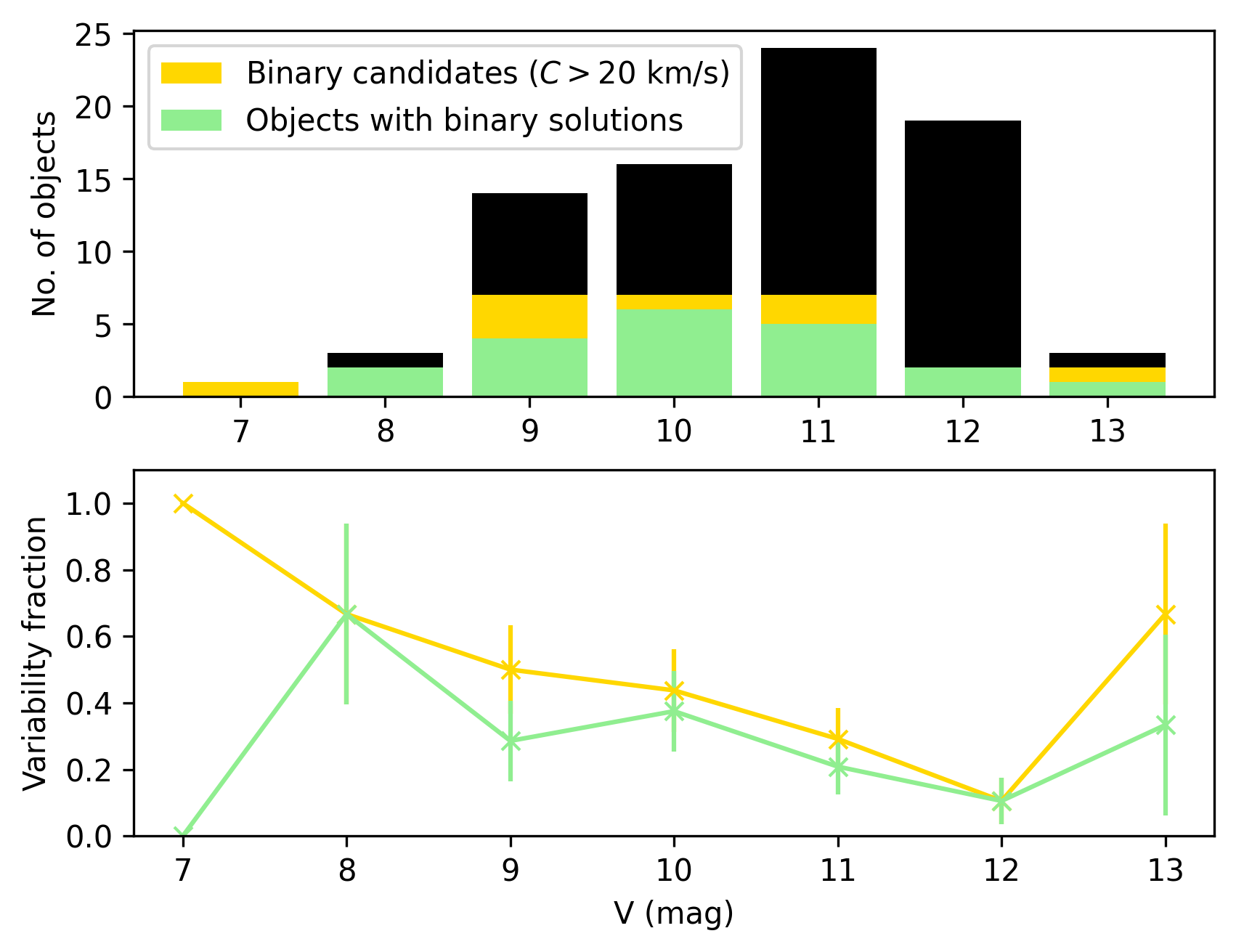}
    \caption{Frequency of magnitudes in the sample (top), where black corresponds to the total number of targets, yellow corresponds to identified binaries that meet the variability criteria in Fig.~\ref{FigBinaryFraction}, and green corresponds to systems that have an orbital solution. The variability and binary fractions as a function of the V magnitude is shown underneath, with their associated binomial counting uncertainties.
         }
    \label{FigFbinVsmag}
\end{figure}

\begin{figure}
\centering
\includegraphics[width=\hsize]{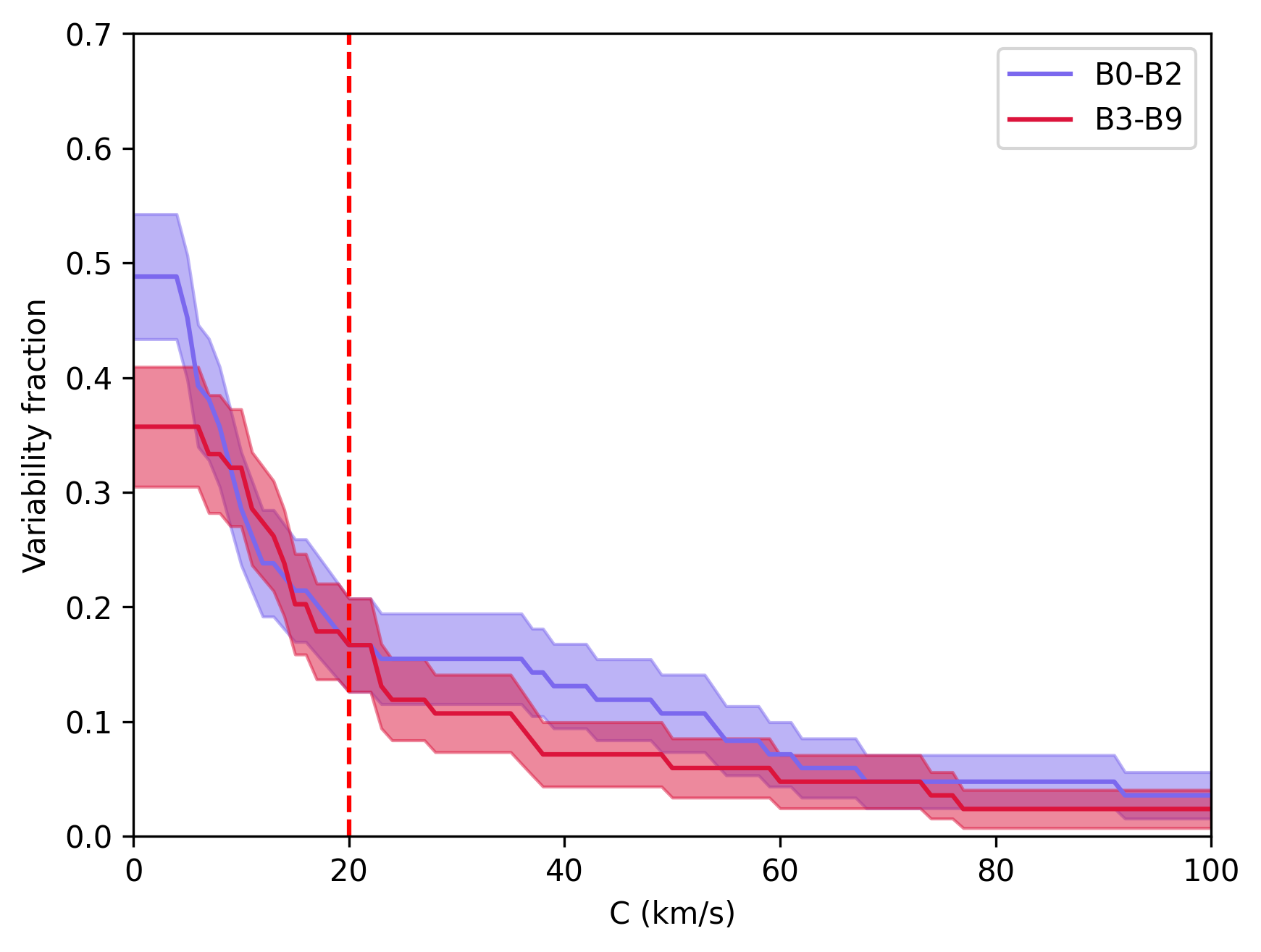}
    \caption{Same as Fig.~\ref{FigBinaryFraction} but for subsamples of earlier and later than spectral type B2.  
         }
    \label{FigBinaryFractionBins}
\end{figure}

The observed spectroscopic binary fraction as a function of spectral subtype (along with the fraction of binary candidates, which includes systems that do not have a constrained orbital solution) is shown in Fig.~\ref{FigSpTDist}. The binary fraction appears to decrease with later spectral types up until about B6, where it suddenly increases again to a considerable fraction of around 80\% for B9 stars. The same uncertainty in spectral types should be taken into account here regarding the lack of B1 binaries for example. One explanation for the considerable increasing fraction of binaries in B7 to B9 stars is that these stars are of an incomplete sample at the edge of the campaign's brightness range, and thus only the brightest systems of these types are in the sample, which are likelier to be objects containing multiple stars. \par

To further verify the potential trend of decreasing binarity with stellar mass, we can also use the apparent brightness of the star as a proxy for mass. The V band magnitude distribution for the sample and the corresponding fractions of binary candidates and constrained binaries are shown in Fig. \ref{FigFbinVsmag}. As mentioned in Section \ref{ss:targetselection}, the V magnitudes are largely taken from the photometric survey by \cite{sung_ubvri_1998}, but 10 targets not found in this survey have their V magnitudes taken from the UCAC4 \citep{zacharias_fourth_2013} and VVV \citep{minniti_vista_2010} surveys. The decreasing binary fraction with stellar mass proxy becomes more pronounced. The aforementioned potential binary selection effect at lower brightnesses still seems present, and the increased number of binary candidates without orbital solutions at the lower edge of the campaign's brightness distribution could result in part from the increased RV scatter from the noisier spectra of these objects. It should be noted that a similar trend is found for the B-type stars in NGC 330 \citep{bodensteiner_young_2020}.

The observed spectroscopic binary fraction as a function of threshold velocity for stars earlier and later than spectral type B2 is shown in Fig.~\ref{FigBinaryFractionBins}. Ideally, we would compare stars above and below spectral type B3 to more accurately probe the multiplicity properties of the CCSNe progenitors against the later type stars, but we choose B2 as the threshold so that we have similar sample sizes for both groups. It appears as the fraction of variable stars is greater overall for the earlier-type systems, but the observed spectroscopic binary fraction at the threshold velocity chosen in Section \ref{Sec:f_bin} ($C = 20$ \kms) are similar for both samples. \par

\begin{table}
\centering
\caption{Results on Kuiper tests between the period and eccentricity distributions of early (B0-B2) and late (B3-B9) NGC 6231 B-type stars, B-type star binaries in 30 Doradus \citepalias{villasenor_b-type_2021} and the B-type stars in Cygnus OB2 \citepalias{kobulnicky_toward_2014}. The spectral sub-type range is given in the brackets next to the ID of each sample.}
\label{tab:kuipertests_subtypes}
\begin{tabular}{llll}
\hline\hline
\multirow{2}{*}{Sample 1}       & \multirow{2}{*}{Sample 2} & \multicolumn{2}{l}{Kuiper test} \\
                            &                           & D             & dpp             \\
\hline
\multicolumn{4}{l}{Period distributions:}   \\
NGC 6231 (B0-B2)              & NGC 6231 (B3-B9)            & 0.55          & 38.7\%           \\
NGC 6231 (B0-B2)              & K14 (B0-B2)               & 0.43          & 30.4\%          \\
NGC 6231 (B3-B9)              & K14 (B0-B2)               & 0.51          & 38.4\%          \\
NGC 6231 (B0-B2)              & BBC (B0-B5)               & 0.42          & 18.7\%          \\
NGC 6231 (B3-B9)              & BBC (B0-B5)                      & 0.53          & 18.1\%          \\
\hline
\multicolumn{4}{l}{Eccentricity distributions:}   \\
NGC 6231 (B0-B2)              & NGC 6231 (B3-B9)            & 0.57          & 32.1\%          \\
NGC 6231 (B0-B2)              & K14 (B0-B2)               & 0.36          & 64.2\%          \\
NGC 6231 (B3-B9)              & K14 (B0-B2)               & 0.31          & 97.5\%          \\
NGC 6231 (B0-B2)              & BBC (B0-B5)               & 0.39          & 26.9\%          \\
NGC 6231 (B3-B9)              & BBC (B0-B5)               & 0.50          & 25.3\%          \\
\hline
\end{tabular}
\end{table}

\begin{figure}
\centering
\includegraphics[width=\hsize]{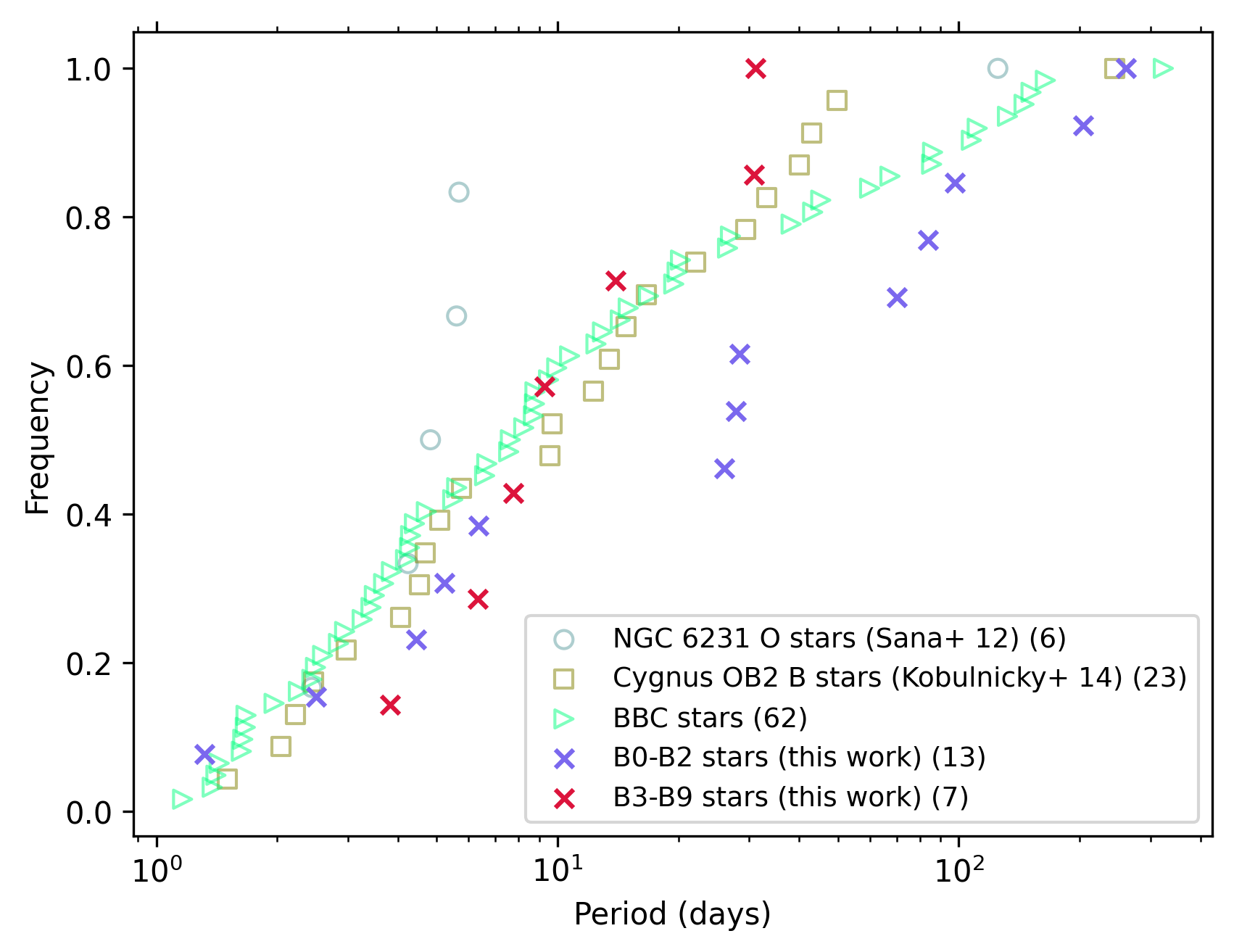}
    \caption{Observed cumulative distribution of periods of the NGC 6231 sample divided into earlier (B0-B2) and later (B3-B9) spectral subtypes, compared with previously mentioned samples seen in Figure \ref{FigPDistObs}.
         }
    \label{FigPDistSpT}
\end{figure}

\begin{figure}
\centering
\includegraphics[width=\hsize]{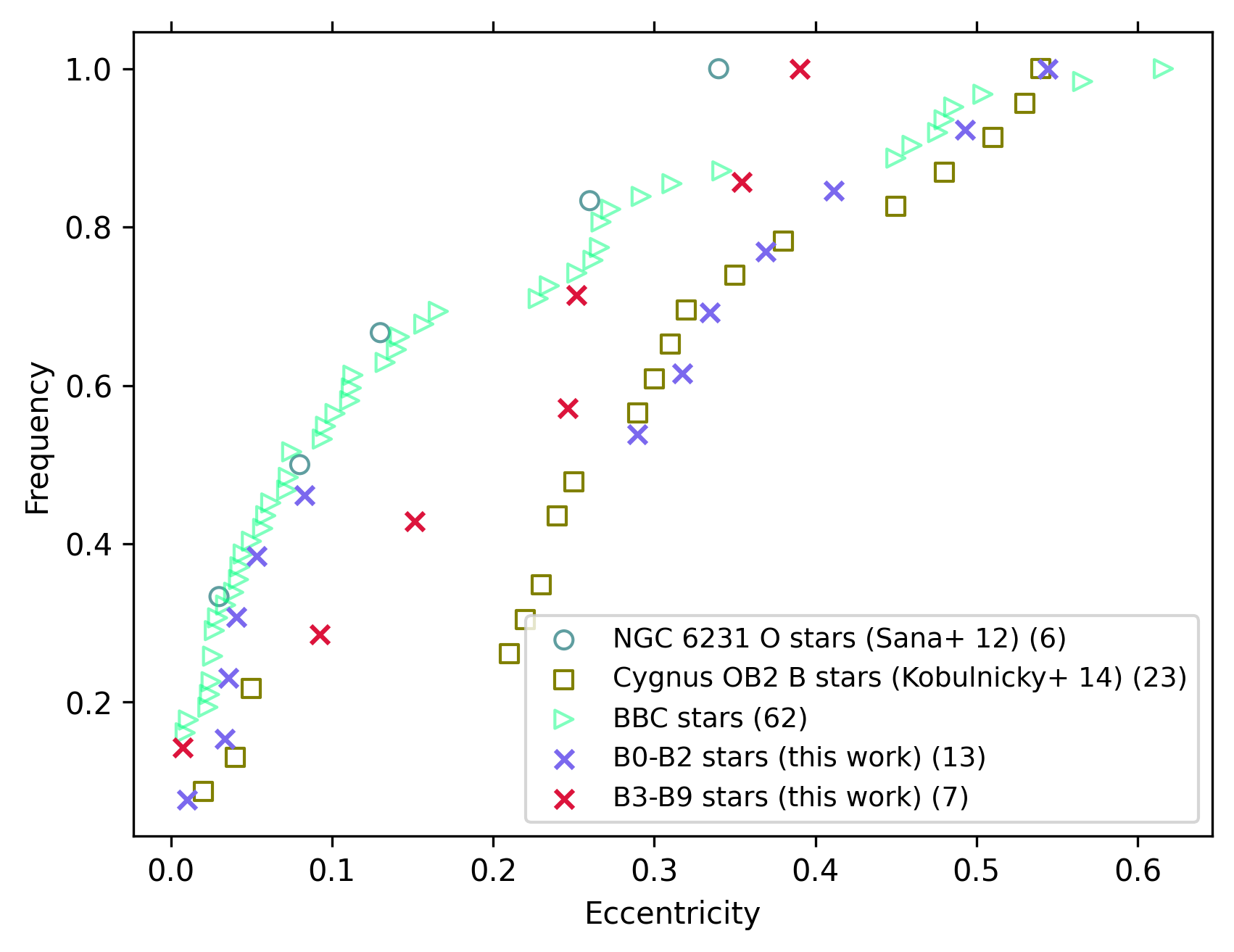}
    \caption{Observed cumulative distribution of eccentricities of the NGC 6231 sample divided into earlier (B0-B2) and later (B3-B9) spectral subtypes, compared with previously mentioned samples seen in Figure \ref{FigEDistObs}.
         }
    \label{FigEDistSpT}
\end{figure}

\subsubsection{Orbital periods by spectral type}

The properties of the early- and late-type sub-samples in our campaign appear to appear to diverge when looking at their orbital parameter distributions. The observed distribution of the periods of the stars above and below spectral type B2 is shown in Fig.~\ref{FigPDistSpT}, compared with the previously described observed period distributions of samples of stars in Cygnus OB2 \citepalias{kobulnicky_toward_2014} and O stars in NGC 6231 \citepalias{sana_binary_2012}. No systems in NGC 6231 with B-type stars later than B2 occupy orbits with periods longer than 30 days or below 3.8 days, whereas systems with B-type stars earlier than B2 occupy the period space between 1.2 and 261.4 days. The two samples appear to to have a similar slope up to periods of $P < 6$ d, and then diverge for longer periods. \par

We again compare with some of the period distributions of other samples discussed in  Section \ref{ss:orbitalperiods}, again only using stars in each sample with orbital periods below a year. The aforementioned similarities largely remain between the samples. 
It appears that both the early and late B-stars of NGC 6231 follow a similar period distribution to the other B-type star samples before the early-type B-stars of NGC 6231 diverge at around $P = 6$ d and the distributions flattens. Interestingly, the later-type sample in NGC 6231 appears to more closely follow the other observed distributions (with the exception of periods over 30 d), which all have more massive stars in them. The latest spectral type in the \citetalias{kobulnicky_toward_2014} sample is B2 III/V, B2.5 V in the \citetalias{villasenor_b-type_2021} sample, and there are the NGC 6231 O stars. That the early-type NGC 6231 period distribution appears to diverge from these samples potentially indicates the lack of influence of stellar mass on the period distribution, though it should be noted that we are comparing small subsamples. \par
Despite these differences, performing K tests between the samples (Table \ref{tab:kuipertests_subtypes}) shows that they are largely not statistically significant. There is a 38.4\% probability that the period distributions of the early- and late-type NGC 6231 samples can originate from the same distribution, which we regard as statistically insignificant. All of the other K tests performed between the various observed samples and the early- and late-type NGC 6231 B-star samples also result in statistically insignificant results. It must be stressed that we are comparing small samples of stars here in some cases, but this further suggests a universality amongst the period distributions of the various samples. We do not perform K tests between the early-/late-B type samples and the O stars in NGC 6231 as the samples become too small.

\subsubsection{Eccentricities by spectral type}

The observed distribution of the eccentricities of the B-type stars in NGC 6231 above and below spectral type B2 is shown in Fig.~\ref{FigEDistSpT}, also shown with the \citetalias{sana_binary_2012}, \citetalias{kobulnicky_toward_2014} and \citetalias{villasenor_b-type_2021} observed eccentricity distributions. Both subsamples of the NGC 6231 B-type stars have around 10\% of their stars with eccentricities below $e < 0.01$, with the two samples diverging strongly at eccentricities in the range $0.03 < e < 0.3$. For example, about 55\% of the early-type B-stars have eccentricities below $e < 0.1$, compared to just $\sim30\%$ for the later-type stars. This could be a consequence of the larger mass of the earlier-type B-stars allowing for more efficient circularisation of orbits, and the similarity to the NGC 6231 O-type star and \citetalias{villasenor_b-type_2021} observed eccentricity distributions in this low range of eccentricities supports this. The samples then generally converge again at eccentricities higher than $e > 0.3$. \par
Computing the Kuiper statistics between the samples, however, shows no statistically significant differences, either between the NGC 6231 B-type star subsamples or with any of the previously observed samples, so it remains unclear whether there is an increased circularisation efficiency observed with increasing mass.

\section{Summary}
\label{s:summary}
We have performed a multi-epoch spectroscopic campaign of 80 B-type stars in NGC 6231. Using 31 epochs of FLAMES/GIRAFFE spectra observed between 2017 and 2018, we have found an observed spectroscopic binary fraction of $33 \pm 5$\%, with an estimated true spectroscopic binary fraction of $52 \pm 8\%$ for systems with periods less than $P < 3000$ d when accounting for the impact of observational biases. This may be evidence that B-stars are less commonly found in binary systems than the more massive O-type stars - for example, a minimum (not corrected) spectroscopic binary fraction of $60 \pm 8$\% for the O-type stars in the same cluster and a bias corrected fraction of $69 \pm 9\%$ for O stars across several Galactic clusters (though this quoted value does not take into account the impact of the apparent reduction of measured RVs with increasing mass ratio and decreasing RV amplitude, unlike the NGC 6231 B-star fraction). This, along with some evidence for an increasing binary fraction as we look at earlier B subtypes, suggests that there may be some mass dependence with the binary fraction of these massive stars. We characterise the orbits of 20 out of 27 candidate binary systems, which include 15 SB1s and 5 SB2s. Out of the other nine, four are found to have a significant periodic signal but no or poorly constrained orbital solution, and five do not show any periodicity. There are also five systems below our detection threshold amplitude of dRV > 20\,\kms that have a significant periodic signal, and we flag some of these as potential long period binary system candidates (see Section \ref{app:longp} for their RV curves). \par

We find the observed distribution of periods of the B-type stars in NGC 6231 to be approximately uniform in log$P$ with small excesses of systems at roughly 7- and 30-d periods, and dearths of systems at roughly 4- and 20-d periods. There are no systems found with periods longer than $P > 260$ d, though this is likely due to the length of the observational campaign (441 d). The obtained distribution of observed periods has no statistically meaningful differences to observed period distributions produced by similar studies of O-type stars in the same cluster \citepalias{sana_binary_2012}, B-type star binary systems in the Galaxy \citepalias{kobulnicky_toward_2014,abt_frequency_1990}, and O- \citepalias{almeida_tarantula_2017} and B-type \citep{villasenor_b-type_2021} stars in the LMC. \par We do not find any late-type (B3-B9) B-type stars in NGC 6231 with a period below $P < 3.8$ d or at periods longer than $P > 30$ d, but the differences between the observed distributions of the early- (B0-B2) and late-type B-stars of NGC 6231 are not statistically significantly different. These latter comparisons are, however, comparisons with small subsamples (13 early-type binaries and 7 late-type binaries) so it is unclear whether the period distribution is truly universal with primary mass. \par 
A first order-of-magnitude correction to the observed period distribution is produced by calculating the detection probability across the period range, and the obtained distribution is less flat than the observed distribution at periods greater than around 100 days. A comparison with the other samples corrected for their respective observational biases associated with their campaigns would validify or nullify the statistical similarity between the period distributions of these stars, but we conclude that both mass and the metallicity of the environment seems to have little effect on shape of the period distributions of these stars in young open clusters within and outside the Milky Way. \par

We also find the observed distribution of eccentricities flattens $e > 0.4$, with no systems found to be more eccentric than $e = 0.55$. Comparison with the same published samples reveals similar observed eccentricity distributions with the Cygnus OB2 B-type stars, the O-type stars in NGC 6231 and the B-type stars of 30 Dor. However, compared to our sample, more eccentric systems appear to be found within the (mostly) O stars of 30 Dor (TMBM), and with a 0.22\% probability that they are drawn from the same distribution. Some evidence is found in the increased efficiency of circularisation with higher mass, as both the early-type B-stars and the O-type stars of NGC 6231 populate orbits with eccentricities $e < 0.3$ more frequently than the later-type B-stars in the same cluster. The Kuiper statistic calculated between these samples, however, does not indicate a statistically significant difference. Again, it should be noted that these two subsamples are small.\par
The results presented in this work provide robust characterisations on the multiplicity of the main progenitors of CCSNe at near-birth conditions, which can be used to inform massive star formation and evolution models along with models of synthesised stellar populations. Continuing to observe B-type stars at various evolutionary stages and metallicity environments is an important exercise in helping to further assess the impact of mass, age and environment on the multiplicity of massive stars, with a larger sample size giving less uncertainty in the binary fraction, for example. There may well also be candidates among the SB1s in this sample that contain a NS- or BH- mass unseen companion, the discovery of which would allow us to have some direct observational constraints on the frequency and nature of GW progenitor systems in populations of massive stars.

\begin{acknowledgements}
G.B.\ and H.S.\ acknowledge support from the European Research  Council  (ERC)  innovation  programme  under  the  European  Union’s DLV-772225-MULTIPLES Horizon 2020 research and innovation programme. L.M.\ thanks the European Space Agency (ESA) and the Belgian Federal Science Policy Office (BELSPO) for their support in the framework of the PRODEX Programme. J.B.\ acknowledges support from the FWO Odysseus program under project G0F8H6N. J.I.V.\ acknowledges support from CONICYT-Becas Chile, ``Doctorado en el extranjero'' programme, Grant No. 72170619. This work has made use of data from the European Space Agency (ESA) mission {\it Gaia} (\url{https://www.cosmos.esa.int/gaia}), processed by the {\it Gaia} Data Processing and Analysis Consortium (DPAC, \url{https://www.cosmos.esa.int/web/gaia/dpac/consortium}). Funding for the DPAC has been provided by national institutions, in particular the institutions participating in the {\it Gaia} Multilateral Agreement.
\end{acknowledgements}

\bibliographystyle{aa} 
\bibliography{ngc6231bib} 

\pagebreak
\begin{sidewaystable*}
\caption{Orbital solutions for confirmed SB1 systems.}             
\label{table:SB1}      
\centering
\resizebox{\textwidth}{!}{%
\begin{tabular}{l l l l l l l l l l l l l}
\hline\hline  
Object name           & V (mag) & SpT  & $\chi_2^{red}$ & $T_0 -2450000$              & $P$ (d)               & $e$               & $\omega$ ($^{\circ}$)            & $K_1$ (\kms)     & $\gamma$ (\kms)   & $a_{1}$sin$i$ ($R_{\odot}$) & $f(M) (M_{\odot})$ & RUWE \\
\hline
NGC 6231 723          & 11.41   & B3V   & 3.4            & 7896.4$\pm$0.4   & 30.93$\pm$0.05    & 0.39$\pm$0.03   & 14$\pm$5    & 15.9$\pm$0.5 & -19.5$\pm$0.4   & 9.0$\pm$0.3                 & 0.010$\pm$0.001 & 23.41     \\
HD 326328             & 10.213  & B2V   & 1.2            & 7888.6$\pm$0.1   & 28.417$\pm$0.01   & 0.49$\pm$0.01   & 85$\pm$2    & 24.4$\pm$0.4 & -27.2$\pm$0.2   & 11.9$\pm$0.2                & 0.028$\pm$0.001 & 0.97     \\
HD 152200             & 8.391   & B0V   & 2.2            & 7901.2$\pm$0.4   & 4.4440$\pm$0.0007 & 0.04$\pm$0.02*   & 190$\pm$30  & 19.5$\pm$0.4 & -2.6$\pm$0.3    & 1.71$\pm$0.03               & 0.0034$\pm$0.0002   & 0.86  \\
NGC 6231 255          & 12.842  & B9V   & 2.6            & 7899.2$\pm$0.8   & 9.271$\pm$0.008   & 0.09$\pm$0.05*   & 50$\pm$30   & 9.3$\pm$0.4  & -25.5$\pm$0.3   & 1.71$\pm$0.07               & 0.0008$\pm$0.0001   & 1.05   \\
V* V946 Sco           & 10.281  & B2V   & 32.6           & 7935$\pm$3       & 97.9$\pm$0.8      & 0.32$\pm$0.07   & 140$\pm$10  & 23$\pm$2     & -27$\pm$1       & 43$\pm$4                    & 0.11$\pm$0.03   & 0.93       \\
CD-41 11030           & 9.457   & B0.5V & 0.8            & 7869.2$\pm$0.3   & 70.06$\pm$0.05    & 0.369$\pm$0.008 & 24$\pm$1    & 31.2$\pm$0.3 & -24.8$\pm$0.2   & 40.1$\pm$0.4                & 0.176$\pm$0.006 & 1.00     \\
NGC 6231 273          & 12.770   & B9V   & 1.9            & 7866.45$\pm$0.09 & 13.948$\pm$0.005  & 0.36$\pm$0.02   & 67$\pm$3    & 25.4$\pm$0.5 & -26.4$\pm$0.3   & 6.5$\pm$0.2                 & 0.019$\pm$0.001 & 0.91     \\
CD-41 11038           & 8.756   & B0V   & 4.5            & 7881$\pm$4       & 83.9$\pm$0.9      & 0.33$\pm$0.07   & 80$\pm$10   & 9.6$\pm$0.8  & -27.9$\pm$0.5   & 15$\pm$1                    & 0.006$\pm$0.002 & 0.86     \\
NGC 6231 78           & 11.585  & B4V   & 5.0            & 7896.2$\pm$0.4   & 31.07$\pm$0.03    & 0.25$\pm$0.03   & 87$\pm$6    & 24.4$\pm$0.7 & -23.1$\pm$0.4   & 14.5$\pm$0.4                & 0.042$\pm$0.004 & 0.81     \\
NGC 6231 225          & 13.062  & B9V   & 8.2            & 7901$\pm$4       & 7.768$\pm$0.002   & 0.01$\pm$0.02*   & 200$\pm$200 & 37.4$\pm$0.7 & -27.1$\pm$0.5   & 5.7$\pm$0.1                 & 0.042$\pm$0.003 & 1.04     \\
V* V1208 Sco          & 9.708   & B0.5V & 1.7            & 7900.1$\pm$0.7   & 5.2193$\pm$0.0003 & 0.010$\pm$0.008* & 140$\pm$50  & 44.7$\pm$0.4 & -26.5$\pm$0.3   & 4.61$\pm$0.04               & 0.048$\pm$0.001 & 0.82     \\
CXOU J165421.3-415536 & 11.178  & B2V   & 2.1            & 7943$\pm$4       & 204$\pm$7         & 0.41$\pm$0.05   & 40$\pm$10   & 9.0$\pm$0.4  & -33.0$\pm$0.4   & 33$\pm$2                    & 0.012$\pm$0.002 & 0.95     \\
CPD-41 7717           & 10.21   & B2V   & 1.4            & 7898.4$\pm$0.2   & 2.5080$\pm$0.0001 & 0.04$\pm$0.02*   & 200$\pm$20  & 18.0$\pm$0.3 & -27.3$\pm$0.2   & 0.89$\pm$0.02               & 0.00152$\pm$0.00008 & 0.94 \\
CPD-41 7722           & 10.017  & B0V   & 1.1            & 7894.4$\pm$0.2   & 27.877$\pm$0.009  & 0.29$\pm$0.01   & 32$\pm$2    & 35.5$\pm$0.4 & -24.5$\pm$0.3   & 18.7$\pm$0.2                & 0.114$\pm$0.004 & 0.83     \\
CPD-41 7746           & 9.24    & B0.2V & 11.8           & 7897.2$\pm$0.6   & 6.3498$\pm$0.0008 & 0.03$\pm$0.01*   & 130$\pm$30  & 62$\pm$1     & -32.3$\pm$0.7   & 7.8$\pm$0.1                 & 0.157$\pm$0.008 & 0.94    \\
\hline
\end{tabular}%
}
\flushleft
\footnotesize * Eccentricities that do not pass the Lucy-Sweeney test ($e/\sigma_{e}\leq\,2.49$)
\end{sidewaystable*}

\begin{sidewaystable*}
\caption{Orbital solutions for confirmed SB2 systems.}
\label{table:SB2}      
\centering
\resizebox{\textwidth}{!}{
\begin{tabular}{l l l l l l l l l l l l l l l l l}
\hline\hline
Object name  & V (mag) & SpT       & $\chi_2^{red}$ & $T_0 -2450000$            & $P$ (d)             & $e$           & $\omega$ ($^{\circ}$)         & $K_1$ (\kms) & $K_2$ (\kms) & $\gamma$ (\kms) & $q$           & $a_{1}$sin$i$ ($R_{\odot}$) & $a_{2}$sin$i$ ($R_{\odot}$) & $m_{1}$sin${^3}i$ ($M_{\odot}$) & $m_{2}$sin${^3}i$ ($M_{\odot}$) & RUWE \\
\hline
HD 326343    & 10.579  & B1.5V + B4V & 16.3           & 7886$\pm$0.1     & 26.06$\pm$0.01      & 0.54$\pm$0.02   & 189$\pm$2   & 59$\pm$2     & 89$\pm$5     & -22.7$\pm$0.8   & 0.67$\pm$0.06 & 40$\pm$2                    & 60$\pm$3                    & 3.1$\pm$0.4                     & 2.1$\pm$0.2 & 0.92                     \\
CPD-41 7727  & 9.424   & B2V + B1III   & 1.8            & 7740$\pm$10      & 260.7$\pm$0.8       & 0.08$\pm$0.01   & 290$\pm$10  & 25$\pm$1     & 34$\pm$3     & -25.3$\pm$0.9   & 0.74$\pm$0.07 & 200$\pm$10                  & 270$\pm$20                  & 3.0$\pm$0.5                     & 2.3$\pm$0.3  & 0.92                     \\
NGC 6231 223 & 11.719  & B7V + B8V   & 56.6           & 7898.6$\pm$0.1   & 6.3200$\pm$0.0009   & 0.15$\pm$0.02   & -16$\pm$7   & 87$\pm$3     & 94$\pm$4     & -25.2$\pm$1.5   & 0.93$\pm$0.04 & 16.9$\pm$0.5                & 18.2$\pm$0.9                & 2.0$\pm$0.2                     & 1.8$\pm$0.1   & 0.93                     \\
V* V1293 Sco & 10.2    & B0.2V + B2V & 132.5          & 7899.34$\pm$0.05 & 1.31771$\pm$0.00003 & 0.05$\pm$0.01   & 100$\pm$10  & 166$\pm$2    & 300$\pm$10   & -34.8$\pm$1.5   & 0.55$\pm$0.07 & 6.76$\pm$0.08               & 12.4$\pm$0.5                & 9.1$\pm$0.8                     & 5.0$\pm$0.3   & 1.23                     \\
NGC 6231 189 & 11.573  & B9V + B9V   & 22.2           & 7899.54$\pm$0.06 & 3.8275$\pm$0.0004   & 0.25$\pm$0.022  & -39$\pm$5   & 75$\pm$2     & 77$\pm$3     & -26.5$\pm$1.3   & 0.97$\pm$0.04 & 8.6$\pm$0.2                 & 8.9$\pm$0.4                 & 0.65$\pm$0.06                   & 0.63$\pm$0.04   & 0.94
                  \\        
\hline
\end{tabular}%
}
\end{sidewaystable*}

\appendix

\section{Target spectra}
\label{app:spectra}

\begin{figure*}
\centering
\includegraphics[width=0.9\textwidth]{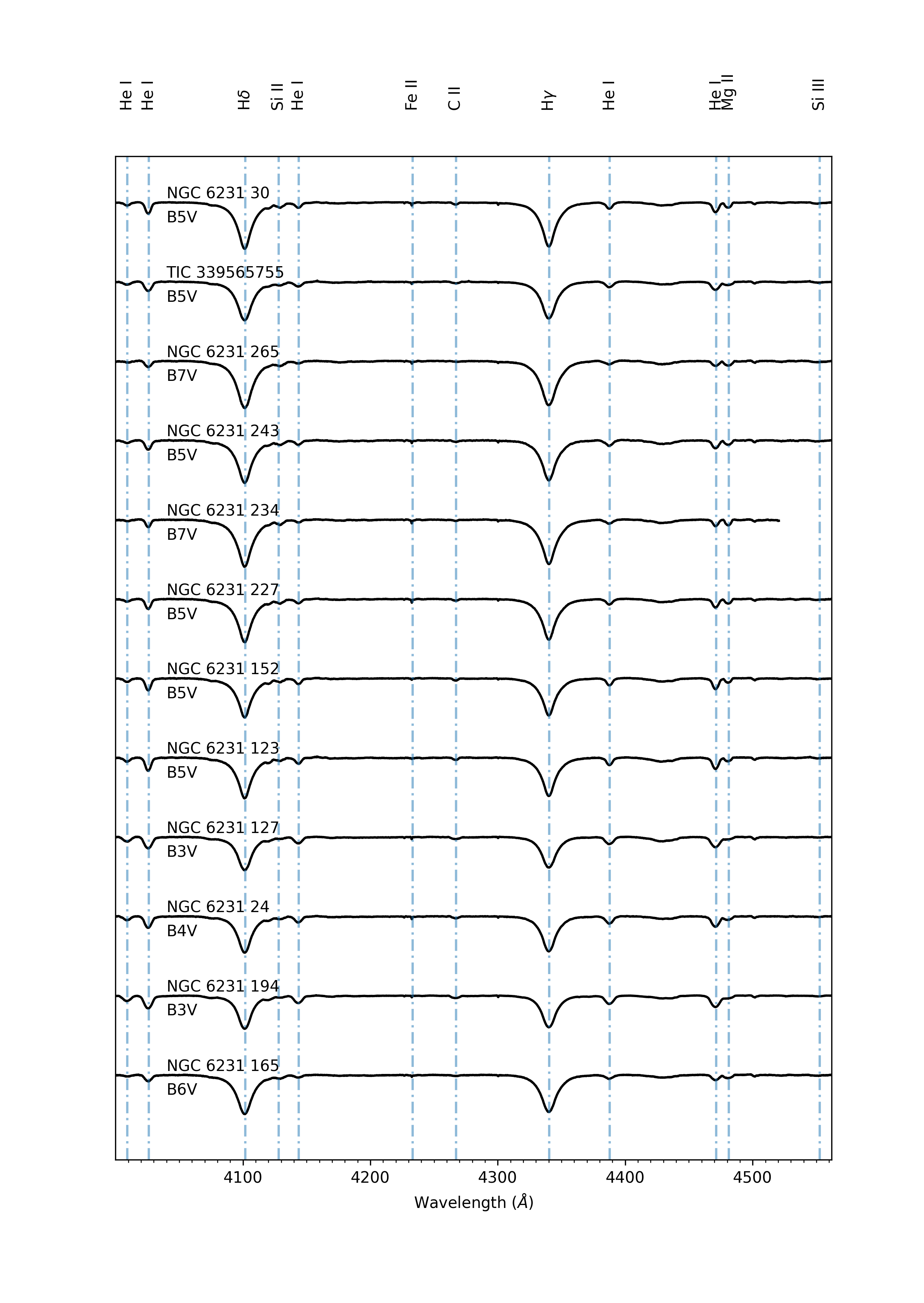}
    \caption{Medians of VLT/FLAMES spectra for the NGC 6231 stars considered stable in RV (Table \ref{table:nonvariable}).
         }
    \label{FigNonVarSpec}
\end{figure*}

\begin{figure*}
\centering
\includegraphics[width=0.9\textwidth]{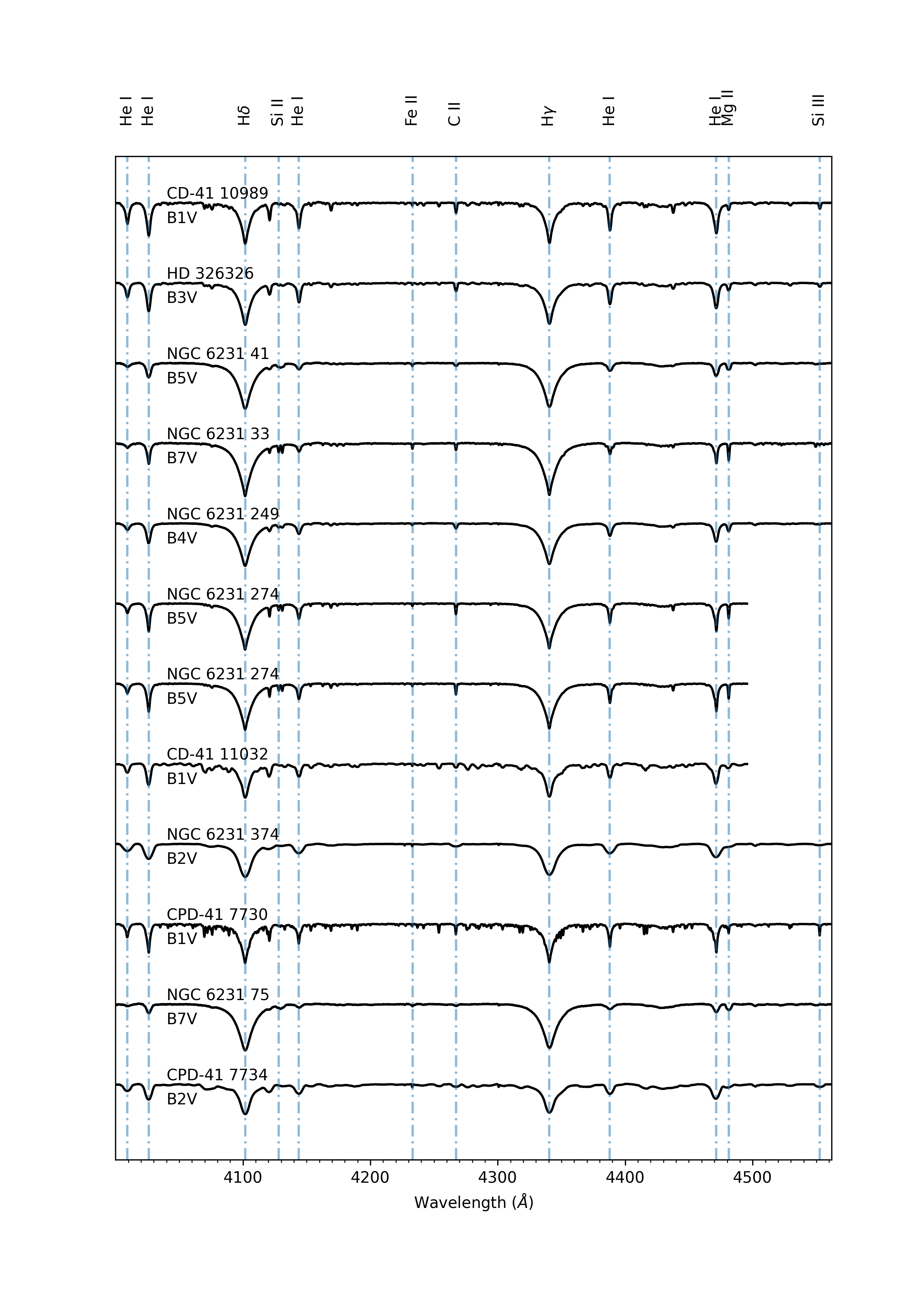}
    \caption{Shift-and-added VLT/FLAMES spectra for the NGC 6231 B-type stars considered variable in RV with no periodicity found (Table \ref{table:variablenoperiod}).}
    \label{FigVarNoPSpec1}
\end{figure*}

\begin{figure*}
\centering
\includegraphics[width=0.9\textwidth]{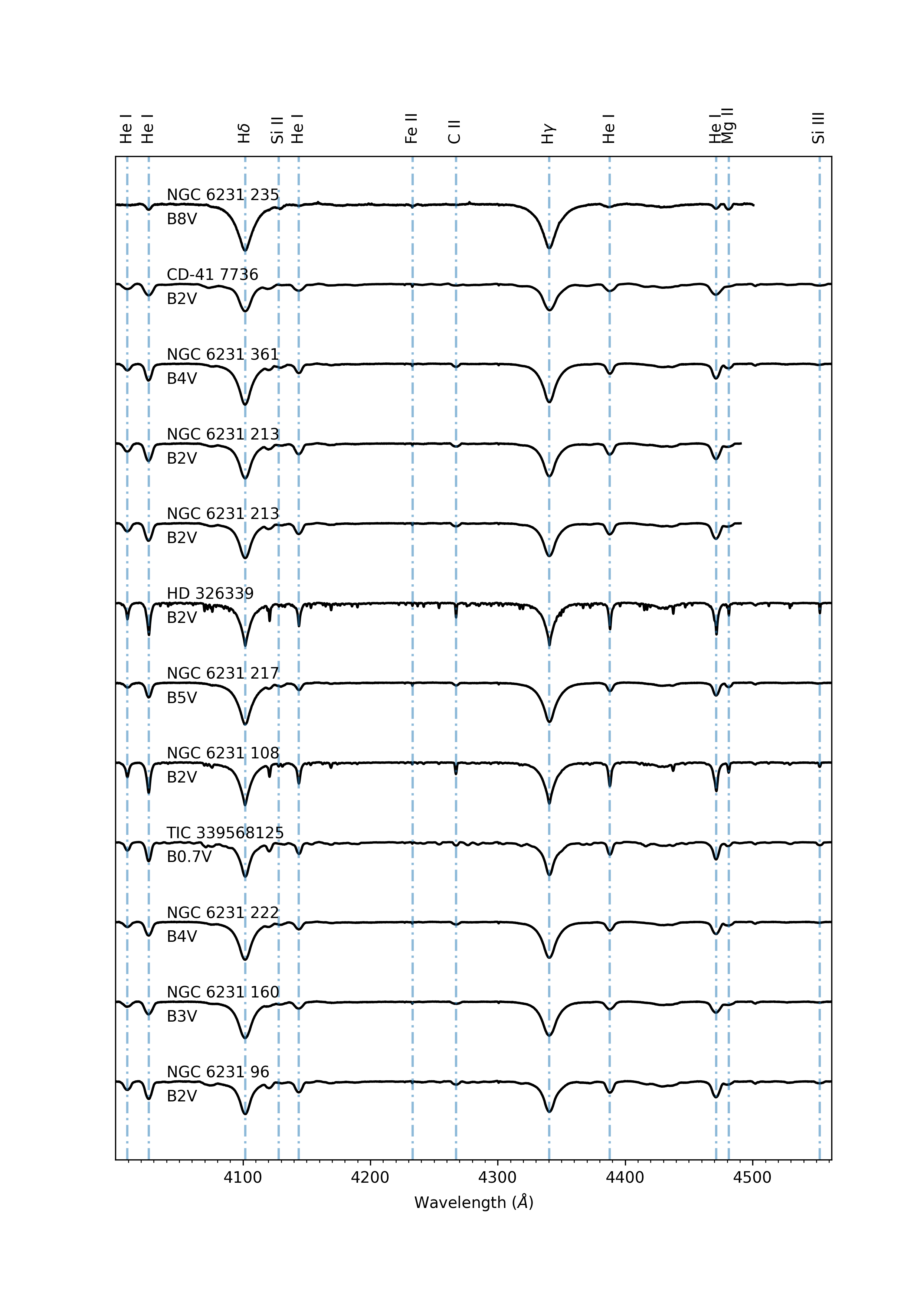}
    \caption{More shift-and-added VLT/FLAMES spectra for the NGC 6231 B-type stars considered variable in RV with no periodicity found (Table \ref{table:variablenoperiod}). 
         }
    \label{FigVarNoPSpec2}
\end{figure*}

\begin{figure*}
\centering
\includegraphics[width=0.9\textwidth]{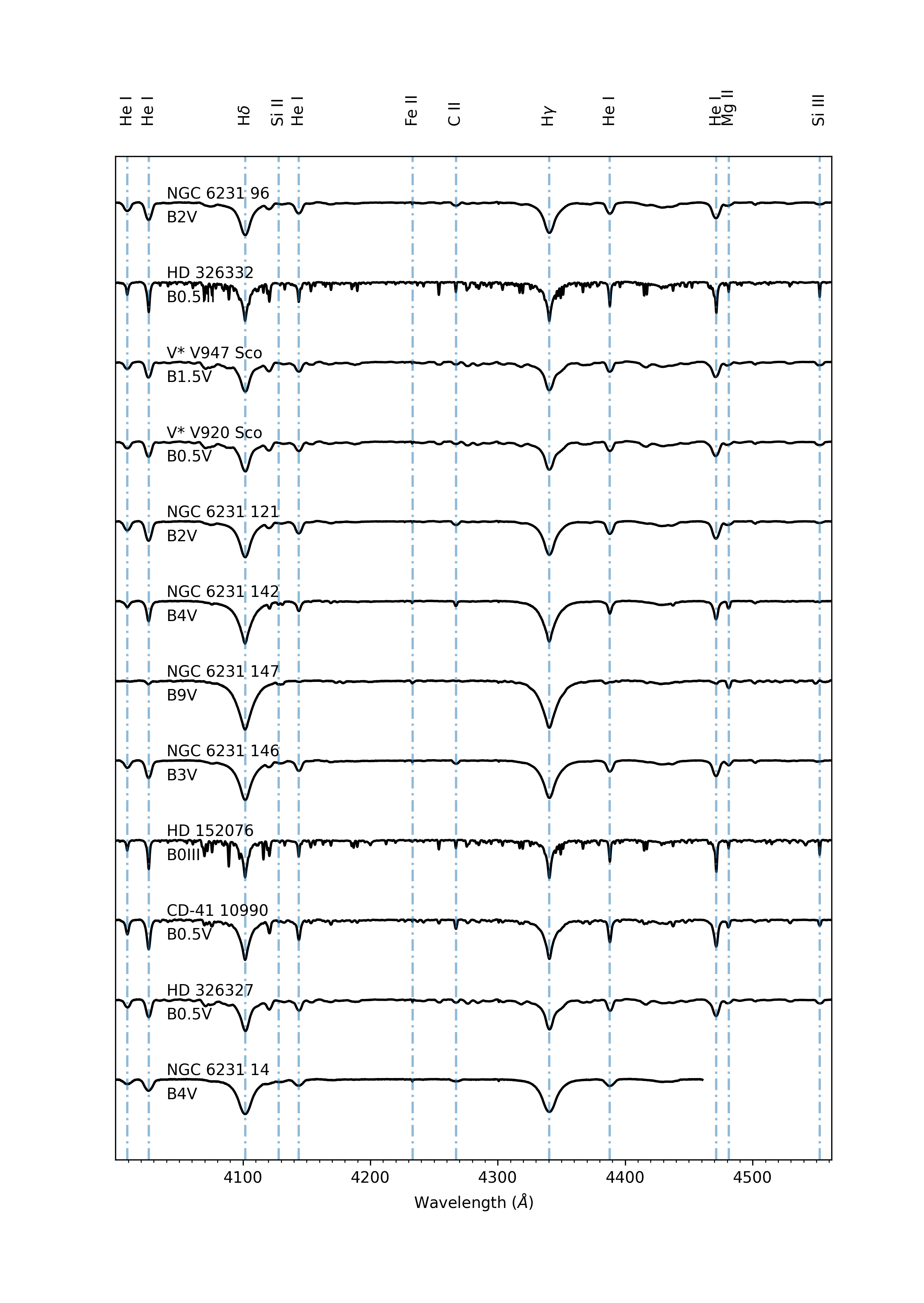}
    \caption{More shift-and-added VLT/FLAMES spectra for the NGC 6231 B-type stars considered variable in RV with no periodicity found (Table \ref{table:variablenoperiod}).}
    \label{FigVarNoPSpec3}
\end{figure*}

\begin{figure*}
\centering
\includegraphics[width=0.9\textwidth]{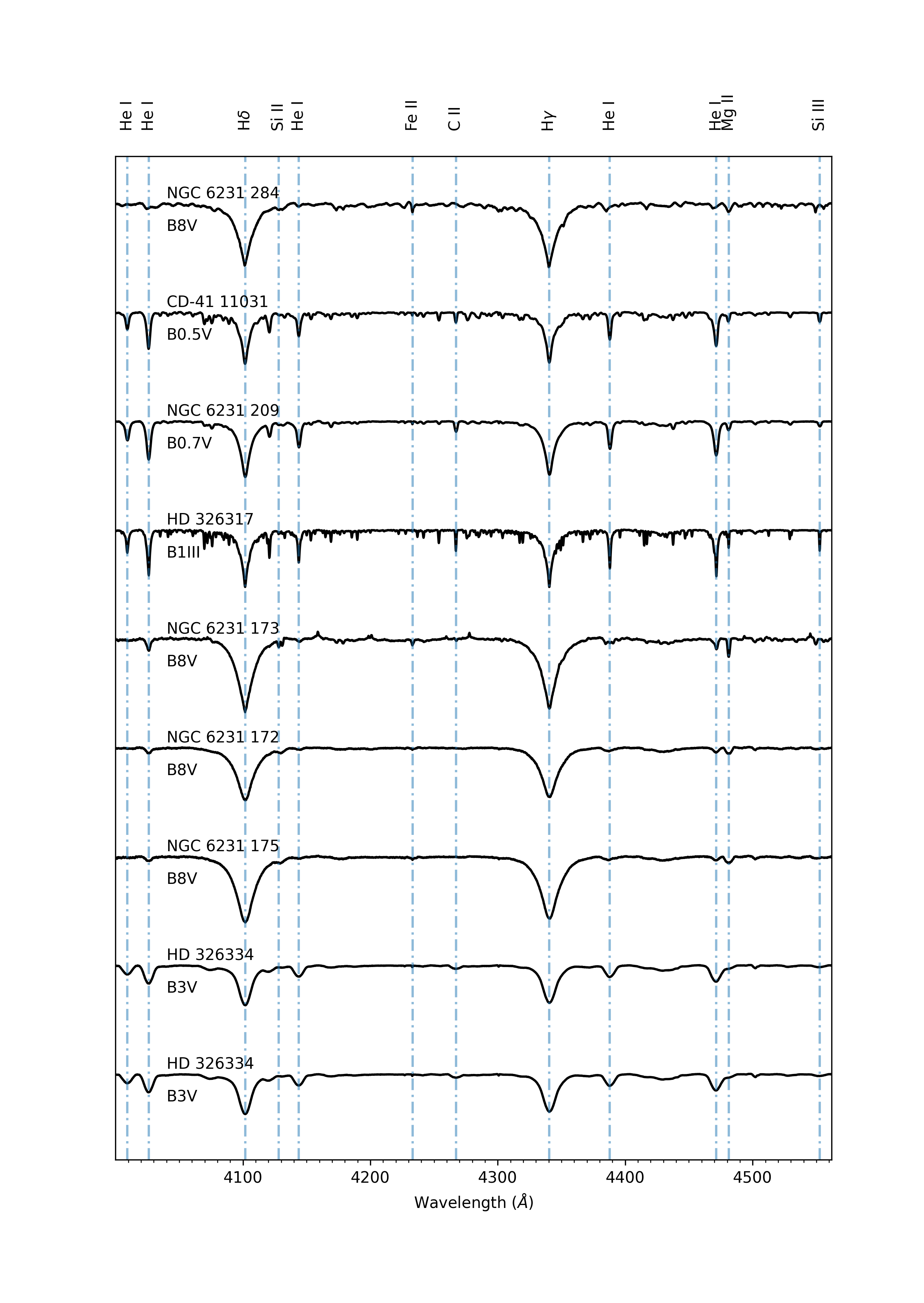}
    \caption{More shift-and-added VLT/FLAMES spectra for the NGC 6231 B-type stars considered variable in RV with no periodicity found (Table \ref{table:variablenoperiod}).
         }
    \label{FigVarNoPSpec4}
\end{figure*}

\begin{figure*}
\centering
\includegraphics[width=0.9\textwidth]{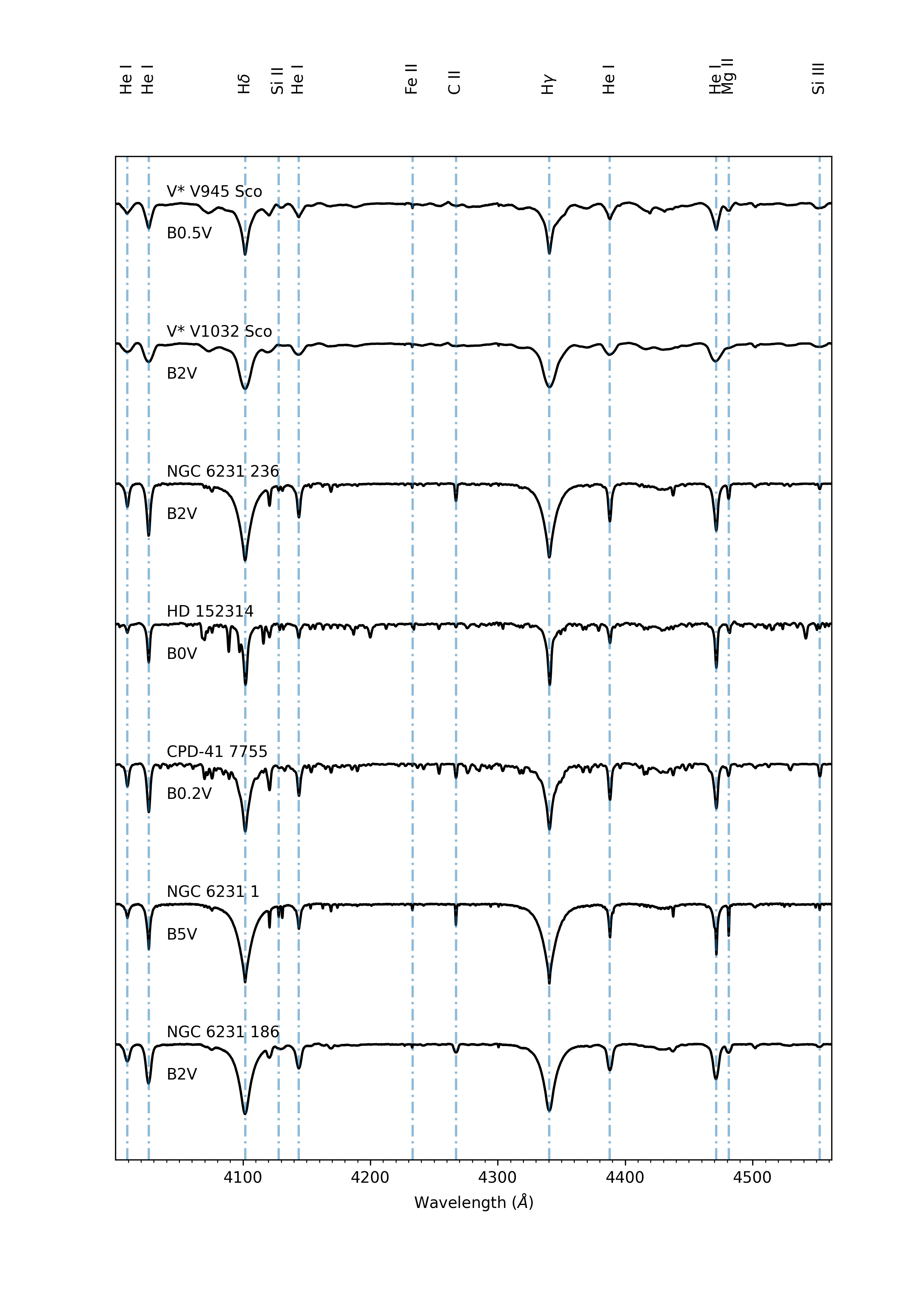}
    \caption{Shift-and-added spectra VLT/FLAMES spectra of targets with  a  significant  periodic  signal  but  no  or  poorly constrained orbital solution (Table \ref{table:periodnoorbsol}).
         }
    \label{FigPNoOrb}
\end{figure*}

\begin{figure*}
\centering
\includegraphics[width=0.9\textwidth]{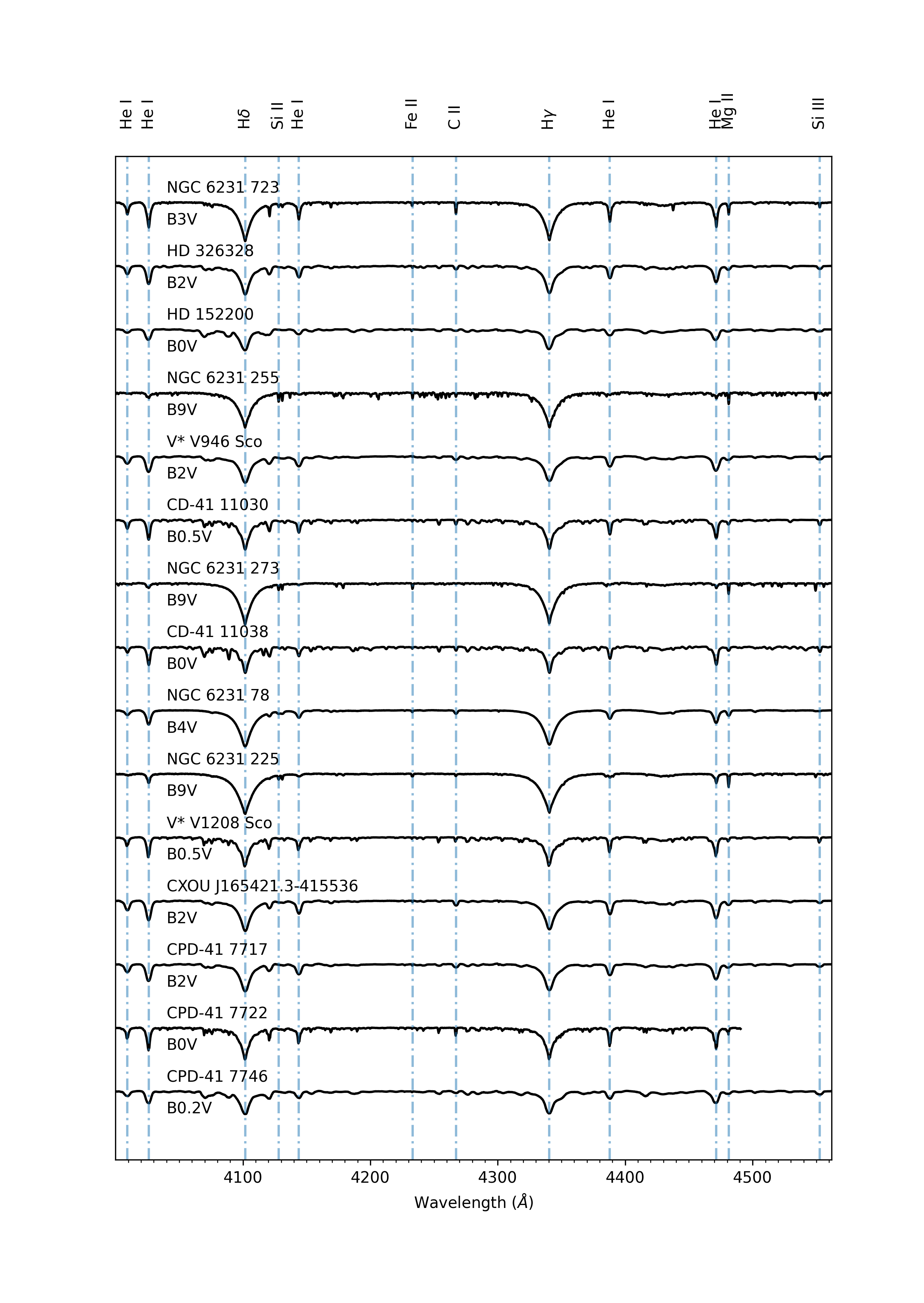}
    \caption{Shift-and-added VLT/FLAMES spectra for the SB1 B-type stars in NGC 6231 (Table \ref{table:SB1}).
         }
    \label{FigSB1Spec}
\end{figure*}

\begin{figure*}
\centering
\includegraphics[width=0.9\textwidth]{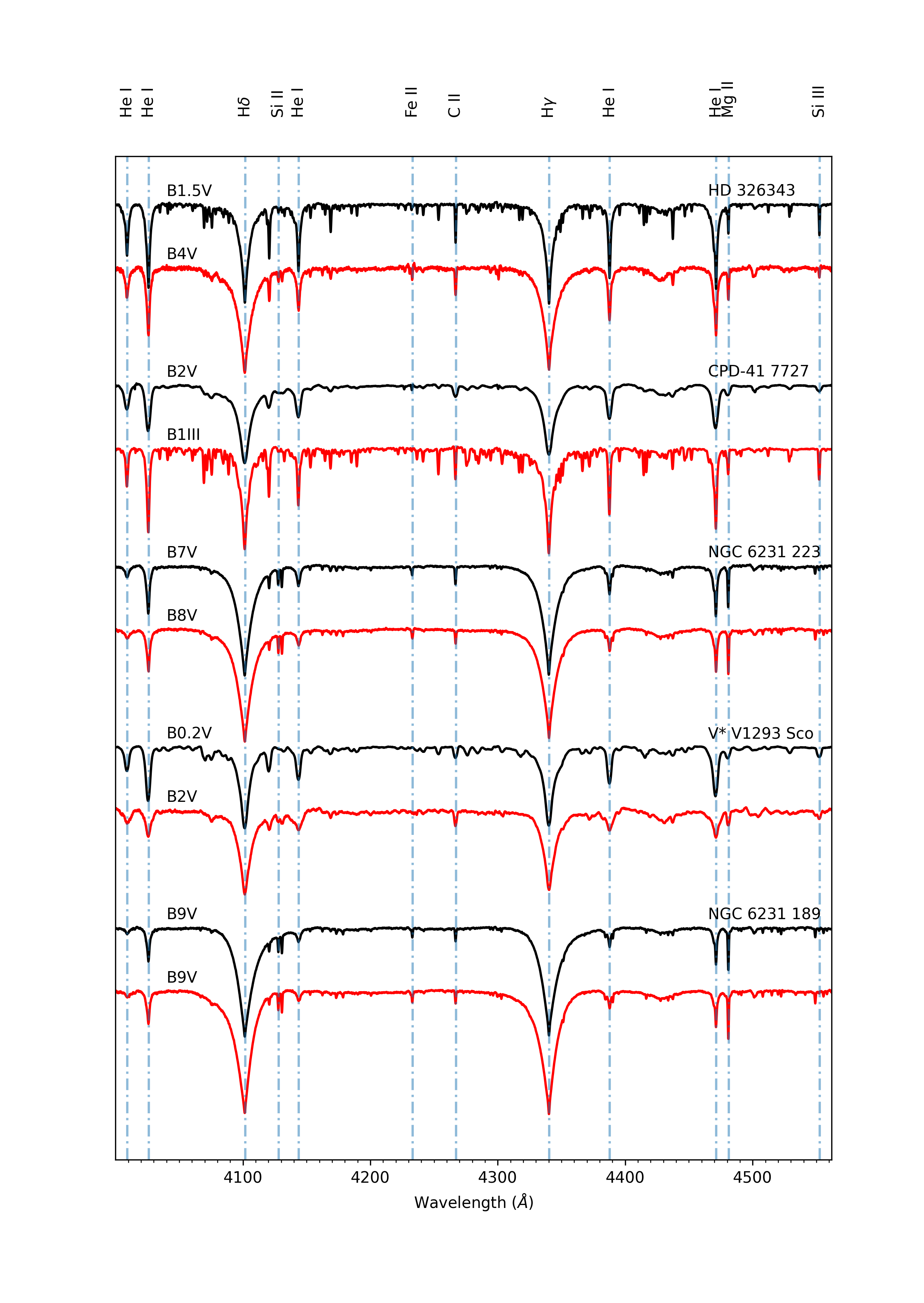}
    \caption{Disentangled VLT/FLAMES spectra for the NGC 6231 B-type SB2s (Table \ref{table:SB2}).
         }
    \label{FigSB2Spec}
\end{figure*}

\pagebreak
\newpage

\section{Long period binary candidates}
\label{app:longp}

\begin{figure*}
\centering
\includegraphics[width=0.9\textwidth]{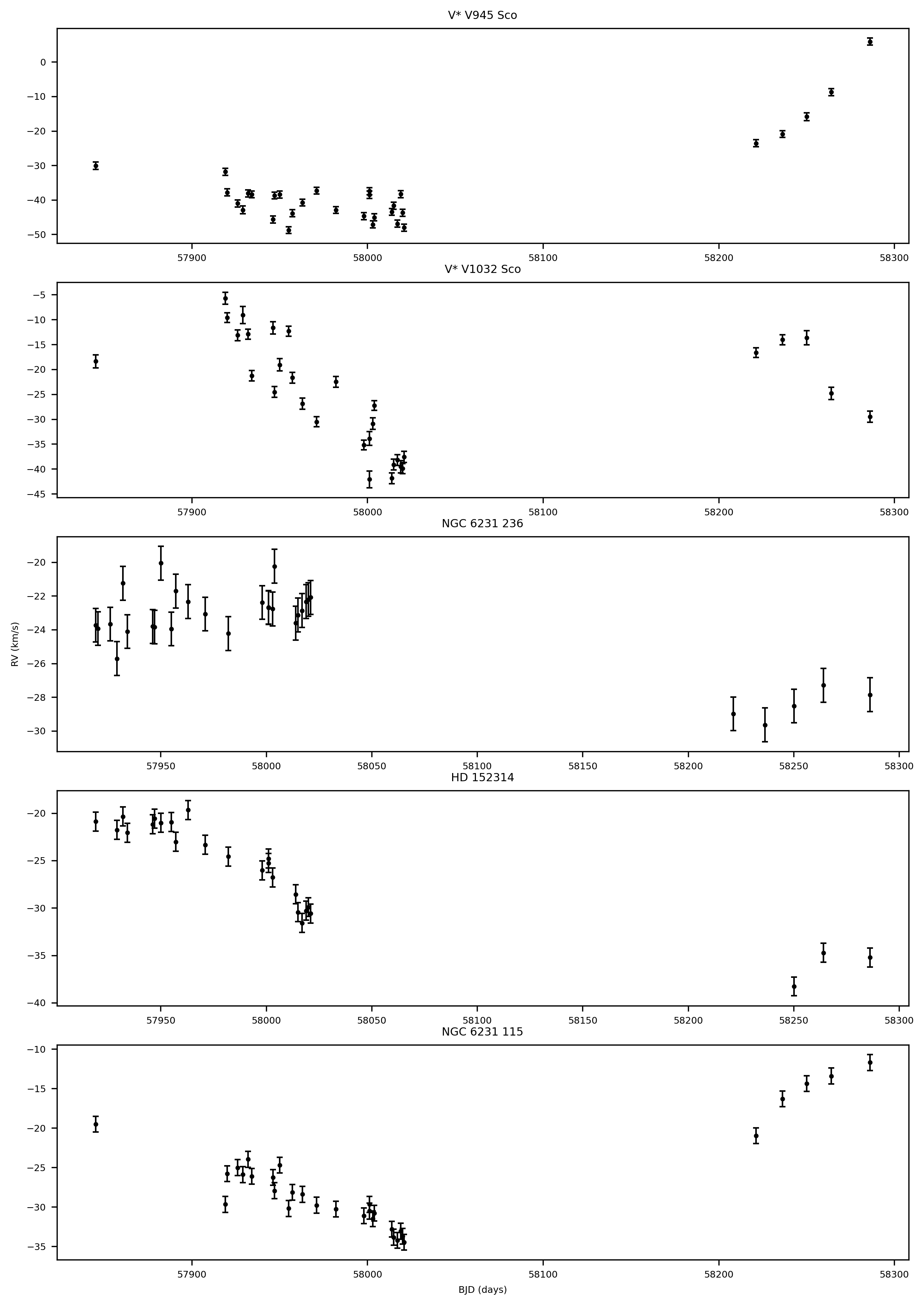}
    \caption{RV curves for B-type stars with identified periodicities but no constrained orbital solution (Table \ref{table:periodnoorbsol}) that appear to modulate in RV smoothly over the duration of the campaign, making them potential binary systems with periods longer than the available baseline of observations.}
         
    \label{FigLongP}
\end{figure*}

\pagebreak
\newpage

\section{RV measurements}
\label{app:rvtables}

\begin{table*}
\centering
\caption{Measured RVs (in \kms) with associated uncertainties for the confirmed SB2 systems (Table \ref{table:SB2}). Dashes (-) indicate RV measurements that were not taken either due to the object not being observed at the given date or due to sub-optimal data quality. (1) designates the primary star and (2) the secondary star in each SB2 system. A full version of this table is available electronically. The first few lines are shown as an example.}
\label{tab:sb2rvs}
\begin{tabular}{lllllll}
\hline\hline
Object Name      & \multicolumn{6}{l}{BJD (days)}                                             \\
\hline
                 & 57845.27 & 57919.16     & 57920.19     & 57926.04     & 57929.19     & ... \\
\hline
HD 326343 (1)    & -        & 8 $\pm$ 1    & -2 $\pm$ 1   & -6 $\pm$ 1   & -8 $\pm$ 1   & ... \\
HD 326343 (2)    & -        & -57 $\pm$ 4  & -38 $\pm$ 7  & -68 $\pm$ 1  & -72 $\pm$ 3  & ... \\
CPD-41 7727 (1)  & -        & -23 $\pm$ 5  & -52 $\pm$ 3  & -51 $\pm$ 2  & -49 $\pm$ 4  & ... \\
CPD-41 7727 (2)  & -        & -35 $\pm$ 1  & 5.8 $\pm$ 1  & 70 $\pm$ 1   & 6 $\pm$ 1    & ... \\
NGC 6231 223 (1) & -        & -            & -35 $\pm$ 1  & 36 $\pm$ 1   & 14 $\pm$ 1   & ... \\
NGC 6231 223 (2) & -        & -            & -12 $\pm$ 1  & -95 $\pm$ 1  & -69 $\pm$ 1  & ... \\
V* V1293 Sco (1) & -        & -112 $\pm$ 1 & 103 $\pm$ 1  & -189 $\pm$ 1 & 105 $\pm$ 1  & ... \\
V* V1293 Sco (2) & -        & 76 $\pm$ 5   & -278 $\pm$ 2 & 282 $\pm$ 5  & -281 $\pm$ 4 & ... \\
NGC 6231 189 (1) & -        & -55 $\pm$ 2  & 54 $\pm$ 2   & -43 $\pm$ 1  & 6 $\pm$ 2    & ... \\
NGC 6231 189 (2) & -        & 13 $\pm$ 3   & -101 $\pm$ 2 & -2 $\pm$ 1   & -60 $\pm$ 2  & ... \\
\hline
\end{tabular}
\end{table*}

\begin{table*}
\centering
\caption{Measured RVs with associated uncertainties for the confirmed SB1 systems (Table \ref{table:SB1}). Dashes (-) indicate RV measurements that were not taken either due to the object not being observed at the given date or due to sub-optimal data quality. A full version of this table is available electronically. The first few lines are shown as an example.}
\label{tab:sb1rvs}
\begin{tabular}{lllllll}
\hline\hline
Object name           & \multicolumn{6}{l}{BJD (days)}                                           \\
\hline
                      & 57845.27 & 57919.16    & 57920.19    & 57926.04      & 57929.19    & ... \\
\hline
NGC 6231 723          & -        & -12 $\pm$ 1 & -           & -21 $\pm$ 1   & -40 $\pm$ 1 & ... \\
HD 326328             & -        & -32 $\pm$ 1 & -51 $\pm$ 1 & -50 $\pm$ 1   & -41 $\pm$ 1 & ... \\
HD 152200             & -        & 16 $\pm$ 1  & -21 $\pm$ 1 & 5 $\pm$ 1     & 10 $\pm$ 1  & ... \\
NGC 6231 255          & -        & -33 $\pm$ 1 & -30 $\pm$ 1 & -32 $\pm$ 1   & -16 $\pm$ 1 & ... \\
V* V946 Sco           & -        & -12 $\pm$ 1 & -           & -             & -44 $\pm$ 1 & ... \\
CD-41 11030           & -        & -33 $\pm$ 1 & -26 $\pm$ 1 & -26 $\pm$ 1   & -13 $\pm$ 1 & ... \\
NGC 6231 273          & -        & -21 $\pm$ 1 & -46 $\pm$ 1 & -52 $\pm$ 1   & -6 $\pm$ 1  & ... \\
CD-41 11038           & -        & -32 $\pm$ 1 & -           & -             & -29 $\pm$ 1 & ... \\
NGC 6231 78           & -        & -12 $\pm$ 1 & -45 $\pm$ 1 & -45 $\pm$ 1   & -34 $\pm$ 1 & ... \\
NGC 6231 225          & -        & -61 $\pm$ 1 & 7 $\pm$ 1   & 1 $\pm$ 1     & -7 $\pm$ 1  & ... \\
V* V1208 Sco          & -        & 5 $\pm$ 1   & 16 $\pm$ 1  & -17.9 $\pm$ 1 & -57 $\pm$ 1 & ... \\
CXOU J165421.3-415536 & -        & -38 $\pm$ 1 & -23 $\pm$ 1 & -23 $\pm$ 1   & -24 $\pm$ 1 & ... \\
CPD-41 7717           & -        & -15 $\pm$ 1 & -38 $\pm$ 1 & -26 $\pm$ 1   & -15 $\pm$ 1 & ... \\
CPD-41 7722           & -        & -4 $\pm$ 1  & -           & -             & -62 $\pm$ 1 & ... \\
CPD-41 7746           & -        & -4 $\pm$ 1  & -15 $\pm$ 1 & 29 $\pm$ 1    & 13 $\pm$ 1  & ... \\
\hline
\end{tabular}
\end{table*}

\begin{table*}
\centering
\caption{Measured RVs (in \kms) with associated uncertainties for the targets with a significant periodic signal but no or poorly constrained orbital solution (Table \ref{table:periodnoorbsol}). Dashes (-) indicate RV measurements that were not taken either due to the object not being observed at the given date or due to sub-optimal data quality. A full version of this table is available electronically. The first few lines are shown as an example.}
\label{tab:periodnoorbsolrvs}
\begin{tabular}{lllllll}
\hline\hline
Object name  & \multicolumn{6}{l}{BJD (days)}                                         \\
\hline
             & 57845.27 & 57919.16    & 57920.19    & 57926.04    & 57929.19    & ... \\
\hline
V* V945 Sco  & -        & -30 $\pm$ 1 & -32 $\pm$ 1 & -38 $\pm$ 1 & -41 $\pm$ 1 & ... \\
V* V1032 Sco & -        & -18 $\pm$ 1 & -6 $\pm$ 1  & -10 $\pm$ 1 & -13 $\pm$ 1 & ... \\
NGC 6231 236 & -        & -           & -24 $\pm$ 1 & -24 $\pm$ 1 & -24 $\pm$ 1 & ... \\
HD 152314 & -        & -           & -           & -           & -21 $\pm$ 1 & ... \\
CPD-41 7755  & -        & -31 $\pm$ 1 & -33 $\pm$ 1 & -38 $\pm$ 1 & -29 $\pm$ 1 & ... \\
NGC 6231 1   & -        & -27 $\pm$ 1 & -26 $\pm$ 1 & -15 $\pm$ 1 & -5 $\pm$ 1  & ... \\
NGC 6231 186 & -        & -29 $\pm$ 1 & -26 $\pm$ 1 & -27 $\pm$ 1 & -30 $\pm$ 1 & ... \\
\hline
\end{tabular}
\end{table*}

\begin{table*}
\centering
\caption{Measured RVs (in \kms) with associated uncertainties for the targets considered RV variables without a significant periodic signal (Table \ref{table:variablenoperiod}). Dashes (-) indicate RV measurements that were not taken either due to the object not being observed at the given date or due to sub-optimal data quality. A full version of this table is available electronically. The first few lines are shown as an example.}
\label{tab:varnoperiodrvs}
\begin{tabular}{lllllll}
\hline\hline
Object name          & \multicolumn{6}{l}{BJD (days)}                                                \\
\hline
                     & 57845.27      & 57919.16     & 57920.19     & 57926.04    & 57929.19    & ... \\
\hline
CD-41 10989          & -             & -            & -            & -24 $\pm$ 1 & -25 $\pm$ 1 & ... \\
HD 326326            & -             & -32 $\pm$ 1  & -32 $\pm$ 1  & -29 $\pm$ 1 & -33 $\pm$ 1 & ... \\
NGC 6231 41          & -             & -36 $\pm$ 2  & -            & -30 $\pm$ 1 & -34 $\pm$ 1 & ... \\
NGC 6231 33          & -             & -30 $\pm$ 1  & -33 $\pm$ 1  & -27 $\pm$ 1 & -29 $\pm$ 1 & ... \\
NGC 6231 249         & -             & -27 $\pm$ 1  & -27 $\pm$ 1  & -23 $\pm$ 1 & -29 $\pm$ 1 & ... \\
NGC 6231 274         & -             & -26 $\pm$ 1  & -21 $\pm$ 1  & -27 $\pm$ 1 & -28 $\pm$ 1 & ... \\
NGC 6231 274         & -24 $\pm$ 1   & -26 $\pm$ 1  & -21 $\pm$ 1  & -27 $\pm$ 1 & -28 $\pm$ 1 & ... \\
CD-41 11032          & -             & -29 $\pm$ 1  & -29 $\pm$ 1  & -29 $\pm$ 1 & -30 $\pm$ 1 & ... \\
NGC 6231 374         & -             & -33 $\pm$ 1  & -30 $\pm$ 1  & -29 $\pm$ 1 & -35 $\pm$ 1 & ... \\
CPD-41 7730          & -             & -25 $\pm$ 1  & -21 $\pm$ 1  & -22 $\pm$ 1 & -22 $\pm$ 1 & ... \\
NGC 6231 75          & -             & -33 $\pm$ 4  & -35 $\pm$ 4  & -39 $\pm$ 2 & -35 $\pm$ 2 & ... \\
CPD-41 7734          & -             & -27 $\pm$ 1  & -20 $\pm$ 1  & -28 $\pm$ 1 & -22 $\pm$ 1 & ... \\
NGC 6231 235         & -             & -35 $\pm$ 20 & -21 $\pm$ 10 & -31 $\pm$ 6 & -24 $\pm$ 8 & ... \\
CD-41 7736           & -             & -38 $\pm$ 3  & -30 $\pm$ 2  & -32 $\pm$ 1 & -36 $\pm$ 2 & ... \\
NGC 6231 361         & -             & -33 $\pm$ 1  & -            & -34 $\pm$ 1 & -29 $\pm$ 1 & ... \\
NGC 6231 213         & -             & -30 $\pm$ 1  & -            & -           & -30 $\pm$ 1 & ... \\
NGC 6231 213         & -31 $\pm$ 1   & -30 $\pm$ 1  & -            & -           & -31 $\pm$ 1 & ... \\
HD 326339            & -             & -39 $\pm$ 1  & -            & -           & -39 $\pm$ 1 & ... \\
NGC 6231 217         & -             & -37 $\pm$ 2  & -34 $\pm$ 1  & -35 $\pm$ 1 & -37 $\pm$ 1 & ... \\
NGC 6231 108         & -             & -28 $\pm$ 1  & -27 $\pm$ 1  & -26 $\pm$ 1 & -28 $\pm$ 1 & ... \\
TIC 339568125 & -             & -32 $\pm$ 1  & -34 $\pm$ 1  & -33 $\pm$ 1 & -34 $\pm$ 1 & ... \\
NGC 6231 222         & -             & -34 $\pm$ 2  & -26 $\pm$ 1  & -31 $\pm$ 1 & -30 $\pm$ 1 & ... \\
NGC 6231 160         & -             & -26 $\pm$ 2  & -27 $\pm$ 2  & -32 $\pm$ 1 & -33 $\pm$ 1 & ... \\
NGC 6231 96          & -             & -34 $\pm$ 2  & -32 $\pm$ 2  & -32 $\pm$ 1 & -35 $\pm$ 1 & ... \\
NGC 6231 96          & -30 $\pm$ 1   & -34 $\pm$ 2  & -32 $\pm$ 2  & -32 $\pm$ 1 & -35 $\pm$ 1 & ... \\
HD 326332            & -             & -23 $\pm$ 1  & -            & -22 $\pm$ 1 & -24 $\pm$ 1 & ... \\
V* V947 Sco          & -             & -25 $\pm$ 1  & -29 $\pm$ 1  & -32 $\pm$ 1 & -31 $\pm$ 1 & ... \\
V* V920 Sco          & -             & -33 $\pm$ 1  & -25 $\pm$ 1  & -30 $\pm$ 1 & -29 $\pm$ 1 & ... \\
NGC 6231 121         & -             & -30 $\pm$ 2  & -24 $\pm$ 2  & -28 $\pm$ 1 & -26 $\pm$ 1 & ... \\
NGC 6231 142  & -             & -24 $\pm$ 2  & -21 $\pm$ 1  & -26 $\pm$ 1 & -24 $\pm$ 1 & ... \\
NGC 6231 147         & -             & -23 $\pm$ 1  & -16 $\pm$ 1  & -14 $\pm$ 1 & -16 $\pm$ 1 & ... \\
NGC 6231 146         & -             & -38 $\pm$ 2  & -35 $\pm$ 2  & -36 $\pm$ 1 & -37 $\pm$ 1 & ... \\
HD 152076            & -             & -32 $\pm$ 1  & -32 $\pm$ 1  & -30 $\pm$ 1 & -31 $\pm$ 1 & ... \\
CD-41 10990          & -             & -28 $\pm$ 1  & -28 $\pm$ 1  & -29 $\pm$ 1 & -28 $\pm$ 1 & ... \\
HD 326327            & -             & -            & -34 $\pm$ 1  & -35 $\pm$ 1 & -           & ... \\
NGC 6231 14          & -             & -31 $\pm$ 3  & -            & -27 $\pm$ 3 & -24 $\pm$ 3 & ... \\
NGC 6231 284         & -             & -25 $\pm$ 3  & -15 $\pm$ 3  & -19 $\pm$ 2 & -30 $\pm$ 4 & ... \\
CD-41 11031          & -             & -21 $\pm$ 1  & -21 $\pm$ 1  & -21 $\pm$ 1 & -23 $\pm$ 1 & ... \\
NGC 6231 209         & -             & -34 $\pm$ 1  & -38 $\pm$ 1  & -36 $\pm$ 1 & -40 $\pm$ 1 & ... \\
HD 326317            & -             & -30 $\pm$ 1  & -29 $\pm$ 1  & -27 $\pm$ 1 & -28 $\pm$ 1 & ... \\
NGC 6231 173         & -             & -            & -33 $\pm$ 1  & -30 $\pm$ 1 & -33 $\pm$ 1 & ... \\
NGC 6231 172         & -             & -            & -32 $\pm$ 5  & -36 $\pm$ 3 & -37 $\pm$ 4 & ... \\
NGC 6231 175         & -             & -32 $\pm$ 10 & -32 $\pm$ 9  & -28 $\pm$ 4 & -38 $\pm$ 7 & ... \\
HD 326334            & -             & -37 $\pm$ 1  & -35 $\pm$ 1  & -35 $\pm$ 1 & -30 $\pm$ 1 & ... \\
HD 326334            & -22.4 $\pm$ 1 & -37 $\pm$ 1  & -35 $\pm$ 1  & -35 $\pm$ 1 & -30 $\pm$ 1 & ... \\
\hline
\end{tabular}
\end{table*}

\begin{table*}
\centering
\caption{Measured RVs (in \kms) with associated uncertainties for the targets with considered stable in RV (Table \ref{table:nonvariable}). Dashes (-) indicate RV measurements that were not taken either due to the object not being observed at the given date or due to sub-optimal data quality. A full version of this table is available electronically. The first few lines are shown as an example.}
\label{tab:nonvarrvs}
\begin{tabular}{lllllll}
\hline\hline
Object name         & \multicolumn{6}{l}{BJD (days)}                                           \\
\hline
                    & 57845.27 & 57919.16    & 57920.19    & 57926.04      & 57929.19    & ... \\
\hline
NGC 6231 30         & -        & -26 $\pm$ 2 & -28 $\pm$ 2 & -28 $\pm$ 1   & -21 $\pm$ 2 & ... \\
TIC 339565755 & -        & -26 $\pm$ 3 & -26 $\pm$ 3 & -23 $\pm$ 1   & -20 $\pm$ 3 & ... \\
NGC 6231 265        & -        & -24 $\pm$ 7 & -43 $\pm$ 8 & -37 $\pm$ 4   & -37 $\pm$ 6 & ... \\
NGC 6231 243        & -        & -36 $\pm$ 6 & -26 $\pm$ 4 & -31 $\pm$ 2   & -27 $\pm$ 4 & ... \\
NGC 6231 234        & -        & -32 $\pm$ 4 & -           & -26 $\pm$ 2   & -30 $\pm$ 2 & ... \\
NGC 6231 227        & -        & -25 $\pm$ 4 & -29 $\pm$ 3 & -31 $\pm$ 2   & -29 $\pm$ 2 & ... \\
NGC 6231 152        & -        & -28 $\pm$ 2 & -22 $\pm$ 2 & -25 $\pm$ 1   & -26 $\pm$ 2 & ... \\
NGC 6231 123        & -        & -46 $\pm$ 4 & -38 $\pm$ 3 & -33.6 $\pm$ 1 & -37 $\pm$ 1 & ... \\
NGC 6231 127        & -        & -24 $\pm$ 4 & -22 $\pm$ 4 & -23 $\pm$ 1   & -25 $\pm$ 2 & ... \\
NGC 6231 24         & -        & -31 $\pm$ 2 & -30 $\pm$ 2 & -26.5 $\pm$ 1 & -25 $\pm$ 2 & ... \\
NGC 6231 194        & -        & -22 $\pm$ 7 & -15 $\pm$ 5 & -24 $\pm$ 3   & -29 $\pm$ 4 & ... \\
NGC 6231 165        & -        & -35 $\pm$ 9 & -           & -45 $\pm$ 7   & -33 $\pm$ 6 & ... \\
\hline
\end{tabular}
\end{table*}

\pagebreak
\newpage

\end{document}